\documentclass{article}

\usepackage{amssymb}
\usepackage{amsthm}
\usepackage{fancyhdr}
\usepackage{authblk}
\usepackage[title]{appendix}
\usepackage{array}
\usepackage[tableposition=top]{caption}
\usepackage{dsfont}
\usepackage{graphicx}
\usepackage{mathtools}
\usepackage{multicol}
\usepackage{multirow}
\usepackage[square]{natbib}
\usepackage{pgfplots}
\usepackage{tikz}
\usepackage{xcolor}
\usepackage{enumitem}
\setlist[itemize]{noitemsep}
\setlist[enumerate]{noitemsep}
\usepackage{arxiv}
\usepackage{booktabs}       % professional-quality tables
\usepackage{nicefrac} 

\usepackage[OT1]{fontenc}

\usepackage{url}
\usepackage{hyperref}
\usepackage[hyphenbreaks]{breakurl}

\usepackage[capitalise]{cleveref}

\usepackage{amsmath}
\usepackage{arxiv}

\pgfplotsset{compat=1.4}

\pgfplotsset{
    jitter/.style={
        x filter/.code={\pgfmathparse{\pgfmathresult+(rnd-.5)*#1}}
    },
    jitter/.default=0.5
}

\usetikzlibrary{hobby, graphs, decorations.markings, arrows, shapes, positioning, mindmap, backgrounds, calc, matrix, patterns, bayesnet, patterns.meta}

\usepgfplotslibrary{statistics, fillbetween, external}

\newtheorem{remark}{Remark}

\newcommand{\ind}{\mathds 1}

\newcommand{\indep}{\perp \!\!\! \perp}

\newcommand{\nA}{n^{ \mathcal A }}

\newcommand{\nB}{n^{ \mathcal B }}

\newcommand{\tA}{\textbf{t}^{ \mathcal A }}

\newcommand{\tB}{\textbf{t}^{ \mathcal B }}

\newcommand{\gAik}{g^{ \mathcal A }_{i, k}}

\newcommand{\gBjk}{g^{ \mathcal B }_{j, k}}

\newcommand{\hAik}{h^{ \mathcal A }_{i, k}}

\newcommand{\hBjk}{h^{ \mathcal B }_{j, k}}

\newcommand{\GAik}{G^{ \mathcal A }_{i, k}}

\newcommand{\GBjk}{G^{ \mathcal B }_{j, k}}

\newcommand{\HAik}{H^{ \mathcal A }_{i, k}}

\newcommand{\HBjk}{H^{ \mathcal B }_{j, k}}

\definecolor{drawcolor}{cmyk}{.80,.29,.05,0}

\begin{document}

\title{ A flexible model for Record Linkage }

\renewcommand\Authsep{, }
\renewcommand\Authands{, and }

\author{Kayan\'{e} Robach\thanks{Corresponding Author: \texttt{k.c.robach@amsterdamumc.nl}}}

\author{St\'{e}phanie L.\ van der Pas}

\author{Mark A.\ van de Wiel}

\author{Michel H.\ Hof}

\affil{Department of Epidemiology and Data Science, \protect\\ Amsterdam UMC location Vrije Universiteit Amsterdam, De Boelelaan 1117, 1081 HV Amsterdam, The Netherlands}
\affil{Amsterdam Public Health, Methodology, The Netherlands}

\maketitle

\abstract{Combining data from various sources empowers researchers to explore innovative questions, for example those raised by conducting healthcare monitoring studies. However, the lack of a unique identifier often poses challenges. Record linkage procedures determine whether pairs of observations collected on different occasions belong to the same individual using partially identifying variables (e.g.\ birth year, postal code). Existing methodologies typically involve a compromise between computational efficiency and accuracy. Traditional approaches simplify this task by condensing information, yet they neglect dependencies among linkage decisions and disregard the one-to-one relationship required to establish coherent links. Modern approaches offer a comprehensive representation of the data generation process, at the expense of computational overhead and reduced flexibility. We propose a flexible method, that adapts to varying data complexities, addressing registration errors and accommodating changes of the identifying information over time. Our approach balances accuracy and scalability, estimating the linkage using a Stochastic Expectation Maximisation algorithm on a latent variable model. We illustrate the ability of our methodology to connect observations using large real data applications and demonstrate the robustness of our model to the linking variables quality in a simulation study. The proposed algorithm \textit{FlexRL} is implemented and available in an open source \texttt{R} package.}

\section{Introduction}

Record linkage aims to combine information of the same individuals from multiple data sources. E.g.\ in medical contexts, those methods offer a comprehensive view of patient histories, treatment outcomes, and disease progression. Although this task is trivial when a unique identifier is available, data often lack such identifier due to privacy regulations or because the data was not originally collected with a view to record linkage. In this situation, the record linkage must rely on partially identifying variables (PIVs) to identify the pairs of observations belonging to the same individual (i.e.\ links) and monitor the data. Examples of these variables are initials, birth year, and place of residence. In general, those variables are prone to errors and have restricted discriminating power due to a limited number of unique values. In addition, some PIVs are unstable and evolve over time (e.g.\ due to moving). Especially when combining longitudinal data, it is important to deal with this instability. 

\subsection*{State-of-the-art}

Record linkage was originally formalised as a mixture model to classify pairs of records as links or non-links based on the comparison of their PIVs, \citep{bookoflife, Newcombe1959, Tepping1968, Fellegi_RL_1969}. While this traditional method is pervasive due to its practicality, \citep{Larsen_mixture_1999, hof_linkage_2012, guha_bay_causal_RL_2022}, it entails a reduction of the information contained in the PIVs, requires a Cartesian product of data sources, which limits scalability and neglects dependencies among linkage decisions. To deal with this last problem, post-hoc methods have been proposed that restrict the linkage by imposing structural constraints into the model \citep{jaro_RL_matching_1989, fortini_bay_RL_2001, sadinle_bay_bipartite_RL_2017}. Furthermore, this approach does not carry the linkage uncertainty over to study outcomes.

A popular approach instead, is to model the processes through which records are generated and cluster the records to the latent entities they depict, \citep{tancredi_bay_RL_2011, steorts2015, smered2016, dblink_2021, marchantsteorts2023}. This framework supports record linkage across multiple sources while also addressing de-duplication. Recent Bayesian methods developed for this task tackle the issues of the traditional approach though they are computationally intensive due to their complexity and require significant memory to process large graphs, \citep{smered2016, marchantsteorts2023}. This load can be mitigated by blocking, contributing to computational efficiency at the expense of some intrinsic flexibility, \citep{blocking_methods, sadinle_bay_bipartite_RL_2017, dblink_2021}. In an extensive review of the field, \citep{binette2022almost} reference a wide array of existing literature, open-source software programs, and data sets.

\subsection*{Motivations and scope}

% what still lack: scalable method / unstable PIVs / robust method to high level of distorsion
The literature methods often illustrate their performance on small real data examples using strong PIVs. Moreover, the open-source software developed require substantial computational resources. Those methodologies are not scalable and may not be applicable in real-life situations. To our knowledge, none of the developed method tackle the issue of dynamic PIVs which can change over time, though they are often the strongest one available (with high discriminating power).

% contributions
As a solution to the remaining problems, we introduce a new frequentist method, based on the PIVs generation mechanism using a Stochastic Expectation Maximisation algorithm. We address the record linkage problem in real settings where one has to rely on weak partial information to pair observations. Thus we accommodate a wide diversity of registration processes, including inaccuracies and missing values and most notably, we introduce a new approach to handle time-varying variables like place of residence, particularly important for establishing links in follow-up data from longitudinal health studies. Our method focuses on linking records across two data sources, relating individual characteristics to their underlying truth and establishing connections between latent representations of both sources through bipartite matching. We address the dual challenge of consistently modelling complex data structures while providing a scalable algorithm that can handle large data sources on a standard computer. By dint of our methodology, researchers can explore risk factors for certain diseases within a family context by linking household and medical data, \citep{pubhealthlinkage, menezes2024hausdorff}. Similarly, by linking prenatal and paediatric records, it becomes possible to investigate maternal and child health dynamics, \citep{hof_RL_survival_2017}.

In order to evaluate a record linkage method, sets containing realistically weak identifying information as well as a unique identifier are required. Many data sets, in particular in healthcare, cannot include patient numbers for confidentiality reasons. We therefore explore the relevance of modelling PIVs dynamics as part of a simulation study that aims to reproduce the real context of longitudinal medical data. We use as a case study the National Long Term Care Survey (NLTCS) with data from 1982 and 1994 and we provide another illustration on the Survey of Household Income and Wealth (SHIW) in the supplementary material. These data sets are often-used in the record linkage literature since they provide a unique identifier, enabling us to compare and evaluate our algorithm with the existing methods, \citep{rl_nltcs_medical, smered2016, dblink_2021, guha_bay_causal_RL_2022, binette2022almost, menezes2024hausdorff, bayRLFabl}. 

We evaluate the record linkage methods using two criteria. First, the error when linking pairs of records, by the False Discovery Rate (FDR) and second, the ability to detect pairs pertaining to the same individual, by the sensitivity. Due to the low discriminating power of the PIVs in real data, there are a lot of similar records that do not belong to the same individuals. Therefore, it is important to build methods that allow control on the incorrectly linked pairs e.g.\ by FDR, as they can ruin subsequent inference, \citep{RLreg2005, RLregcox2010, RLreg2012}. 

In a nutshell, we build a scalable and robust record linkage method that can handle unstable PIVs, provides acceptable FDR and allows for uncertainty propagation.

\subsection*{Plan}

We provide an overview of the problem and introduce the notations in \cref{sec2: problem statement and model formulation}. We illustrate the relationships between our model components in \cref{fig_PGM_FlexRL} and delve into the details of our statistical model in \cref{sec3: pivs} and in \cref{sec4: linkage}. We present the Stochastic Expectation Maximisation algorithm we develop to estimate the model parameters by Maximum Likelihood and perform the record linkage task in \cref{sec5: the stem algorithm}. Finally, we demonstrate the effectiveness of our method through simulations and real data applications in \cref{sec6: simulations and applications}.

\section{Problem statement and model formulation}\label{sec2: problem statement and model formulation}

Suppose we have access to two files, $\mathcal A$ and $\mathcal B$, containing two overlapping random samples of size $\nA$ and $\nB$, coming from the same population. We assume that no unique identifier is available in both files to perfectly identify the entity to whom a record belongs to. Instead, $K$ Partially Identifying Variables (PIVs) have been registered in both files. Examples of these PIVs are birth year or postal code. 

For record $i$ from $\mathcal{A}$ and record $j$ from $\mathcal{B}$, we denote the registered values of the PIVs as \smash{$\textbf{G}_{i}^{\mathcal{A}} = \big\{ G_{i, 1}^{\mathcal{A}}, G_{i, 2}^{\mathcal{A}}, \dots, G_{i, K}^{\mathcal{A}} \big\}$} and \smash{$\textbf{G}_{j}^{\mathcal{B}} = \big\{ G_{j, 1}^{\mathcal{B}}, G_{j, 2}^{\mathcal{B}}, \dots, G_{j, K}^{\mathcal{B}} \big\}$}. We consider the registered PIVs to be distorted versions of the underlying true values, analogously denoted by $\textbf{H}_{i}^{\mathcal{A}}$ and $\textbf{H}_{j}^{\mathcal{B}}$. Due to errors and missing values, the true and registered values can differ. 

Without loss of generality, we assume that file $\mathcal{B}$ contains more observations than file $\mathcal{A}$, i.e.\ $\nB \geq \nA$. Therefore, for each record in $\mathcal{A}$ we seek for a potential record in $\mathcal{B}$ to form a link with. 

To determine whether pairs of records belong to the same entity of not, we define $\boldsymbol{\Delta}$ as the latent indicator linkage matrix of size $(\nA \times \nB)$. This matrix, also known as `matching indicator' or `matching matrix' in the literature, is given by $$\boldsymbol{\Delta}=
\begin{pmatrix}
 \Delta_{1,1} & \Delta_{1,2} & \dots & \Delta_{1,\nB} \\
 \Delta_{2,1} & \Delta_{2,2} & \dots & \Delta_{2,\nB} \\
 \vdots & \vdots & \ddots & \vdots \\
 \Delta_{\nA,1} & \Delta_{\nA,2} & \dots & \Delta_{\nA,\nB} 
\end{pmatrix},$$ where $\Delta_{i,j}=1$ if the $i^{th}$ record from file $\mathcal A$ and the $j^{th}$ record from file $\mathcal B$ belong to the same entity (i.e.\ a link) and $\Delta_{i,j}=0$ if they belong to different individuals (i.e.\ a non-link). Note that this matrix is unobserved and its estimation is of primary interest with record linkage.

In most situations, each entity maximally has one observation in each file. This constraint, often observed in record linkage \citep{tancredi_bay_RL_2011, sadinle_bay_bipartite_RL_2017}, implies that each observation can be part of at most one link. Given this feature, the possible configurations of $\boldsymbol{\Delta}$ are given by the set:
\begin{equation}\label{def: oneone}
\mathcal D = \Big\{ \boldsymbol{\Delta}: \Delta_{i,j} \in \{0,1\}, \sum_{i=1}^{\nA} \Delta_{i,j} \leq 1 \text{ for all } j \in \{1, \dots, \nB\} \text{ and} \sum_{j=1}^{\nB} \Delta_{i,j} \leq 1 \text{ for all } i \in \{1, \dots, \nA\} \Big\}.
\end{equation}

We denote by $\textbf{G}^{\mathcal{A}}, \textbf{G}^{\mathcal{B}}$ (and respectively $\textbf{H}^{\mathcal{A}}, \textbf{H}^{\mathcal{B}}$) the vectors of the registered values (latent true values) of all records. In addition, let $\tA = \{t^{\mathcal A}_1, \ldots, t^{\mathcal A}_{n^{\mathcal A}} \}$ and $\tB = \{t^{\mathcal B}_1, \ldots, t^{\mathcal B}_{n^{\mathcal B}} \}$ be the dates of registration of all records. These dates can be used in the conditional distribution of true values for linked records, enabling to model dynamics of the PIVs over time.

\begin{figure}[!ht]
    \centering
    \begin{tikzpicture}
        % Nodes
        \node[draw, minimum size=1cm] (gamma) at (0,4) {$\gamma$};
        \node[shape=circle, dashed, draw, minimum size=1cm] (delta) at (0,2) {$\boldsymbol{\Delta}$};
        \node[draw, minimum size=1cm] (eta) at (0,0) {$\boldsymbol{\eta}$};
        \node[draw, minimum size=1cm] (alpha) at (0,-2) {$\boldsymbol{\alpha}$};
        \node[shape=circle, dashed, draw, minimum size=1cm] (HA) at (-3,-2) {$\textbf{H}^{\mathcal{A}}$};
        \node[shape=circle, dashed, draw, minimum size=1cm] (HB) at (3,-2) {$\textbf{H}^{\mathcal{B}}$};
        \node[draw, minimum size=1cm] (phi) at (0,-4) {$\boldsymbol{\phi}$};
        \node[shape=circle, draw, minimum size=1cm] (GA) at (-4.5,-4) {$\textbf{G}^{\mathcal{A}}$};
        \node[shape=circle, draw, minimum size=1cm] (GB) at (4.5,-4) {$\textbf{G}^{\mathcal{B}}$};
        % Directed edges
        \path [-stealth] (gamma) edge (delta);
        \path [-stealth] (delta) edge (HA);
        \path [-stealth] (delta) edge (HB);
        \path [-stealth] (eta) edge (HA);
        \path [-stealth] (eta) edge (HB);
        \path [-stealth] (alpha) edge (HA);
        \path [-stealth] (alpha) edge (HB);
        \path [-stealth] (HA) edge (GA);
        \path [-stealth] (HB) edge (GB);
        \path [-stealth] (phi) edge (GA);
        \path [-stealth] (phi) edge (GB);
        % Plates
        \plate [inner sep=.5cm, yshift=.2cm] {data A} {(HA)(GA)} {$i = 1, \dots, \nA$};
        \plate [inner sep=.5cm, yshift=.2cm] {data B} {(HB)(GB)} {$j = 1, \dots, \nB$};
        \plate [inner sep=.25cm, yshift=.2cm] {data linked} {(HA)(alpha)(HB)} {$(i,j)$};
    \end{tikzpicture}
    \caption{Probabilistic graphical model for the decomposition of the data generation process illustrating the record linkage problem. Circles refer to random variables while squares are reserved for parameters. Dotted lines indicate unobserved latent variables and solid lines observables. The three plates represent data from $\mathcal{A}$, data from $\mathcal{B}$ and their overlapping set.}
    \label{fig_PGM_FlexRL}
\end{figure}
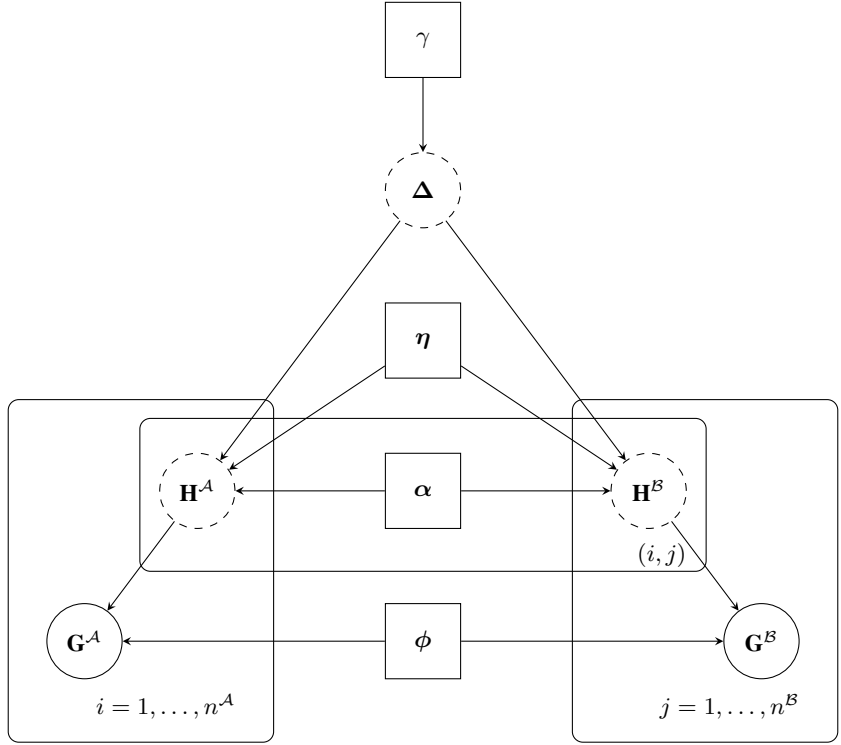

To specify the corresponding complete data likelihood, we assume that the registered values of the partially identifying variables are independent of the linkage decisions given its true values. In addition, we assume that the registration processes in data sources $\mathcal{A}$ and $\mathcal{B}$ are independent of each other. Given these assumptions, we have \begin{align*}
    \mathcal{L}_{\boldsymbol{\theta}} \big( \textbf{G}^{\mathcal{A}}, \textbf{G}^{\mathcal{B}}, \textbf{H}^{\mathcal{A}}, \textbf{H}^{\mathcal{B}}, \textbf{t}^{\mathcal{A}}, \textbf{t}^{\mathcal{B}}, \boldsymbol{\Delta} \big) = \mathcal{L}_{\boldsymbol{\phi}} \big( \textbf{G}^{\mathcal{A}} \bigm| \textbf{H}^{\mathcal{A}} \big) \times \mathcal{L}_{\boldsymbol{\phi}} \big( \textbf{G}^{\mathcal{B}} \bigm| \textbf{H}^{\mathcal{B}} \big)
    \times \mathcal{L}_{\boldsymbol{\alpha}} \big( \textbf{H}^{\mathcal{B}} \bigm| \textbf{H}^{\mathcal{A}}, \textbf{t}^{\mathcal{A}}, \textbf{t}^{\mathcal{B}}, \boldsymbol{\Delta} \big) \times \mathcal{L}_{\boldsymbol{\eta}} \big( \textbf{H}^{\mathcal{A}} \big)
    \times \mathcal{L}_{\gamma} \big( \boldsymbol{\Delta} \big),
\end{align*} where $\boldsymbol{\theta} \coloneqq \big\{ \gamma, \boldsymbol{\eta}, \boldsymbol{\alpha}, \boldsymbol{\phi} \big\}$ gather the model parameters. See Figure \cref{fig_PGM_FlexRL} for a graphical representation of the model. All four parts of the complete data likelihood function are now discussed in more detail. 

\section{Partially identifying variables}
\label{sec3: pivs}

\subsection{Modeling PIVs distribution: assumptions and pre-processing} 
\label{subsec30: modeling pivs distribution: assumptions and pre-processing}

We assume each PIV indexed by $k$ to be categorical with $n_{k}$ possible values (all unique values observed in $\mathcal{A}$ or $\mathcal{B}$). We henceforth map their categorical support to the set of natural numbers $\{1, \dots, n_k\}$. For instance, if the variable represents names with four possible values $\{\text{kayan\'{e}}, \text{st\'{e}phanie}, \text{mark}, \text{michel}\}$, we assign this set to the numerical range $\{1, 2, 3, 4\}$. 

In the context of text content, an additional pre-processing step is necessary. By encoding text-based values into numerical values, some (valuable) information that is contained in the text is lost. To mitigate this loss, we propose to map the text-based values to numerical values using soundex code. With this additional step, similar semantic content like `mark' and `marc', are mapped to the same value \citep{RusselSoundexCode1, RusselSoundexCode2}.

We assume the true values of the PIVs to be independent of each other, since it is unlikely that variables such as gender, place of residence and birth year have dependent distributions. In \cref{remark: correlation PIVs}, we elaborate further on this assumption. In addition, since both files are assumed to be overlapping random samples from the same population, we assume that their distributions are the same in both files. Based on these assumptions, it is possible to characterise their distributions with the vector $\boldsymbol{\eta} \coloneqq \big( \boldsymbol{\eta}_1, \boldsymbol{\eta}_2, \dots, \boldsymbol{\eta}_{K} \big)$, where the distribution of the $k^{\text{th}}$ PIV is represented by the vector $\boldsymbol{\eta}_{k}$ of length $n_k$. The probability of observing a value $\ell \in \{1, \dots, n_k\}$ in the $k^{\text{th}}$ PIV is denoted $\eta_{k,\ell}$: \begin{align} \label{model: h}
    \mathbb{P} \Big( \HAik = \ell ; \boldsymbol{\eta} \Big) = \mathbb{P} \Big( \HBjk = \ell ; \boldsymbol{\eta} \Big) = \eta_{k,\ell}.
\end{align}

\begin{remark}\label{remark: correlation PIVs}
As strong dependencies between PIVs may exist in practice, we conducted a sensitivity analysis (see Appendix \cref{app3: assumptions deviations}) to assess the impact of deviations from the independence assumption. The analysis shows a gradual decline in performance as deviations increase. The results also suggest that incorporating PIVs dynamics improves the method robustness to dependencies among PIVs. However, strong associations between variables can pose challenges when sampling latent variables. For instance, generating data that indicates a person lives in New York City while their state is listed as Texas can compromise the model and increase the FDR substantially. A solution to deal with such strong hierarchical relationship is to combine these variables into a single PIV.
\end{remark}

\subsection{True values dynamics for unstable variables} 
\label{subsec31: true values dynamics for unstable variables}

It is natural to think that if two records from $\mathcal{A}$ and from $\mathcal{B}$ belong to the same individual, their information should match. However, some PIVs may change over time; place of residence or marital status for instance can evolve inducing different true values for each file. Thus, a PIV indexed by $k$ is defined as unstable when \begin{align*}
    \forall (i,j) \in \{1, \dots, \nA\} \times \{1, \dots, \nB\} \text{ such that } \Delta_{i,j}=1, \mathbb{P}\big(\HAik \neq \HBjk\big) > 0.
\end{align*} Otherwise, the probability of true latent values being equal is one and the $k^{\text{th}}$ PIV is considered stable. 

Therefore, in each of the two records pertaining to the same individual, the unstable PIV indexed by $k$ may have distinct values depending on the time elapsed between the two data collections. It is natural to parametrized the probability that those true values for a pair of linked records $(i,j)$ coincide using a survival function. We recommend keeping the model simple, as changes over time are latent and therefore challenging to capture accurately. For example, it is possible to use an exponential survival model with a positive constant baseline hazard \smash{$\lambda_k(t) = \exp(\alpha_k)$}: \begin{equation} \label{def: survival}
    S_{\alpha_k}(t_{i,j}) \coloneqq \mathbb{P}(\HAik = \HBjk \mid t_{i,j}, \Delta_{i,j} = 1 ; \boldsymbol{\alpha}) = \exp \bigg\{- \int_0^{t_{i,j}} \lambda_k(t) dt \bigg\} = \exp\big\{ - \exp(\alpha_k) \, t_{i,j} \big\},
\end{equation} where the registration time difference between the compared records is denoted $t_{i,j} \coloneqq | t_j^{\mathcal{B}} - t_i^{\mathcal{A}} |$%, using the vectors of registration times $\textbf{t}^{\mathcal{A}}$ and $\textbf{t}^{\mathcal{B}}$. 
That way, the probability that the values are the same is maximal when the registration time difference is null and goes to zero as the time difference increases. By modelling $\alpha_k$ in the developed method, we model the log baseline hazard. 

\begin{remark}\label{remark: survival with covariates}
    Assuming proportional hazards, it is straightforward to extend this model with covariates $\boldsymbol{X}$ registered in one of both files with the following conditional hazard function: $$\int_0^{t_{i,j}} \lambda_k(t \mid \boldsymbol{X}_{i,j} = \boldsymbol{x}_{i,j}) dt = \exp(\boldsymbol{x}_{i,j}^T \boldsymbol{\beta}_k) \int_0^{t_{i,j}} \lambda_k(t) dt = \exp(\alpha_k + \boldsymbol{x}_{i,j}^T \boldsymbol{\beta}_k) \, t_{i,j}.$$ If one wants to model relocation flows to deal with the instability of the variable postal code, the observed covariates associated to a pair of linked records, denoted by $\boldsymbol{x}_{i,j}$, may include as an example the income, the number of children, or the age. 
\end{remark}

When values change through time, we assume they have substituted to one of the $n_{k}-1$ other equally likely possible values in the support. Additional knowledge on the possible changes can be used to consider non-uniform weights across the support following \cref{remark: newweights}. Moreover, we assume the distribution of the new values to remain unchanged. For a linked pair of records $(i,j)$ with latent PIVs $\hAik, \hBjk \in \{1, \dots, n_k\}$ and registration time difference $t_{i,j}$, we therefore have the following decomposition of the joint distribution: \begin{align} \label{model: hh}
& \mathbb{P} \Big( \HAik = \hAik, \HBjk = \hBjk \bigm| t_{i,j}, \Delta_{i,j} = 1 ; \boldsymbol{\alpha}, \boldsymbol{\eta} \Big) \\
& \: = \mathbb{P} \Big( \HAik = \hAik ; \boldsymbol{\eta} \Big) \cdot \mathbb{P} \Big( \HBjk = \hBjk \bigm| \HAik = \hAik, t_{i,j}, \Delta_{i,j} = 1 ; \boldsymbol{\alpha} \Big) \nonumber \\
& \: \: \: \: = \bigg\{ \eta_{k,\ell}^{\ind{\{\hAik = \ell\}}} \cdot {\ind{\{\hAik = \hBjk\}}} \bigg\}^{\ind{\{k^{\text{th}} \text{ PIV stable}\}}} \bigg\{ \eta_{k,\ell}^{\ind{\{\hAik = \ell\}}} \cdot {S_{\alpha_k}(t_{i,j}) \vphantom{\bigg)}}^{\ind{\{\hAik=\hBjk\}}} \bigg( \frac{ 1 - S_{\alpha_k}(t_{i,j})}{n_{k}-1} \bigg)^{ \ind{\{\hAik \neq \hBjk\}} } \bigg\}^{\ind{\{k^{\text{th}} \text{ PIV unstable}\}}}. \nonumber \end{align} The stability of a PIV requires matching information in the linked records, ensured through a blocking mechanism on the latent level. Thus, only records within the same (latent) block, characterized by identical latent true values for stable PIVs, are compared. This strategy separates potential links from non-links, \citep{jaro_RL_matching_1989, sadinle_bay_bipartite_RL_2017, dblink_2021}. In the literature, blocking is often used on unstable PIVs assuming no registration errors, \citep{jaro_proba_linkage_1995, tancredi_bay_RL_2011}. By accommodating unstable PIVs instead, we provide an alternative to traditional blocking, enhancing the flexibility of our model.

Using the latent true values models in \cref{model: h} and in \cref{model: hh}, we compute the likelihood contribution of the three independent subgroups of observations: linked pairs, non-linked units from $\mathcal{A}$ and non-linked units from $\mathcal{B}$ as follows: \begin{align*}
    \mathcal{L}_{\boldsymbol{\eta}} \big( \textbf{H}^{\mathcal{A}} \big) \times \mathcal{L}_{\boldsymbol{\alpha}} & \big( \textbf{H}^{\mathcal{B}} \bigm| \textbf{H}^{\mathcal{A}}, \textbf{t}^{\mathcal{A}}, \textbf{t}^{\mathcal{B}}, \boldsymbol{\Delta} \big)\\
    = \prod\limits_{i = 1}^{\nA} \prod\limits_{j = 1}^{\nB} \bigg[ \prod\limits_{k = 1}^{K} \bigg\{ & \prod\limits_{\ell = 1}^{n_{k}} \eta_{k,\ell}^{\ind{\{\hAik = \ell\}}} \cdot {\ind{\{\hAik = \hBjk\}}} \bigg\}^{\ind{\{k^{\text{th}} \text{ PIV stable}\}}}\\
    \qquad \qquad \cdot \bigg\{ & \prod\limits_{\ell = 1}^{n_{k}} \eta_{k,\ell}^{\ind{\{\hAik = \ell\}}} \cdot {S_{\alpha_k}(t_{i,j}) \vphantom{\bigg)}}^{\ind{\{\hAik=\hBjk\}}} \bigg( \frac{ 1 - S_{\alpha_k}(t_{i,j})}{n_{k}-1} \bigg)^{ \ind{\{\hAik \neq \hBjk\}} } \bigg\}^{\ind{\{k^{\text{th}} \text{ PIV unstable}\}}} \bigg]^{\Delta_{i,j}} \\
    \times \prod\limits_{i = 1}^{\nA} \bigg[ \prod\limits_{k = 1}^{K} \hphantom{\bigg\{} & \prod\limits_{\ell = 1}^{n_{k}} \eta_{k,\ell}^{\ind{\{\hAik = \ell\}}} \bigg]^{1 - \sum_{j = 1}^{\nB} \Delta_{i,j}}\\
    \times \prod\limits_{j = 1}^{\nB} \bigg[ \prod\limits_{k = 1}^{K} \hphantom{\bigg\{} & \prod\limits_{\ell = 1}^{n_{k}} \eta_{k,\ell}^{\ind{\{\hBjk = \ell\}}} \bigg]^{1 - \sum_{i = 1}^{\nA} \Delta_{i,j}},
\end{align*} which determines the candidate pairs of records eligible for linkage.

\subsection{Registration errors} 
\label{subsec32: registration errors}

Similarly to \citep{tancredi_bay_RL_2011}, \citep{smered2016}, and \citep{marchantsteorts2023}, we treat registered values of PIVs as distorted versions of the true unobserved values. These distortions capture registration errors, referring to any disagreement between an observed value and its underlying truth. Our approach processes categorical PIVs and treats all numerical PIVs as discrete. Text content variables require a preliminary processing step. For such data, we apply a soundex code transformation, \citep{RusselSoundexCode1, RusselSoundexCode2}. This transformation ensures that similar values are associated with the same code and eliminates typographical errors. Subsequently, the support of true values should align with the support constructed from the encoded registered values.

Hence we distinguish between two types of registration errors: missing values and mistakes. Treating missing values as mistakes would underestimate the probability of records forming a link and overestimate the discriminating strength of the PIVs. Therefore, we define the support of true values with a set of natural numbers starting at `1' and we encode missing values with a `0'. For any PIV indexed by $k$ and any record $i$, a missing value corresponds to \smash{$\big\{ \GAik = 0 \big\}$} while a mistake squares with \smash{$\big\{ \GAik \neq 0 \big\} \cap \big\{ \GAik \neq \HAik \big\}$} (and similarly for any record $j$ in $\mathcal{B}$). Such discrepancy between registered and true values illustrates a scenario where the encoded observed value would totally differ from the truth, as typographical errors are already addressed in a preliminary step. Although the method limits the type of errors we can handle, it is rarely necessary in a real setting where one has only access to categorical data and, is shows good results in the RLData500 application available in the supplementary material.

When conditioning on true values, for each PIV we assume the possible mistakes to be equally likely and, we presume that the probability of mistake is identical in both data sets. We postulate that missing values among PIVs happen completely at random though the probability of missing values may differ among the different PIVs and data sources. If the registered value is not missing, either it agrees with the true latent value drawn or not. If not, the registered value---which is a distorted version of the truth---has substituted to one of the $n_{k}-1$ other possible values in the support (that we suppose equiprobable). The parameter $\boldsymbol{\phi}$ governs the possible distortion mechanisms at the source of the differences between observed and true values; we index its coordinates with `missing' or `mistake' to specify the kind of registration error considered. Thus we can explicitly model the relationship between observed value $\gAik \in \{0, 1, \dots, n_k\}$ and latent true value $\hAik \in \{1, \dots, n_k\}$ for a record $i$ in $\mathcal{A}$ by: \begin{align} \label{model: g|h}
    & \mathbb{P} \Big( \GAik = \gAik \bigm| \HAik = \hAik ; \boldsymbol{\phi} \Big)\\
    & \qquad = \bigg\{ \phi_{k,\text{missing}}^{\mathcal{A}} \bigg\}^{\ind{\{\gAik = 0\}}} \cdot \bigg\{ \bigg( 1 - \phi_{k,\text{missing}}^{\mathcal{A}} \bigg) \cdot \bigg( 1 - \phi_{k,\text{mistake}}^{\mathcal{A}} \bigg)^{\ind{\{\gAik = \hAik\}}} \cdot \bigg( \frac{\phi_{k,\text{mistake}}^{\mathcal{A}}}{n_{k}-1} \bigg)^{\ind{\{\gAik \neq \hAik\}}} \bigg\}^{\ind{\{\gAik \neq 0\}}} \nonumber
\end{align} and similarly for $\mathbb{P}( \GBjk = \gBjk \bigm| \HBjk = \hBjk ; \boldsymbol{\phi} )$ in file $\mathcal{B}$ since registration processes are the same.  

From the registered values model in \cref{model: g|h} for two independent subgroups: observations from $\mathcal{A}$ and observations from $\mathcal{B}$, we derive the likelihood contribution: \begin{align*}
    & \mathcal{L}_{\boldsymbol{\phi}} \big( \textbf{G}^{\mathcal{A}} \bigm| \textbf{H}^{\mathcal{A}} \big) \times \mathcal{L}_{\boldsymbol{\phi}} \big( \textbf{G}^{\mathcal{B}} \bigm| \textbf{H}^{\mathcal{B}} \big)\\
    & \qquad = \prod\limits_{i=1}^{\nA} \prod\limits_{k=1}^{K} \bigg\{ \phi_{k,\text{missing}}^{\mathcal{A}} \bigg\}^{\ind{\{\gAik = 0\}}} \cdot \bigg\{ \bigg( 1 - \phi_{k,\text{missing}}^{\mathcal{A}} \bigg) \cdot \bigg( 1 - \phi_{k,\text{mistake}}^{\mathcal{A}} \bigg)^{\ind{\{\gAik = \hAik\}}} \cdot \bigg( \frac{\phi_{k,\text{mistake}}^{\mathcal{A}}}{n_{k}-1} \bigg)^{\ind{\{\gAik \neq \hAik\}}} \bigg\}^{\ind{\{\gAik \neq 0\}}} \\
    & \qquad \times \prod\limits_{j=1}^{\nB} \prod\limits_{k=1}^{K} \bigg\{ \phi_{k,\text{missing}}^{\mathcal{B}} \bigg\}^{\ind{\{\gBjk = 0\}}} \cdot \bigg\{ \bigg( 1 - \phi_{k,\text{missing}}^{\mathcal{B}} \bigg) \cdot \bigg( 1 - \phi_{k,\text{mistake}}^{\mathcal{B}} \bigg)^{\ind{\{\gBjk = \hBjk\}}} \cdot \bigg( \frac{\phi_{k,\text{mistake}}^{\mathcal{B}}}{n_{k}-1} \bigg)^{\ind{\{\gBjk \neq \hBjk\}}} \bigg\}^{\ind{\{\gBjk \neq 0\}}}.
\end{align*}

\begin{remark}\label{remark: newweights}
    One can specify joint probabilities for \smash{$\mathbb{P} \big(\HAik, \HBjk | \HAik \neq \HBjk, \Delta_{i,j} = 1 \big)$} that can be used in \cref{model: hh} instead of the uniform weights that we assign to the possible changes between the true values through time. Similarly, this can be done for the possible mistakes in \cref{model: g|h}. We studied the sensitivity of our modelling against actual non-uniform changes in the data. We observe that the FDR level is well maintained, while sensitivity decreases in the misspecified case but remains higher than that of existing methods (see Appendix \cref{app3: assumptions deviations} for more detail).
\end{remark}

Note that, the instability of a PIV indexed by $k$ results in differences between the true latent values generated along the method, i.e.\ for a pair of records $(i,j)$ each true value in $\mathcal{A}$ or $\mathcal{B}$ matches the registered value \smash{$\gAik = \hAik$} and \smash{$\gBjk = \hBjk$}, while true values between files $\mathcal{A}$ and $\mathcal{B}$ changed \smash{$\hAik \neq \hBjk$}. In contrast, mistakes occur when the values registered (when not missing) differ from the true latent values associated. It happens when for any record $i$ in $\mathcal{A}$: \smash{$\gAik \neq \hAik$} or for any record $j$ in $\mathcal{B}$: \smash{$\gBjk \neq \hBjk$}. The observable consequence of those two mechanisms is the same: registered values of the pair of records disagree.

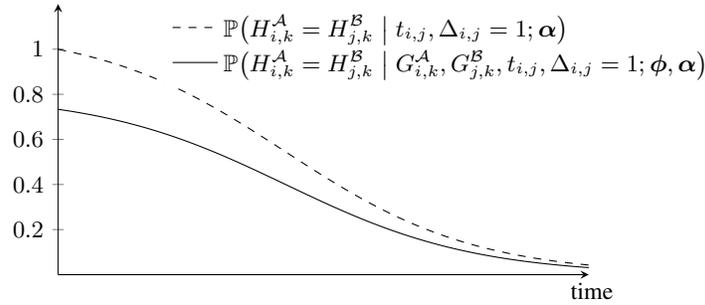
\begin{figure}[!ht]
  \centering
  \begin{tikzpicture}
  \pgfplotsset{
    width=0.5\textwidth,
    height=0.3\textwidth}
    \begin{axis}[
      legend entries={${ \mathbb{P} \big( \HAik = \HBjk \bigm| t_{i,j}, \Delta_{i,j} = 1 ; \boldsymbol{\alpha} \big) }$, ${ \mathbb{P} \big( \HAik = \HBjk \bigm| \GAik, \GBjk, t_{i,j}, \Delta_{i,j} = 1 ; \boldsymbol{\phi}, \boldsymbol{\alpha} \big) }$},
      xmin=0, xmax=7,
      ymin=0, ymax=1.2,
      axis lines=middle,
      xtick=\empty,
      ytick={0,0.2,0.4,0.6,0.8,1},
      xlabel style={at={(0.95,0)},below right},
      xlabel={time},
      legend style={draw=none, at={(1.25,1)}},
      legend cell align={left}
      ]
      \addplot[domain=0:7, samples=50, smooth, dashed] {(1.09)/(1 + exp(0.8*(x-3)))};
      \addplot[domain=0:7, samples=50, smooth] {(0.8)/(1 + exp(0.8*(x-3)))};
    \end{axis}
  \end{tikzpicture}
  \caption{Example of survival function $S_{\alpha_k}$ representing the probability a priori that true values of an unstable PIV indexed by $k$ between linked record $(i,j)$ coincide (dashed curve). In practice, the conditional probability a posteriori that registered values of an unstable PIV coincide is lower due to mistakes (solid curve).}
  \label{fig_survival_illustration_latent_coincide}
\end{figure}

This interplay between instability and mistakes translates into a gap between the distribution of the latent true values a priori and a posteriori---see \cref{fig_survival_illustration_latent_coincide}. Both parameters involved $\boldsymbol{\phi}$ and $\boldsymbol{\alpha}$ are asymptotically identifiable but are not guaranteed to be estimable with a finite number of records, i.e.\ the likelihood given the data may not have a unique optimum. Specific data requirements allow for distinguishing errors from changes, for instance a sufficient number of links with time gaps near zero or two main groups of links with distinct time gaps. Otherwise, we recommend to privilege the most likely dominating process (i.e.\ either mistakes or changes). Since none of the current approaches can accommodate the dynamics of PIVs, they all attribute disagreements to mistakes. However, a time process may be more appropriate for modelling disagreements in certain PIVs (i.e.\ place of residence). This may require to fix the mistakes parameter to a certain level, which is less restrictive than assuming no mistake as is typically done in blocking strategies \citep{tancredi_bay_RL_2011}. When fixing the parameter for mistakes to a given level, the parameter for changes across time adapts and performance of our method remains stable (see Appendix \cref{app3: assumptions deviations}). Moreover, the algorithm converges faster when fixing the mistakes parameter.

\section{Linkage} 
\label{sec4: linkage}

%The record linkage task is to estimate the latent matrix $\boldsymbol{\Delta}$ using the partially identifying information provided in $\mathcal{A}$ and $\mathcal{B}$. Despite the one-to-one assignment constraint, $\boldsymbol{\Delta}$ has a lot of possible designs. %
%It is computationally easier to focus on rows configuration rather than columns ones given that $\nB \geq \nA$; this approach also has the advantage to restrict the number of possible designs for $\boldsymbol{\Delta}$.

To model the latent matrix $\boldsymbol{\Delta}$, we assume that the rows sums configuration of $\boldsymbol{\Delta}$ is an independent and identically distributed (i.i.d.\@) sample of Bernoulli random variables \smash{$\sum_{j=1}^{\nB} \Delta_{1,j}, \sum_{j=1}^{\nB} \Delta_{2,j}, \dots, \sum_{j=1}^{\nB} \Delta_{\nA,j}$} from $\mathcal{D}$. The rows sums point the records in $\mathcal{A}$ which form pairs with records in $\mathcal{B}$. Given a particular row sums configuration, the number of possible designs for $\boldsymbol{\Delta}$ is given by \smash{$\nB \cdot (\nB - 1) \cdot (\nB - 2) \dots (\nB - (\sum_{i,j} \Delta_{i,j} - 1)) = {\nB!}/{( \nB-\sum_{i,j} \Delta_{i,j} )!}$}. This formula reflects the fact that there are $\nB$ options for the first link to be made, $\nB -1$ for the second, and so on, until the \smash{${\sum_{i,j} \Delta_{i,j}}^{th}$} link. This formula corresponds to the number of arrangements of \smash{$\sum_{i,j} \Delta_{i,j}$} items from $\nB$ objects, sometimes called partial permutation or k-permutation.

We define the probability that a record $i$ from $\mathcal{A}$ forms a link with an observation in $\mathcal{B}$ as \smash{$\gamma \coloneqq \mathbb{P}( \sum_{j=1}^{\nB} \Delta_{i,j} = 1 )$}. This probability may be augmented depending on the field of application, for a time to event estimation for instance, \citep{hof_RL_survival_2017}, or it can be estimated using a prior on the number of entities in a Bayesian graphical entity resolution model, \citep{tancredi_bay_RL_2011, smered2016, marchantsteorts2023}. To manage event data encountered in medical follow-up studies, the chronological ordering of records is critical to form links and, one would need to add a time constraint in $\gamma$ to ensure consistency in the linkage. In that case, the probability for a record in $\mathcal{A}$ to form a link in $\mathcal{B}$ would depend on registration time information $\tA, \tB$. As our intention is to present our method in a comprehensive and adaptable manner, we do not elaborate on this aspect here. 

As previously stated, we presume that files $\mathcal{A}$ and $\mathcal{B}$ are overlapping samples from the same population, inducing that there is a non-zero probability to be in file $\mathcal{B}$ when you are in file $\mathcal{A}$ and, we assume there are no duplicates within the data sets. Those hypotheses are reasonable if we think about studies in a medical setting. 

Although we assume an i.i.d.\ sample of Bernoulli to model $\boldsymbol{\Delta}$, this assumption relies on infinite sample size and may be inconsistent in some settings, \citep{hermann_indep_capture}. For instance, it requires no interference between individuals, thereby disregarding the potential influence of a family member entry in a study under specific disease suspicion. Nevertheless, when $\mathcal{B}$ is large compared to the links set, the dependencies between rows sums are negligible and our method performs as well in scenario where all records are linked---see Appendix \cref{app1: rows sums independence}.

Using independence on rows sums we can express their joint distribution as a product: \begin{align} \label{model: delta} \mathbb{P}\Big( \sum_{j=1}^{\nB} \Delta_{1,j}, \sum_{j=1}^{\nB} \Delta_{2,j}, \dots, \sum_{j=1}^{\nB} \Delta_{\nA,j} ; \gamma \Big) = \prod\limits_{i = 1}^{\nA} \phantom{(} \gamma \phantom{)} ^{ \big(\sum_{j=1}^{\nB} \Delta_{i,j}\big) } \cdot \big( 1 - \gamma \big)^{ \big(1 - \sum_{j=1}^{\nB} \Delta_{i,j}\big) }\end{align} and, because all configurations are equally likely the model for the linkage develops into a uniform distribution over the probability of rows sums configurations. We therefore derive the likelihood of the linkage matrix as follows: \begin{align*}
    \mathcal{L}_{\gamma} \big( \boldsymbol{\Delta} \big) = \ind{\big\{ \boldsymbol{\Delta} \in \mathcal{D} \big\}} \dfrac{ \big( \nB-\sum_{i,j} \Delta_{i,j} \big)! }{ \nB! } \prod\limits_{i = 1}^{\nA} \phantom{(} \gamma \phantom{)} ^{ \big(\sum_{j=1}^{\nB} \Delta_{i,j}\big) } \cdot \big( 1 - \gamma \big)^{ \big(1 - \sum_{j=1}^{\nB} \Delta_{i,j}\big) }. \end{align*}

\section{The StEM algorithm} 
\label{sec5: the stem algorithm}

Let $\boldsymbol{\theta}^\star \in \boldsymbol{\Theta}$ be the true unknown set of parameters of our statistical model, where $\boldsymbol{\Theta}$ is the parameter space. The maximum likelihood principle consists in estimating \smash{$\boldsymbol{\theta}^\star$} with \smash{$\boldsymbol{\hat\theta}_{\text{ML}}$} which maximises the observed data likelihood: \begin{align*}
    \mathcal{L}_{\boldsymbol{\theta}} \big( \textbf{G}^{\mathcal{A}}, \textbf{G}^{\mathcal{B}}, \textbf{t}^{\mathcal{A}}, \textbf{t}^{\mathcal{B}} \big) = \sum\limits_{\textbf{H}^{\mathcal{A}}} \sum\limits_{ \textbf{H}^{\mathcal{B}}} \sum\limits_{\vphantom{\textbf{H}^{\mathcal{B}}}\boldsymbol{\Delta}} \mathcal{L}_{\boldsymbol{\theta}} \big( \textbf{G}^{\mathcal{A}}, \textbf{G}^{\mathcal{B}}, \textbf{H}^{\mathcal{A}}, \textbf{H}^{\mathcal{B}}, \textbf{t}^{\mathcal{A}}, \textbf{t}^{\mathcal{B}}, \boldsymbol{\Delta} \big).
\end{align*} We assume the data provide sufficient information to estimate the parameters (see our comment on the parameters identifiability at the end of \cref{subsec32: registration errors}). In addition, we assume that a unique maximum likelihood estimator (MLE) exists. 

Calculating the observed data likelihood requires exploring the entire latent space of possible linkage matrices from \cref{def: oneone} and values of the PIVs. As it implies summing over all possible values of the unobserved variables \smash{$\textbf{H}^{\mathcal{A}}, \textbf{H}^{\mathcal{B}}, \boldsymbol{\Delta}$}, this is computationally infeasible. As an alternative, we propose a Stochastic Expectation Maximisation (SEM or StEM) approach to find the MLE \citep{celeux_diebolt_SEM, stem_nielsen_2000}. The StEM is an iterative procedure based on the EM algorithm, in which the E-step is approximated with simulation techniques to lower its computational burden in contexts involving a high-dimensional integration of the complete data likelihood. 

We set initial values of the parameters $\boldsymbol{\theta}(0)$ such that the probabilities of mistakes in registrations and changes in the potential unstable PIVs are low ($0.05$), there is a low proportion of linked records ($0.05$) and, the PIVs distributions are uniform. In each iteration $v = 1, 2, \dots$, the algorithm produces an update of the estimate $\boldsymbol{\theta}(v)$ by performing the following two steps---which are described in more detail in \cref{subsec51: stE-step} and \cref{subsec52: M-step}. The Stochastic E-step uses a Gibbs sampler to simulate $Z$ sets of latent data \smash{$\big\{ \textbf{H}^{\mathcal{A}}(v,z), \textbf{H}^{\mathcal{B}}(v,z), \boldsymbol{\Delta}(v,z) \big\}_{z}$} for $z = 1, 2, \dots, Z$, from their posterior distribution \smash{$\mathbb{P}\big(\textbf{H}^{\mathcal{A}}, \textbf{H}^{\mathcal{B}}, \boldsymbol{\Delta} | \textbf{G}^{\mathcal{A}}, \textbf{G}^{\mathcal{B}}, \textbf{t}^{\mathcal{A}}, \textbf{t}^{\mathcal{B}};\boldsymbol{\theta}(v-1)\big)$} given in \cref{subsec51: stE-step}. The M-step uses the simulated sets to compute a new estimate $\boldsymbol{\theta}(v)$.

Under some regularity assumptions, the Markov chain $\{ {\boldsymbol{\theta}}(v) \}_{v}$ produced along the algorithm converges with the number of iterations towards an MLE, \citep[Paragraph 4.2]{celeux_diebolt_SEM}, \citep[Section 3.1]{stem_nielsen_2000}. Hence the convergence and asymptotic normality of the StEM estimator through the asymptotic properties of an MLE (as the sample size tends to infinity). 

\begin{remark}\label{remark: MLE properties}
    Let $\{ {\boldsymbol{\theta}}(v) \}_{v}$ be a sequence of estimators built on $n$ observations, converging towards ${\boldsymbol{\tilde\theta}}$ as $v$ increases. ${\boldsymbol{\tilde\theta}}$ is asymptotically unbiased when \smash{$\lim_{n \rightarrow \infty} \mathbb{E}[{\boldsymbol{\tilde\theta}}] = \boldsymbol{\theta}^\star$} and it is consistent when p\smash{$\lim_{n \rightarrow \infty} {\boldsymbol{\tilde\theta}} = \boldsymbol{\theta}^\star$} with a convergence in probability. The asymptotic normality of ${\boldsymbol{\tilde\theta}}$ means that \smash{$\lim_{n \rightarrow \infty} \sqrt{n} \big( {\boldsymbol{\tilde\theta}} - \boldsymbol{\theta}^\star \big) = \mathcal{N}\big( 0, \Sigma_{\boldsymbol{\theta}^\star} \big)$} with a convergence in distribution. In that case we say that ${\boldsymbol{\tilde\theta}}$ is asymptotically efficient if \smash{$\Sigma_{\boldsymbol{\theta}^\star}$} achieves the Cramér Rao lower bound: the inverse Fisher information \smash{${I_{\boldsymbol{\theta}^\star}}^{-1}$}. An MLE is asymptotically efficient.
\end{remark}

We run the StEM for $V=V_0+V_1$ iterations, which generate a Markov chain of estimates $\{\boldsymbol{\theta}(1), \dots, \boldsymbol{\theta}(V)\}$. An estimate of $\boldsymbol{\theta}^\star$ can then be derived by averaging the $V_1$ last elements in the sequence and discarding the first $V_0$ iterations as burn-in to remove the estimate dependency on initial values of the parameters: \begin{equation} \label{thetaHat}
    \boldsymbol{\hat\theta} = \frac{1}{V_1} \sum_{v=V_0+1}^{V_0+V_1} \boldsymbol{\theta}(v).
\end{equation}

The covariance matrix of the limiting multivariate normal distribution of the StEM estimator is given by \smash{${I_{\boldsymbol{\theta}^\star}}^{-1} + 1/V_1 \Psi_{\boldsymbol{\theta}^\star}(Z)$} where \smash{${I_{\boldsymbol{\theta}^\star}}^{-1}$} denotes the reciprocal Fisher information and \smash{$\Psi_{\boldsymbol{\theta}^\star}(Z)$} the additional variance introduced by the Gibbs sampler simulation noise. As the number of iterations $V_1$ increases, this residual variance vanishes. As a result, the final estimate is asymptotically efficient. Additionally, the complete data likelihood conditional expectation can be more accurately estimated by increasing the number $Z$ of latent variables sampled in each StEM iteration. Furthermore, the StEM estimator can be improved by averaging the last $V_1$ iterations of the Markov chain as suggested in \cref{thetaHat}, \citep[section 4]{stem_nielsen_2000}.

\subsection{Stochastic Expectation, the E-step}
\label{subsec51: stE-step}

In iteration $v \in \{1, \dots, V\}$ of the StEM, we use the previous value of the parameters $\boldsymbol{\theta}(v-1)$ to run a Gibbs sampler of $Z=Z_0+Z_1$ iterations to simulate latent variables. The initial values of the latent data are carefully chosen such that the complete data likelihood is positive. Therefore we initialise \smash{$\textbf{H}^{\mathcal{A}}$} and \smash{$\textbf{H}^{\mathcal{B}}$} with the registered values and $\boldsymbol{\Delta}$ as a zero matrix: $$\textbf{H}^{\mathcal{A}}(v,0) = \textbf{G}^{\mathcal{A}}, \textbf{H}^{\mathcal{B}}(v,0) = \textbf{G}^{\mathcal{B}}, \boldsymbol{\Delta}(v,0) = O_{\nA,\nB}.$$ We discard the first $Z_0$ samples as burn-in to get rid of the influence of the initial state and we keep the $Z_1$ subsequent sets of latent data \smash{$\big\{ \textbf{H}^{\mathcal{A}}(v,z), \textbf{H}^{\mathcal{B}}(v,z), \boldsymbol{\Delta}(v,z) \big\}_{z > Z_0}$} to later estimate the observed data likelihood in the M-step. 

In iteration $z \in \{1, \dots, Z\}$ of the Gibbs sampler, we can update the true values for non-linked records from each file separately from the linked records thanks to the factorisation of the PIVs submodels of \cref{sec3: pivs} using the previous linkage matrix $\boldsymbol{\Delta}(v,z-1)$. We then update the linkage matrix accordingly. The computation details are provided in Appendix \cref{app2.0: update of the latent variables}.

\subsubsection*{True values for non-linked records} We use the models developed in \cref{model: h} and in \cref{model: g|h} to update true values of a PIV indexed by $k$ for non-linked records from $\mathcal{A}$ or $\mathcal{B}$. If a value \smash{$\gAik \in \{1, \dots, n_k\}$} has been recorded we account for potential mistakes using $\boldsymbol{\phi}_{k}$, if not we simply generate a value based on $\boldsymbol{\eta}_{k}$, thus for any true value \smash{$\hAik \in \{1, \dots, n_k\}$} we have: \begin{align}
    & \mathbb{P} \Big( \HAik(v,z) = \hAik \bigm| \GAik = \gAik, \textstyle\sum_{j=1}^{\nB} \Delta_{i,j}(v,z-1)=0 ; \boldsymbol{\theta}(v-1) \Big) \\
    & \qquad \propto \mathbb{P} \Big( \GAik = \gAik \bigm| \HAik(v,z) = \hAik ; \boldsymbol{\theta}(v-1) \Big) \cdot \mathbb{P} \Big( \HAik(v,z) = \hAik ; \boldsymbol{\theta}(v-1) \Big), \nonumber
\end{align} and similarly for $\mathbb{P} \big( \HBjk(v,z) = \hBjk \bigm| \GBjk = \gAik, \sum_{j=1}^{\nB} \Delta_{i,j}(v,z-1)=0 ; \boldsymbol{\theta}(v-1) \big)$.

\subsubsection*{True values for linked records} We use the quantities given by \cref{model: hh} and \cref{model: g|h} to update true values of a PIV indexed by $k$ for records in $\mathcal{A}$ which form a link with records in $\mathcal{B}$. We take into consideration 3 scenarios. If both registrations are missing, we sample a pair of values using $\boldsymbol{\eta}_{k}$ and potentially $S_{\alpha_k}$. If one value is registered, we account for errors on this value with $\boldsymbol{\phi}_{k}$ and we generate the other one based on $\boldsymbol{\eta}_{k}$ and potentially $S_{\alpha_k}$. When both values are registered we account for errors on both values and possibly change one. Thus, for the registered values \smash{$\gAik, \gBjk \in \{0, 1, \dots, n_k\}$} and any true value \smash{$\hAik, \hBjk \in \{1, \dots, n_k\}$} we have: \begin{align*}
    & \mathbb{P} \Big( \HAik(v,z) = \hAik, \HBjk(v,z) = \hBjk \bigm| \GAik = \gAik, \GBjk = \gBjk, {t}_{i,j}, \Delta_{i,j}(v,z-1) = 1 ; \boldsymbol{\theta}(v-1) \Big)\\
    & \qquad \propto \mathbb{P} \Big( \HAik(v,z) = \hAik, \HBjk(v,z) = \hBjk, \GAik = \gAik, \GBjk = \gBjk \bigm| {t}_{i,j}, \Delta_{i,j}(v,z-1) = 1 ; \boldsymbol{\theta}(v-1) \Big)
\end{align*}

\subsubsection*{Linkage indicators} The linkage matrix is updated sequentially so that we give an explicit formula to update value $\Delta_{i,j}(v,z)$ given the elements of the matrix updated so far and the elements of the matrix which are not yet updated. We gather those elements with the notation $\boldsymbol{\Delta}_{-( i,j)}(v,z-1,z)$ where we highlight the iterative updating process of the linkage matrix in which precedent elements are new while subsequent ones are old, hence the dependence on $z-1$ and on $z$: \begin{align*}
    & \mathbb{P} \Big( \Delta_{i,j}(v,z) = 1 \bigm| \boldsymbol{\Delta}_{-(i,j)}(v,z-1,z), \textbf{H}_{i}^{\mathcal{A}}(v,z), \textbf{H}_{j}^{\mathcal{B}}(v,z), \textbf{G}_{i}^{\mathcal{A}}, \textbf{G}_{j}^{\mathcal{B}}, {t}_{i,j}; \boldsymbol{\theta}(v-1) \Big)\\
    & \qquad \propto \mathbb{P} \Big( \Delta_{i,j}(v,z) = 1, \boldsymbol{\Delta}_{-(i,j)}(v,z-1,z), \textbf{H}_{i}^{\mathcal{A}}(v,z), \textbf{H}_{j}^{\mathcal{B}}(v,z), \textbf{G}_{i}^{\mathcal{A}}, \textbf{G}_{j}^{\mathcal{B}}, {t}_{i,j}; \boldsymbol{\theta}(v-1) \Big).
\end{align*}

Accordingly we simulate latent data in each iteration of the Gibbs sampler, we discard the first $Z_0$ iterations as burn-in and keep the $Z_1$ last iterations for the M-step.

\subsection{Maximisation, the M-step}
\label{subsec52: M-step}

The M-step then seeks for $\boldsymbol{\theta}(v)$ maximising: \begin{equation*}
    \sum_{z=Z_0+1}^{Z_0+Z_1} \log \mathcal{L}_{\boldsymbol{\theta}} \big( \textbf{G}^{\mathcal{A}}, \textbf{G}^{\mathcal{B}}, \textbf{H}^{\mathcal{A}}(v,z), \textbf{H}^{\mathcal{B}}(v,z), \textbf{t}^{\mathcal{A}}, \textbf{t}^{\mathcal{B}}, \boldsymbol{\Delta}(v,z) \big).
\end{equation*} The StEM maximises the complete data log-likelihood in the M-step, which is derived from both observed and simulated data in the presence of missing data (true values of the PIVs are latent). Due to the decomposition of the likelihood we can update the parameters separately at each iteration $v$, some details are provided in Appendix \cref{app2: parameters update}. For each PIV indexed by $k$ we have \begin{align*}
    \phi_{k,\text{mistake}}^{\mathcal{A}}(v) = \: & \underset{\phi_{k,\text{mistake}}^{\mathcal{A}}}{\text{argmax}} \, \sum_{z=Z_0+1}^{Z_0+Z_1} \, \log \mathcal{L}_{\boldsymbol{\phi}} \big( \textbf{G}^{\mathcal{A}} \bigm| \textbf{H}^{\mathcal{A}}(v,z) \big), \text{likewise for }\phi_{k,\text{mistake}}^{\mathcal{B}}(v),\\
    \alpha_k(v) = \: & \underset{\alpha_k}{\text{argmax}} \, \sum_{z=Z_0+1}^{Z_0+Z_1} \, \log \mathcal{L}_{\boldsymbol{\alpha}} \big( \textbf{H}^{\mathcal{B}}(v,z) \bigm| \textbf{H}^{\mathcal{A}}(v,z), \textbf{t}^{\mathcal{A}}, \textbf{t}^{\mathcal{B}}, \boldsymbol{\Delta}(v,z) \big),\\
    \boldsymbol{\eta}_{k}(v) = \: & \underset{{\boldsymbol{\eta}}_{k}}{\text{argmax}} \, \sum_{z=Z_0+1}^{Z_0+Z_1}\, \log \mathcal{L}_{\boldsymbol{\eta}} \big( \textbf{H}^{\mathcal{A}}(v,z) \big) \text{ subject to } \sum_\ell \eta_{k,\ell}(v) = 1,\\
    \gamma(v) = \: & \underset{\gamma}{\text{argmax}} \, \sum_{z=Z_0+1}^{Z_0+Z_1} \, \log \mathcal{L}_{\gamma} \big( \boldsymbol{\Delta}(v,z) \big).
\end{align*}

The update for \smash{$\phi_{k,\text{mistake}}^{\mathcal{A}}$} and $\phi_{k,\text{mistake}}^{\mathcal{B}}$ are thereby simply given by the proportion of disagreements between registered and true values among all records in $\mathcal{A}$ and respectively in $\mathcal{B}$ for which the registered value is not missing. Note that \smash{$\phi_{k,\text{missing}}^{\mathcal{A}}$} and \smash{$\phi_{k,\text{missing}}^{\mathcal{B}}$} are fixed by the data and correspond to the proportion of missing data in each file. The update for $\eta_{k,\ell}$ corresponds to the occurrence of value $\ell \in \{1, \dots, n_k\}$ among the true values of all records in $\mathcal{A}$ and $\mathcal{B}$. The update for $\alpha_k$ is more complex to derive, it incorporates the proportion of disagreements between true values of linked records for the potential unstable $k^{\text{th}}$ PIV; we solve it using a computational optimisation algorithm. Finally we update $\gamma$ with the proportion of linked records as a fraction of the smallest file. We can bound $\boldsymbol{\phi}$ to address potential identifiability issues, especially for stable PIVs the probability of a mistake rarely exceeds 10\%.

\begin{remark}\label{remark: Gibbs samples nbr linkrec} 
    Although there is no universal rule to choose the number of iterations necessary for the algorithm to converge nor for the burn-in period, we can use convergence diagnostics to assess whether the chains have reached a stationary distribution over the StEM iterations. In the Gibbs sampler within each iteration of the StEM, we can determine the appropriate burn-in period in an exploratory approach by monitoring the number of linked pairs along the iterations. It may be deemed sufficient when the number of linked pairs stabilises.
\end{remark}

Although we do not emphasise it here with the notations, note that each $\boldsymbol{\theta}(v)$ is dependent on the number of simulated data in the Gibbs sampler, as is our final estimate.

\subsection{Estimate the linkage} \label{subsec53: estimate the linkage}

After obtaining the estimate $\boldsymbol{\hat\theta}$ using \cref{thetaHat}, we sample multiple sets of true values \smash{$\big\{ \textbf{H}^{\mathcal{A}}(1), \textbf{H}^{\mathcal{B}}(1) \big\}, \dots, \big\{ \textbf{H}^{\mathcal{A}}(n_{\text{sim}}), \textbf{H}^{\mathcal{B}}(n_{\text{sim}}) \big\}$} and the linkage matrices associated \smash{$\boldsymbol{\Delta}(1), \dots, \boldsymbol{\Delta}(n_{\text{sim}})$} from their posterior distribution \smash{$\mathbb{P}\big(\textbf{H}^{\mathcal{A}}, \textbf{H}^{\mathcal{B}}, \boldsymbol{\Delta} | \textbf{G}^{\mathcal{A}}, \textbf{G}^{\mathcal{B}}, \textbf{t}^{\mathcal{A}}, \textbf{t}^{\mathcal{B}};\boldsymbol{\hat\theta}\big)$} given in \cref{subsec51: stE-step}. An estimate of $\boldsymbol{\Delta}$ can then be derived by averaging the $n_{\text{sim}}$ matrices simulated. Thus we obtain a marginal probability for each observation pair to be linked: \begin{equation*}
    \hat\Delta_{i,j} = \frac{1}{n_{\text{sim}}} \sum_{\ell=1}^{n_{\text{sim}}} \Delta_{i,j}(\ell).
\end{equation*}

We can use these probabilities to quantify the uncertainty of linked pairs for subsequent inference or to build a set of linked pairs and evaluate it. When focusing on the latter, we need a threshold $\xi$ on the linkage probabilities to select a final set of pairs \smash{$\big\{ (i,j) ; \hat\Delta_{i,j} > \xi \big\}$}. Then we can assess the performance of our methodology using the partial confusion matrix detailing true positives $TP$, false positives $FP$ and false negative $FN$. In addition, we use the False Discovery Rate (FDR) representing the error rate when linking pairs, and the sensitivity to judge our ability to detect links, as well as the F1-score which compares $TP$ to any error, $FP$ or $FN$. Then, we have: $$\text{FDR}(\xi) = \mathbb E \bigg[ \frac{FP(\xi)}{TP(\xi) + FP(\xi)} \bigg],\: \text{Sensitivity}(\xi) = \mathbb E \bigg[ \frac{TP(\xi)}{TP(\xi) + FN(\xi)} \bigg],\: \text{F1-Score}(\xi) = \mathbb E \bigg[ \frac{TP(\xi)}{TP(\xi) + 1/2(TP(\xi) + FN(\xi))} \bigg].$$

We expect the posterior probabilities for pairs of records to be linked to have a bimodal distribution representing a mixture between non-linked records, with probability mass concentrated at $0$, and linked records, with probability mass concentrated at a higher level (which would depend on the weakness of the PIVs). The threshold $\xi \in [0.5;1]$ should separate those modes and maintain a one-to-one assignment constraint in the set of linked records, \citep[Theorem 4.1]{tancredi_bay_RL_2011}, \citep[Corollary 1.1]{sadinle_bay_bipartite_RL_2017}. A natural way to set $\xi$ is by controlling for the estimated FDR level, \citep{fdr_genom}, such that it would not exceed a certain level $\overline{\text{FDR}}$: $$\widehat{\text{FDR}} = 1 - \frac{ \sum_{i,j} \hat\Delta_{i,j} \cdot \ind \big\{ \hat\Delta_{i,j} > \xi \big\} }{ \sum_{i,j} \ind \big\{ \hat\Delta_{i,j} > \xi \big\} } < \overline{\text{FDR}}.$$

The choice of PIVs may affect the linkage estimation. Our method scales linearly with the number of PIVs, therefore the estimation will mostly benefit from more information, yielding less stochasticity in the chain of estimates and faster convergence. The intrinsic limited reliability of the PIVs, on the other hand, hinders the process. Thus, one should select variables that offer valuable information while minimising missing data, error rates, instability and dependencies, which are obstacles to the linkage.

\section{Simulations and applications} 
\label{sec6: simulations and applications}

We performed a simulation study to investigate the impact of our model contributions. In particular, we explored the added value of incorporating dynamics of the PIVs in our modelling. To show the scalability of the method and review its performance we conducted an empirical study on a large real data application, the National Long Term Care Survey (NLTCS): a longitudinal study on the elderly population health in the United States. Additional applications, on the often-used SHIW, \citep{smered2016, guha_bay_causal_RL_2022, menezes2024hausdorff}, and the RLData500, \citep{RLdata, steorts2015, Gbrl_2018, blockingER_2020, dedup_randomforest_2022, fscore_RL_2023, dedup_2024}, can be found in the supplementary material. The open-source software developed is available online, as well as the data sets used and a vignette to reproduce experiments, \citep{flexrlpackage}.

\subsubsection*{Baseline methods} 

After reviewing the available methods for software availability and relevance to the record linkage task, we chose to compare our method \textit{FlexRL}, with two recent state-of-the-art methods, \textit{BRL} and \textit{Exchanger}. 

The \textit{BRL} program from \citep{brlpackage}, a Bayesian bipartite record linkage method, addresses the limitations of the foundational mixture model from Fellegi and Sunter incorporating dependencies among the linkage decisions, \citep{Fellegi_RL_1969, sadinle_bay_bipartite_RL_2017}. In contrast the \textit{Exchanger} program from \citep{marchantsteorts2023}, a Bayesian graphical entity resolution method, models a latent population of individuals which records are clustered to, taking into account different distortion mechanisms of the data generation process. This approach is inspired by the seminal work of \citep{tancredi_bay_RL_2011}, which also serves as main motivation for our work, and other Bayesian models, \citep{steorts2015, smered2016, dblink_2021}. 

Moreover, we also compared those methodologies with a simplistic approach linking the records for which all the PIVs match exactly, regardless of the one-to-one assignment constraint. In this last method, the linked pairs ($TP$ and $FP$) will thus always agree at 100\% in all their values. This approach enables to judge the task difficulty by computing the false positive rate for pairs of records with identical information. It is expected that any compared method would provide more refined results than the simplistic approach.

\subsubsection*{Computational considerations and implementations} 

We conducted the simulations and applications on \texttt{R} version 4.3.2 using a standard machine (8-core CPU, M1 chip, 32 GB of memory). Our open-source \texttt{R} package is written in \texttt{R} and \texttt{C++} thanks to the Rcpp interface, as it is the case for \textit{Exchanger}. \textit{BRL} is written in \texttt{R}. Our method stands out for its low memory footprint, it is capable of processing large data sets on this standard computer. This contrasts with the other methods, which encounter memory limitations when running on large real data applications. 

We used the default parametrizations for the baseline methods, which can be found on the \textit{Exchanger} vignette: 20 000 iterations, discarding the first 10 000 as burn-in, with thinning interval at 10, \citep{exchangervignette}, and the \textit{BRL} documentation: 1 000 iterations, discarding the first 100 as burn-in, \citep{brlpackage}. 

For \textit{FlexRL} we ran 100 iterations of the StEM, discarding the first 75 ones as burn-in. Convergence of the parameters can be checked visually on diagnostic plots. We adapted the number of Gibbs samples according to the \cref{remark: Gibbs samples nbr linkrec} and ran 200 iterations, discarding the first 100 as burn-in within each StEM iteration. 

\subsubsection*{Model evaluation} 

\textit{Exchanger} uses most probable maximal matching set to produce a final set of linked records, \citep{smered2016}. This is an optimal strategy, \citep{tancredi_bay_RL_2011, smered2016}, as is the one employed in \textit{BRL} selecting links for which the posterior linkage probability is superior to $0.5$, \citep{sadinle_bay_bipartite_RL_2017}. 

In order to compare \textit{FlexRL} to the literature we therefore selected linked records using a threshold $\xi = 0.5$ on the linkage posterior. When running \textit{FlexRL} on the large real data sets we also provide results when estimating the set of linked records using a threshold $\xi$ such that the estimated FDR would not exceed $10\%$.

Note that, as an entity resolution method, \textit{Exchanger} may also incorrectly detect duplicates, accounted for in the $FP$; we subtracted them to fairly compare the methods.

\subsection{Simulations} 
\label{subsec61: simulations}

Files $\mathcal{A}$ and $\mathcal{B}$ gather respectively 800 and 1000 patient records, among which 500 are common to both sources. We use five PIVs sampled from five independent discrete non-uniform distributions: $$\mathbb{P}(\HAik=h) = \mathbb{P}(\HBjk=h) = \frac{\exp\{0.25 h\}}{\sum_{\ell=1}^{n_k} \exp\{0.25 \ell\}},$$ where $k$ is the index of the PIV, which can take any value $h \in \{1, \dots, n_k\}$. We generate weak PIVs in that they have low discriminating strength due to a low number of unique values. Those variables could represent the sex, postal code, birth year, education level, type of occupation or household size for example. 

We distort the values of all PIVs in each file with 2\% of mistakes (except the $5^{\text{th}}$ one, for which no mistake is added) and 0.7\% of missing values. We generate instability in the $5^{\text{th}}$ PIV representing the postal code. For each of the pairs of records $(i,j)$ referring to the same individual, we compute the registration time difference \smash{$t_{i,j} = |t_j^{\mathcal{B}} - t_i^{\mathcal{A}}|$}, using registration dates $\textbf{t}^{\mathcal{A}}$ and $\textbf{t}^{\mathcal{B}}$ generated with continuous uniform distributions in $[0,3]$ and in $[3,6]$ respectively. The probability that true values for a link coincide is defined by \smash{$S_{\alpha_5}(t_{i,j}) = \exp\big\{ - 0.28\, t_{i,j} \big\}$}, following the theory developed in \cref{subsec31: true values dynamics for unstable variables}. Then, for each link we change the value of the $5^{\text{th}}$ PIV based on a Bernoulli with probability of success $1 - S_{\alpha_5}(t_{i,j})$ to one of the other $n_5 - 1$ possible values to represent patients migration in the database. In practice, this scenario gives approximately 95\% of agreements between registered values of the links for stable PIVs and 46\% for the unstable PIV and, 1.8\% of missing values in the links for all PIVs as we observe in \cref{tab_Simu_Story_And_Results}. As mentioned in \cref{subsec32: registration errors}, both mistakes and changes parameters may not be estimable from the data without further requirement. In some cases a time process may more accurately model disagreements between records of the same individual and, as online forms usually incorporate address verification checks or city cross-referencing, one can reasonably assume a null probability of mistake in postal codes. Moreover, we show the robustness of FlexRL to wrongly fixed mistakes parameter in Appendix \cref{app3: assumptions deviations}. We illustrate such scenario in the simulations and, to explore the importance of modelling PIVs instability in the record linkage process, we compare the baseline methods with two versions of our method: one correctly taking account of relocation flows and one incorrectly considering all PIVs as stable. This simulation framework is particularly suited for healthcare applications, where PIVs are categorical with limited discriminating power. In such contexts, variables such as place of residence are often the strongest discriminators, albeit unstable ones.

% The independence of rows sums configuration of $\boldsymbol{\Delta}$ required by our model was discussed in \cref{sec4: linkage} and in \cref{app1: rows sums independence} of the appendix. Although this assumption does not properly hold in a setting where file $\mathcal{B}$ comprises 1 000 records among which 500 are shared with file $\mathcal{A}$ (the overlapping set of units is only 50\% smaller than $\mathcal{B}$), the proposed methodology is not affected in terms of performance. We chose such simulation setting to make computational time reasonable while estimating the model parameters on a sufficient amount of links.

\begin{table}[!ht]
    \centering
    \begin{tabular}{ccccccc}
        \hline
        \noalign{\vskip\doublerulesep \vskip-\arrayrulewidth}
        \multicolumn{2}{c}{\textbf{Registrations}} & \textbf{V1} & \textbf{V2} & \textbf{V3} & \textbf{V4} & \textbf{V5} \\
        \hline
        \noalign{\vskip\doublerulesep \vskip-\arrayrulewidth}
        % \noalign{\vskip\doublerulesep \vskip-\arrayrulewidth}
        % \hline
        \multirow{2}{*}{\textbf{Data}} & \textbf{Unique} & $6$ & $7$ & $8$ & $9$ & $15$ \\
        % \cline{2-7}
        & \textbf{Missing} & $.02 (.00)$ & $.01 (.01)$ & $.01 (.00)$ & $.02 (.01)$ & $.02 (.00)$ \\
        \cline{2-2} 
        \noalign{\vskip\doublerulesep \vskip-\arrayrulewidth}
        % \hline
        \multirow{2}{*}{\textbf{True Links}} & \textbf{Agree} & $.95 (.01)$ & $.95 (.01)$ & $.94 (.01)$ & $.95 (.01)$ & $.46 (.02)$ \\
        % \cline{2-7}
        & \textbf{Missing} & $.02 (.01)$ & $.01 (.01)$ & $.01 (0.01)$ & $.02 (.01)$ & $.01 (.01)$ \\
        % \hline
        % \noalign{\vskip\doublerulesep \vskip-\arrayrulewidth}
        \hline
        \noalign{\vskip\doublerulesep \vskip-\arrayrulewidth}
        \multicolumn{1}{c}{\textbf{Methods}} & \multicolumn{2}{c}{\textbf{Linked Records}} & \multirow{2}{*}{\hphantom{vity}\textbf{FN}\hphantom{vity}} & \multirow{2}{*}{\hphantom{i}\textbf{F1-Score}\hphantom{i}} & \multirow{2}{*}{\hphantom{ity}\textbf{FDR}\hphantom{ity}} & \multirow{2}{*}{\textbf{Sensitivity}}\\
        \cline{2-3}
        \noalign{\vskip\doublerulesep \vskip-\arrayrulewidth}
        \multicolumn{1}{c}{} & \hphantom{sens}\textbf{TP}\hphantom{vity} & \hphantom{sens}\textbf{FP}\hphantom{vity} & & & &  \\
        \hline
        \noalign{\vskip\doublerulesep \vskip-\arrayrulewidth}
        {\textbf{\textit{FlexRL} with instability}} & $290 (11)$ & $74 (10)$ & $209 (11)$ & $.67 (.02)$ & $.20 (.02)$ & $.58 (.02)$\\
        % \hline
        {\textbf{\textit{FlexRL} all stable}} & $272 (12)$ & $72 (10)$ & $227 (12)$ & $.64 (.02)$ & $.21 (.02)$ & $.55 (.02)$\\
        % \hline
        {\textbf{\textit{Exchanger}}} & $152 (9)$ & $61 (15)$ & $348 (9)$ & $.42(.03)$ & $.29 (.04)$ & $.30 (.02)$\\
        % \hline
        {\textbf{\textit{BRL}}} & $203 (36)$ & $43 (14)$ & $297 (36)$ & $.54 (.07)$ & $.17 (.03)$ & $.41 (.07)$\\
        % \hline
        {\textbf{Simplistic approach}} & $204 (9)$ & $110 (14)$ & $296 (9)$ & $.50 (.02)$ & $.35 (.03)$ & $.41 (.02)$\\
        \hline
        \noalign{\vskip\doublerulesep \vskip-\arrayrulewidth}
    \end{tabular}
    \caption{Characteristics of the PIVs in the simulated data with 800 and 1000 records and, level of agreement among the 500 links referring to the same individuals over the 500 simulations (mean proportions and standard deviation in parentheses). The $5^{\text{th}}$ PIV is unstable without mistake: the 54\% of disagreements generated are due to changes (except for some missing values).\\Below, performance of the compared methods over the 500 simulations (mean values of the metrics and standard deviation in parentheses).}
    \label{tab_Simu_Story_And_Results}
\end{table}

\begin{figure}[!h]
    \centering
    \begin{tikzpicture}
    % \node at (7.8,-1.2) {\textit{FlexRL} with instability};
    % \node at (0,-3.5) {\scalebox{0.7}{\input{Figures/Fig3 phi/phi V1 stable}}};
    % \node at (1.5,-2.4) {$1^{\text{st}}$ PIV};
    % \node at (0,-7.35) {\scalebox{0.7}{\input{Figures/Fig3 phi/phi V5 stable}}};
    % \node at (1.5,-6.24) {$5^{\text{th}}$ PIV};
    % \node at (7.5,-3.5) {\scalebox{0.7}{\input{Figures/Fig3 phi/phi V1 instability}}};
    % \node at (9,-2.4) {$1^{\text{st}}$ PIV};
    % \node at (7.5,-7.35) {\scalebox{0.7}{\input{Figures/Fig3 phi/phi V5 instability}}};
    % \node at (9,-6.24) {$5^{\text{th}}$ PIV};
    % \node at (0,-11.2) {\scalebox{0.7}{\input{Figures/Fig4 gamma/gamma stable}}};
    % \node at (7.5,-11.2) {\scalebox{0.7}{\input{Figures/Fig4 gamma/gamma instability}}};
    % \node at (-3.4,-11.2) {$\gamma$};
    % \node[rotate=90] at (-3.4,-5.425) {$\boldsymbol{\phi}_{\text{mistake}}$};
    % \node[rotate=90] at (11.6,-5.425) {};
    % \node at (4.1,-13.65) {StEM iterations};
    \node at (7.2,-1.3) {$\gamma$};
    \node at (-2.8,-1.3) {$\boldsymbol{\phi}_{1,\text{mistake}}$};
    \node at (2.2,-1.3) {$\boldsymbol{\phi}_{5,\text{mistake}}$};
    \node[rotate=90] at (-5.8,-3.3) {\textit{FlexRL} all stable};
    \node[rotate=90] at (-5.8,-7.3) {\textit{FlexRL} with instability};
    \node at (-3,-3.5) {\scalebox{0.7}{\begin{tikzpicture}[x=1pt,y=1pt]
\definecolor{fillColor}{RGB}{255,255,255}

\begin{scope}
\path[clip] ( 49.20, 61.20) rectangle (209.68,185.68);
\definecolor{fillColor}{RGB}{190,190,190}
\path[fill=fillColor,fill opacity=0.60] 
        ( 53.7,71.6) --
        ( 55.74, 83.17) --
	( 56.64, 82.26) --
	( 58.15, 81.23) --
	( 59.65, 80.10) --
	( 61.15, 78.99) --
	( 62.65, 77.97) --
	( 64.15, 76.98) --
	( 65.65, 76.05) --
	( 67.15, 75.30) --
	( 68.65, 74.65) --
	( 70.15, 74.06) --
	( 71.65, 73.56) --
	( 73.15, 73.12) --
	( 74.66, 72.70) --
	( 76.16, 72.31) --
	( 77.66, 71.98) --
	( 79.16, 71.68) --
	( 80.66, 71.41) --
	( 82.16, 71.18) --
	( 83.66, 70.95) --
	( 85.16, 70.75) --
	( 86.66, 70.59) --
	( 88.16, 70.45) --
	( 89.66, 70.32) --
	( 91.17, 70.18) --
	( 92.67, 70.05) --
	( 94.17, 69.91) --
	( 95.67, 69.77) --
	( 97.17, 69.63) --
	( 98.67, 69.54) --
	(100.17, 69.45) --
	(101.67, 69.37) --
	(103.17, 69.28) --
	(104.67, 69.22) --
	(106.17, 69.16) --
	(107.68, 69.09) --
	(109.18, 69.03) --
	(110.68, 68.99) --
	(112.18, 68.93) --
	(113.68, 68.87) --
	(115.18, 68.83) --
	(116.68, 68.78) --
	(118.18, 68.75) --
	(119.68, 68.71) --
	(121.18, 68.66) --
	(122.68, 68.61) --
	(124.19, 68.57) --
	(125.69, 68.54) --
	(127.19, 68.52) --
	(128.69, 68.48) --
	(130.19, 68.46) --
	(131.69, 68.42) --
	(133.19, 68.41) --
	(134.69, 68.38) --
	(136.19, 68.34) --
	(137.69, 68.31) --
	(139.19, 68.25) --
	(140.70, 68.23) --
	(142.20, 68.20) --
	(143.70, 68.17) --
	(145.20, 68.14) --
	(146.70, 68.13) --
	(148.20, 68.13) --
	(149.70, 68.10) --
	(151.20, 68.10) --
	(152.70, 68.08) --
	(154.20, 68.08) --
	(155.70, 68.07) --
	(157.21, 68.04) --
	(158.71, 68.03) --
	(160.21, 68.01) --
	(161.71, 68.00) --
	(163.21, 67.98) --
	(164.71, 67.97) --
	(166.21, 67.93) --
	(167.71, 67.92) --
	(169.21, 67.90) --
	(170.71, 67.88) --
	(172.21, 67.86) --
	(173.72, 67.84) --
	(175.22, 67.84) --
	(176.72, 67.85) --
	(178.22, 67.83) --
	(179.72, 67.80) --
	(181.22, 67.79) --
	(182.72, 67.80) --
	(184.22, 67.80) --
	(185.72, 67.79) --
	(187.22, 67.78) --
	(188.72, 67.78) --
	(190.23, 67.78) --
	(191.73, 67.78) --
	(193.23, 67.78) --
	(194.73, 67.77) --
	(196.23, 67.79) --
	(197.73, 67.79) --
	(199.23, 67.79) --
	(200.73, 67.77) --
	(202.23, 67.78) --
	(203.73, 67.77) --
	(203.73, 71.24) --
	(202.23, 71.27) --
	(200.73, 71.29) --
	(199.23, 71.27) --
	(197.73, 71.29) --
	(196.23, 71.31) --
	(194.73, 71.34) --
	(193.23, 71.33) --
	(191.73, 71.36) --
	(190.23, 71.39) --
	(188.72, 71.43) --
	(187.22, 71.45) --
	(185.72, 71.50) --
	(184.22, 71.52) --
	(182.72, 71.55) --
	(181.22, 71.60) --
	(179.72, 71.60) --
	(178.22, 71.58) --
	(176.72, 71.56) --
	(175.22, 71.60) --
	(173.72, 71.63) --
	(172.21, 71.66) --
	(170.71, 71.66) --
	(169.21, 71.66) --
	(167.71, 71.69) --
	(166.21, 71.71) --
	(164.71, 71.66) --
	(163.21, 71.69) --
	(161.71, 71.69) --
	(160.21, 71.72) --
	(158.71, 71.72) --
	(157.21, 71.74) --
	(155.70, 71.76) --
	(154.20, 71.78) --
	(152.70, 71.80) --
	(151.20, 71.79) --
	(149.70, 71.83) --
	(148.20, 71.83) --
	(146.70, 71.87) --
	(145.20, 71.90) --
	(143.70, 71.89) --
	(142.20, 71.91) --
	(140.70, 71.92) --
	(139.19, 71.95) --
	(137.69, 71.96) --
	(136.19, 72.02) --
	(134.69, 72.04) --
	(133.19, 72.08) --
	(131.69, 72.15) --
	(130.19, 72.17) --
	(128.69, 72.21) --
	(127.19, 72.24) --
	(125.69, 72.30) --
	(124.19, 72.35) --
	(122.68, 72.40) --
	(121.18, 72.44) --
	(119.68, 72.50) --
	(118.18, 72.56) --
	(116.68, 72.64) --
	(115.18, 72.69) --
	(113.68, 72.78) --
	(112.18, 72.83) --
	(110.68, 72.88) --
	(109.18, 73.01) --
	(107.68, 73.09) --
	(106.17, 73.15) --
	(104.67, 73.23) --
	(103.17, 73.30) --
	(101.67, 73.38) --
	(100.17, 73.47) --
	( 98.67, 73.57) --
	( 97.17, 73.68) --
	( 95.67, 73.83) --
	( 94.17, 73.94) --
	( 92.67, 74.07) --
	( 91.17, 74.22) --
	( 89.66, 74.38) --
	( 88.16, 74.53) --
	( 86.66, 74.68) --
	( 85.16, 74.84) --
	( 83.66, 75.04) --
	( 82.16, 75.24) --
	( 80.66, 75.52) --
	( 79.16, 75.80) --
	( 77.66, 76.12) --
	( 76.16, 76.44) --
	( 74.66, 76.85) --
	( 73.15, 77.26) --
	( 71.65, 77.67) --
	( 70.15, 78.11) --
	( 68.65, 78.70) --
	( 67.15, 79.33) --
	( 65.65, 80.06) --
	( 64.15, 80.81) --
	( 62.65, 81.63) --
	( 61.15, 82.55) --
	( 59.65, 83.47) --
	( 58.15, 84.27) --
	( 56.64, 84.91) --
	( 54.14, 85.15) --
        ( 53.2,71.6) --
	cycle;
\definecolor{drawColor}{cmyk}{.80,.29,.05,0}
\path[draw=drawColor, loosely dotted, line width= 1.2pt,line join=round,line cap=round] (  0.00, 80.97) -- (234.88, 80.97);
\end{scope}

\begin{scope}
\definecolor{drawColor}{RGB}{0,0,0}
\path[draw=drawColor,line width= 0.4pt,line join=round,line cap=round] 
        ( 53.5,72) --
        ( 55.14, 84.16) --
	( 56.64, 83.58) --
	( 58.15, 82.75) --
	( 59.65, 81.78) --
	( 61.15, 80.77) --
	( 62.65, 79.80) --
	( 64.15, 78.90) --
	( 65.65, 78.05) --
	( 67.15, 77.32) --
	( 68.65, 76.67) --
	( 70.15, 76.08) --
	( 71.65, 75.62) --
	( 73.15, 75.19) --
	( 74.66, 74.77) --
	( 76.16, 74.38) --
	( 77.66, 74.05) --
	( 79.16, 73.74) --
	( 80.66, 73.46) --
	( 82.16, 73.21) --
	( 83.66, 73.00) --
	( 85.16, 72.80) --
	( 86.66, 72.64) --
	( 88.16, 72.49) --
	( 89.66, 72.35) --
	( 91.17, 72.20) --
	( 92.67, 72.06) --
	( 94.17, 71.92) --
	( 95.67, 71.80) --
	( 97.17, 71.65) --
	( 98.67, 71.56) --
	(100.17, 71.46) --
	(101.67, 71.37) --
	(103.17, 71.29) --
	(104.67, 71.23) --
	(106.17, 71.15) --
	(107.68, 71.09) --
	(109.18, 71.02) --
	(110.68, 70.93) --
	(112.18, 70.88) --
	(113.68, 70.82) --
	(115.18, 70.76) --
	(116.68, 70.71) --
	(118.18, 70.65) --
	(119.68, 70.60) --
	(121.18, 70.55) --
	(122.68, 70.51) --
	(124.19, 70.46) --
	(125.69, 70.42) --
	(127.19, 70.38) --
	(128.69, 70.35) --
	(130.19, 70.32) --
	(131.69, 70.29) --
	(133.19, 70.25) --
	(134.69, 70.21) --
	(136.19, 70.18) --
	(137.69, 70.14) --
	(139.19, 70.10) --
	(140.70, 70.07) --
	(142.20, 70.05) --
	(143.70, 70.03) --
	(145.20, 70.02) --
	(146.70, 70.00) --
	(148.20, 69.98) --
	(149.70, 69.97) --
	(151.20, 69.95) --
	(152.70, 69.94) --
	(154.20, 69.93) --
	(155.70, 69.91) --
	(157.21, 69.89) --
	(158.71, 69.87) --
	(160.21, 69.87) --
	(161.71, 69.85) --
	(163.21, 69.84) --
	(164.71, 69.82) --
	(166.21, 69.82) --
	(167.71, 69.80) --
	(169.21, 69.78) --
	(170.71, 69.77) --
	(172.21, 69.76) --
	(173.72, 69.74) --
	(175.22, 69.72) --
	(176.72, 69.71) --
	(178.22, 69.71) --
	(179.72, 69.70) --
	(181.22, 69.70) --
	(182.72, 69.67) --
	(184.22, 69.66) --
	(185.72, 69.64) --
	(187.22, 69.62) --
	(188.72, 69.61) --
	(190.23, 69.58) --
	(191.73, 69.57) --
	(193.23, 69.56) --
	(194.73, 69.55) --
	(196.23, 69.55) --
	(197.73, 69.54) --
	(199.23, 69.53) --
	(200.73, 69.53) --
	(202.23, 69.52) --
	(203.73, 69.50);
\end{scope}

\begin{scope}
\definecolor{drawColor}{RGB}{0,0,0}

\path[draw=drawColor,line width= 0.4pt,line join=round,line cap=round] ( 53.64, 61.20) -- (203.73, 61.20);

\path[draw=drawColor,line width= 0.4pt,line join=round,line cap=round] ( 53.64, 61.20) -- ( 53.64, 55.20);

\path[draw=drawColor,line width= 0.4pt,line join=round,line cap=round] ( 83.66, 61.20) -- ( 83.66, 55.20);

\path[draw=drawColor,line width= 0.4pt,line join=round,line cap=round] (113.68, 61.20) -- (113.68, 55.20);

\path[draw=drawColor,line width= 0.4pt,line join=round,line cap=round] (143.70, 61.20) -- (143.70, 55.20);

\path[draw=drawColor,line width= 0.4pt,line join=round,line cap=round] (173.72, 61.20) -- (173.72, 55.20);

\path[draw=drawColor,line width= 0.4pt,line join=round,line cap=round] (203.73, 61.20) -- (203.73, 55.20);

\node[text=drawColor,anchor=base,inner sep=0pt, outer sep=0pt, scale=  1.00] at ( 53.64, 39.60) {0};

\node[text=drawColor,anchor=base,inner sep=0pt, outer sep=0pt, scale=  1.00] at ( 83.66, 39.60) {20};

\node[text=drawColor,anchor=base,inner sep=0pt, outer sep=0pt, scale=  1.00] at (113.68, 39.60) {40};

\node[text=drawColor,anchor=base,inner sep=0pt, outer sep=0pt, scale=  1.00] at (143.70, 39.60) {60};

\node[text=drawColor,anchor=base,inner sep=0pt, outer sep=0pt, scale=  1.00] at (173.72, 39.60) {80};

\node[text=drawColor,anchor=base,inner sep=0pt, outer sep=0pt, scale=  1.00] at (203.73, 39.60) {100};

\path[draw=drawColor,line width= 0.4pt,line join=round,line cap=round] ( 49.20, 65.81) -- ( 49.20,181.07);

\path[draw=drawColor,line width= 0.4pt,line join=round,line cap=round] ( 49.20, 65.81) -- ( 43.20, 65.81);

\path[draw=drawColor,line width= 0.4pt,line join=round,line cap=round] ( 49.20, 88.86) -- ( 43.20, 88.86);

\path[draw=drawColor,line width= 0.4pt,line join=round,line cap=round] ( 49.20,111.91) -- ( 43.20,111.91);

\path[draw=drawColor,line width= 0.4pt,line join=round,line cap=round] ( 49.20,134.96) -- ( 43.20,134.96);

\path[draw=drawColor,line width= 0.4pt,line join=round,line cap=round] ( 49.20,158.02) -- ( 43.20,158.02);

\path[draw=drawColor,line width= 0.4pt,line join=round,line cap=round] ( 49.20,181.07) -- ( 43.20,181.07);

\node[text=drawColor,rotate= 90.00,anchor=base,inner sep=0pt, outer sep=0pt, scale=  1.00] at ( 34.80, 65.81) {0.0};

\node[text=drawColor,rotate= 90.00,anchor=base,inner sep=0pt, outer sep=0pt, scale=  1.00] at ( 34.80, 94.625) {0.1};

\node[text=drawColor,rotate= 90.00,anchor=base,inner sep=0pt, outer sep=0pt, scale=  1.00] at ( 34.80,123.44) {0.2};

\node[text=drawColor,rotate= 90.00,anchor=base,inner sep=0pt, outer sep=0pt, scale=  1.00] at ( 34.80,152.255) {0.3};

\node[text=drawColor,rotate= 90.00,anchor=base,inner sep=0pt, outer sep=0pt, scale=  1.00] at ( 34.80,181.07) {0.4};

\path[draw=drawColor,line width= 0.4pt,line join=round,line cap=round] ( 49.20, 61.20) --
	(209.68, 61.20) --
	(209.68,185.68) --
	( 49.20,185.68) --
	cycle;
\end{scope}

\end{tikzpicture}}};
    \node at (-3,-7.5) {\scalebox{0.7}{\begin{tikzpicture}[x=1pt,y=1pt]
\definecolor{fillColor}{RGB}{255,255,255}

\begin{scope}
\path[clip] ( 49.20, 61.20) rectangle (209.68,185.68);
\definecolor{fillColor}{RGB}{190,190,190}
\path[fill=fillColor,fill opacity=0.60] 
        ( 53.7,71.6) --
        ( 55.64, 80.56) --
	( 58.15, 79.11) --
	( 59.65, 77.80) --
	( 61.15, 76.81) --
	( 62.65, 75.91) --
	( 64.15, 75.27) --
	( 65.65, 74.78) --
	( 67.15, 74.31) --
	( 68.65, 73.91) --
	( 70.15, 73.47) --
	( 71.65, 73.12) --
	( 73.15, 72.83) --
	( 74.66, 72.54) --
	( 76.16, 72.26) --
	( 77.66, 72.08) --
	( 79.16, 71.89) --
	( 80.66, 71.69) --
	( 82.16, 71.41) --
	( 83.66, 71.26) --
	( 85.16, 71.09) --
	( 86.66, 70.97) --
	( 88.16, 70.85) --
	( 89.66, 70.75) --
	( 91.17, 70.65) --
	( 92.67, 70.52) --
	( 94.17, 70.45) --
	( 95.67, 70.36) --
	( 97.17, 70.23) --
	( 98.67, 70.14) --
	(100.17, 70.05) --
	(101.67, 69.96) --
	(103.17, 69.92) --
	(104.67, 69.86) --
	(106.17, 69.81) --
	(107.68, 69.70) --
	(109.18, 69.66) --
	(110.68, 69.56) --
	(112.18, 69.51) --
	(113.68, 69.44) --
	(115.18, 69.39) --
	(116.68, 69.35) --
	(118.18, 69.32) --
	(119.68, 69.28) --
	(121.18, 69.23) --
	(122.68, 69.18) --
	(124.19, 69.14) --
	(125.69, 69.12) --
	(127.19, 69.10) --
	(128.69, 69.09) --
	(130.19, 69.05) --
	(131.69, 69.03) --
	(133.19, 69.03) --
	(134.69, 69.01) --
	(136.19, 68.97) --
	(137.69, 68.93) --
	(139.19, 68.90) --
	(140.70, 68.86) --
	(142.20, 68.83) --
	(143.70, 68.81) --
	(145.20, 68.78) --
	(146.70, 68.75) --
	(148.20, 68.70) --
	(149.70, 68.68) --
	(151.20, 68.64) --
	(152.70, 68.61) --
	(154.20, 68.57) --
	(155.70, 68.58) --
	(157.21, 68.55) --
	(158.71, 68.55) --
	(160.21, 68.53) --
	(161.71, 68.52) --
	(163.21, 68.49) --
	(164.71, 68.47) --
	(166.21, 68.46) --
	(167.71, 68.43) --
	(169.21, 68.45) --
	(170.71, 68.42) --
	(172.21, 68.40) --
	(173.72, 68.36) --
	(175.22, 68.31) --
	(176.72, 68.30) --
	(178.22, 68.27) --
	(179.72, 68.26) --
	(181.22, 68.25) --
	(182.72, 68.24) --
	(184.22, 68.25) --
	(185.72, 68.22) --
	(187.22, 68.20) --
	(188.72, 68.17) --
	(190.23, 68.16) --
	(191.73, 68.16) --
	(193.23, 68.14) --
	(194.73, 68.12) --
	(196.23, 68.11) --
	(197.73, 68.10) --
	(199.23, 68.09) --
	(200.73, 68.08) --
	(202.23, 68.07) --
	(203.73, 68.07) --
	(203.73, 75.06) --
	(202.23, 75.07) --
	(200.73, 75.06) --
	(199.23, 75.02) --
	(197.73, 74.99) --
	(196.23, 74.96) --
	(194.73, 74.96) --
	(193.23, 74.94) --
	(191.73, 74.90) --
	(190.23, 74.92) --
	(188.72, 74.94) --
	(187.22, 74.96) --
	(185.72, 74.97) --
	(184.22, 74.96) --
	(182.72, 75.00) --
	(181.22, 75.01) --
	(179.72, 75.02) --
	(178.22, 75.04) --
	(176.72, 75.04) --
	(175.22, 75.05) --
	(173.72, 75.09) --
	(172.21, 75.12) --
	(170.71, 75.12) --
	(169.21, 75.12) --
	(167.71, 75.16) --
	(166.21, 75.14) --
	(164.71, 75.15) --
	(163.21, 75.16) --
	(161.71, 75.15) --
	(160.21, 75.14) --
	(158.71, 75.11) --
	(157.21, 75.09) --
	(155.70, 75.06) --
	(154.20, 75.07) --
	(152.70, 75.02) --
	(151.20, 75.01) --
	(149.70, 74.99) --
	(148.20, 75.00) --
	(146.70, 75.06) --
	(145.20, 75.12) --
	(143.70, 75.17) --
	(142.20, 75.25) --
	(140.70, 75.29) --
	(139.19, 75.30) --
	(137.69, 75.29) --
	(136.19, 75.28) --
	(134.69, 75.23) --
	(133.19, 75.21) --
	(131.69, 75.26) --
	(130.19, 75.28) --
	(128.69, 75.26) --
	(127.19, 75.30) --
	(125.69, 75.37) --
	(124.19, 75.45) --
	(122.68, 75.41) --
	(121.18, 75.45) --
	(119.68, 75.47) --
	(118.18, 75.49) --
	(116.68, 75.49) --
	(115.18, 75.49) --
	(113.68, 75.51) --
	(112.18, 75.52) --
	(110.68, 75.61) --
	(109.18, 75.68) --
	(107.68, 75.70) --
	(106.17, 75.67) --
	(104.67, 75.63) --
	(103.17, 75.64) --
	(101.67, 75.66) --
	(100.17, 75.73) --
	( 98.67, 75.79) --
	( 97.17, 75.77) --
	( 95.67, 75.74) --
	( 94.17, 75.72) --
	( 92.67, 75.76) --
	( 91.17, 75.82) --
	( 89.66, 75.86) --
	( 88.16, 75.92) --
	( 86.66, 75.97) --
	( 85.16, 76.00) --
	( 83.66, 76.10) --
	( 82.16, 76.15) --
	( 80.66, 76.31) --
	( 79.16, 76.44) --
	( 77.66, 76.56) --
	( 76.16, 76.62) --
	( 74.66, 76.81) --
	( 73.15, 77.06) --
	( 71.65, 77.31) --
	( 70.15, 77.49) --
	( 68.65, 77.80) --
	( 67.15, 78.09) --
	( 65.65, 78.41) --
	( 64.15, 78.71) --
	( 62.65, 79.20) --
	( 61.15, 79.88) --
	( 59.65, 80.68) --
	( 58.15, 81.74) --
	( 56.64, 82.81) --
	( 54.14, 83.47) --
        ( 53.2,71.6) --
	cycle;
\definecolor{drawColor}{cmyk}{.80,.29,.05,0}
\path[draw=drawColor, loosely dotted, line width= 1.2pt,line join=round,line cap=round] (  0.00, 80.97) -- (234.88, 80.97);
\end{scope}

\begin{scope}
\definecolor{drawColor}{RGB}{0,0,0}
\path[draw=drawColor,line width= 0.4pt,line join=round,line cap=round] 
        ( 53.5,72) --
	( 55.14, 82.65) --
	( 56.64, 81.69) --
	( 58.15, 80.42) --
	( 59.65, 79.24) --
	( 61.15, 78.34) --
	( 62.65, 77.56) --
	( 64.15, 76.99) --
	( 65.65, 76.59) --
	( 67.15, 76.20) --
	( 68.65, 75.85) --
	( 70.15, 75.48) --
	( 71.65, 75.22) --
	( 73.15, 74.95) --
	( 74.66, 74.67) --
	( 76.16, 74.44) --
	( 77.66, 74.32) --
	( 79.16, 74.17) --
	( 80.66, 74.00) --
	( 82.16, 73.78) --
	( 83.66, 73.68) --
	( 85.16, 73.55) --
	( 86.66, 73.47) --
	( 88.16, 73.38) --
	( 89.66, 73.31) --
	( 91.17, 73.24) --
	( 92.67, 73.14) --
	( 94.17, 73.09) --
	( 95.67, 73.05) --
	( 97.17, 73.00) --
	( 98.67, 72.97) --
	(100.17, 72.89) --
	(101.67, 72.81) --
	(103.17, 72.78) --
	(104.67, 72.75) --
	(106.17, 72.74) --
	(107.68, 72.70) --
	(109.18, 72.67) --
	(110.68, 72.59) --
	(112.18, 72.51) --
	(113.68, 72.48) --
	(115.18, 72.44) --
	(116.68, 72.42) --
	(118.18, 72.40) --
	(119.68, 72.38) --
	(121.18, 72.34) --
	(122.68, 72.30) --
	(124.19, 72.30) --
	(125.69, 72.25) --
	(127.19, 72.20) --
	(128.69, 72.18) --
	(130.19, 72.16) --
	(131.69, 72.15) --
	(133.19, 72.12) --
	(134.69, 72.12) --
	(136.19, 72.12) --
	(137.69, 72.11) --
	(139.19, 72.10) --
	(140.70, 72.07) --
	(142.20, 72.04) --
	(143.70, 71.99) --
	(145.20, 71.95) --
	(146.70, 71.91) --
	(148.20, 71.85) --
	(149.70, 71.83) --
	(151.20, 71.83) --
	(152.70, 71.82) --
	(154.20, 71.82) --
	(155.70, 71.82) --
	(157.21, 71.82) --
	(158.71, 71.83) --
	(160.21, 71.84) --
	(161.71, 71.84) --
	(163.21, 71.83) --
	(164.71, 71.81) --
	(166.21, 71.80) --
	(167.71, 71.80) --
	(169.21, 71.78) --
	(170.71, 71.77) --
	(172.21, 71.76) --
	(173.72, 71.73) --
	(175.22, 71.68) --
	(176.72, 71.67) --
	(178.22, 71.65) --
	(179.72, 71.64) --
	(181.22, 71.63) --
	(182.72, 71.62) --
	(184.22, 71.60) --
	(185.72, 71.59) --
	(187.22, 71.58) --
	(188.72, 71.56) --
	(190.23, 71.54) --
	(191.73, 71.53) --
	(193.23, 71.54) --
	(194.73, 71.54) --
	(196.23, 71.54) --
	(197.73, 71.55) --
	(199.23, 71.55) --
	(200.73, 71.57) --
	(202.23, 71.57) --
	(203.73, 71.56);
\end{scope}

\begin{scope}
\definecolor{drawColor}{RGB}{0,0,0}

\path[draw=drawColor,line width= 0.4pt,line join=round,line cap=round] ( 53.64, 61.20) -- (203.73, 61.20);

\path[draw=drawColor,line width= 0.4pt,line join=round,line cap=round] ( 53.64, 61.20) -- ( 53.64, 55.20);

\path[draw=drawColor,line width= 0.4pt,line join=round,line cap=round] ( 83.66, 61.20) -- ( 83.66, 55.20);

\path[draw=drawColor,line width= 0.4pt,line join=round,line cap=round] (113.68, 61.20) -- (113.68, 55.20);

\path[draw=drawColor,line width= 0.4pt,line join=round,line cap=round] (143.70, 61.20) -- (143.70, 55.20);

\path[draw=drawColor,line width= 0.4pt,line join=round,line cap=round] (173.72, 61.20) -- (173.72, 55.20);

\path[draw=drawColor,line width= 0.4pt,line join=round,line cap=round] (203.73, 61.20) -- (203.73, 55.20);

\node[text=drawColor,anchor=base,inner sep=0pt, outer sep=0pt, scale=  1.00] at ( 53.64, 39.60) {0};

\node[text=drawColor,anchor=base,inner sep=0pt, outer sep=0pt, scale=  1.00] at ( 83.66, 39.60) {20};

\node[text=drawColor,anchor=base,inner sep=0pt, outer sep=0pt, scale=  1.00] at (113.68, 39.60) {40};

\node[text=drawColor,anchor=base,inner sep=0pt, outer sep=0pt, scale=  1.00] at (143.70, 39.60) {60};

\node[text=drawColor,anchor=base,inner sep=0pt, outer sep=0pt, scale=  1.00] at (173.72, 39.60) {80};

\node[text=drawColor,anchor=base,inner sep=0pt, outer sep=0pt, scale=  1.00] at (203.73, 39.60) {100};

\path[draw=drawColor,line width= 0.4pt,line join=round,line cap=round] ( 49.20, 65.81) -- ( 49.20,181.07);

\path[draw=drawColor,line width= 0.4pt,line join=round,line cap=round] ( 49.20, 65.81) -- ( 43.20, 65.81);

\path[draw=drawColor,line width= 0.4pt,line join=round,line cap=round] ( 49.20, 88.86) -- ( 43.20, 88.86);

\path[draw=drawColor,line width= 0.4pt,line join=round,line cap=round] ( 49.20,111.91) -- ( 43.20,111.91);

\path[draw=drawColor,line width= 0.4pt,line join=round,line cap=round] ( 49.20,134.96) -- ( 43.20,134.96);

\path[draw=drawColor,line width= 0.4pt,line join=round,line cap=round] ( 49.20,158.02) -- ( 43.20,158.02);

\path[draw=drawColor,line width= 0.4pt,line join=round,line cap=round] ( 49.20,181.07) -- ( 43.20,181.07);

\node[text=drawColor,rotate= 90.00,anchor=base,inner sep=0pt, outer sep=0pt, scale=  1.00] at ( 34.80, 65.81) {0.0};

\node[text=drawColor,rotate= 90.00,anchor=base,inner sep=0pt, outer sep=0pt, scale=  1.00] at ( 34.80, 94.625) {0.1};

\node[text=drawColor,rotate= 90.00,anchor=base,inner sep=0pt, outer sep=0pt, scale=  1.00] at ( 34.80,123.44) {0.2};

\node[text=drawColor,rotate= 90.00,anchor=base,inner sep=0pt, outer sep=0pt, scale=  1.00] at ( 34.80,152.255) {0.3};

\node[text=drawColor,rotate= 90.00,anchor=base,inner sep=0pt, outer sep=0pt, scale=  1.00] at ( 34.80,181.07) {0.4};

\path[draw=drawColor,line width= 0.4pt,line join=round,line cap=round] ( 49.20, 61.20) --
	(209.68, 61.20) --
	(209.68,185.68) --
	( 49.20,185.68) --
	cycle;
\end{scope}

\end{tikzpicture}}};
    \node at (2,-3.5) {\scalebox{0.7}{\begin{tikzpicture}[x=1pt,y=1pt]
\definecolor{fillColor}{RGB}{255,255,255}

\begin{scope}
\path[clip] ( 49.20, 61.20) rectangle (209.68,185.68);
\definecolor{fillColor}{RGB}{190,190,190}
\path[fill=fillColor,fill opacity=0.60] 
        ( 53.7,69.9) --
        ( 55.44, 83.02) --
	( 56.64, 86.66) --
	( 58.15, 90.95) --
	( 59.65, 95.34) --
	( 61.15, 99.80) --
	( 62.65,104.29) --
	( 64.15,108.79) --
	( 65.65,113.03) --
	( 67.15,116.90) --
	( 68.65,120.36) --
	( 70.15,123.40) --
	( 71.65,126.02) --
	( 73.15,128.27) --
	( 74.66,130.00) --
	( 76.16,131.51) --
	( 77.66,132.74) --
	( 79.16,133.74) --
	( 80.66,134.59) --
	( 82.16,135.20) --
	( 83.66,135.73) --
	( 85.16,136.16) --
	( 86.66,136.48) --
	( 88.16,136.77) --
	( 89.66,137.00) --
	( 91.17,137.12) --
	( 92.67,137.29) --
	( 94.17,137.36) --
	( 95.67,137.48) --
	( 97.17,137.57) --
	( 98.67,137.62) --
	(100.17,137.70) --
	(101.67,137.78) --
	(103.17,137.75) --
	(104.67,137.75) --
	(106.17,137.77) --
	(107.68,137.81) --
	(109.18,137.82) --
	(110.68,137.80) --
	(112.18,137.75) --
	(113.68,137.75) --
	(115.18,137.77) --
	(116.68,137.78) --
	(118.18,137.80) --
	(119.68,137.81) --
	(121.18,137.82) --
	(122.68,137.82) --
	(124.19,137.77) --
	(125.69,137.77) --
	(127.19,137.83) --
	(128.69,137.84) --
	(130.19,137.84) --
	(131.69,137.88) --
	(133.19,137.91) --
	(134.69,137.89) --
	(136.19,137.89) --
	(137.69,137.87) --
	(139.19,137.88) --
	(140.70,137.82) --
	(142.20,137.78) --
	(143.70,137.75) --
	(145.20,137.78) --
	(146.70,137.79) --
	(148.20,137.81) --
	(149.70,137.77) --
	(151.20,137.84) --
	(152.70,137.86) --
	(154.20,137.83) --
	(155.70,137.78) --
	(157.21,137.77) --
	(158.71,137.76) --
	(160.21,137.72) --
	(161.71,137.74) --
	(163.21,137.71) --
	(164.71,137.77) --
	(166.21,137.82) --
	(167.71,137.86) --
	(169.21,137.88) --
	(170.71,137.86) --
	(172.21,137.87) --
	(173.72,137.91) --
	(175.22,138.00) --
	(176.72,137.99) --
	(178.22,138.02) --
	(179.72,138.03) --
	(181.22,137.95) --
	(182.72,137.98) --
	(184.22,137.89) --
	(185.72,137.82) --
	(187.22,137.77) --
	(188.72,137.78) --
	(190.23,137.77) --
	(191.73,137.77) --
	(193.23,137.75) --
	(194.73,137.73) --
	(196.23,137.74) --
	(197.73,137.79) --
	(199.23,137.79) --
	(200.73,137.77) --
	(202.23,137.74) --
	(203.73,137.74) --
	(203.73,148.34) --
	(202.23,148.28) --
	(200.73,148.34) --
	(199.23,148.31) --
	(197.73,148.35) --
	(196.23,148.32) --
	(194.73,148.31) --
	(193.23,148.35) --
	(191.73,148.30) --
	(190.23,148.29) --
	(188.72,148.36) --
	(187.22,148.39) --
	(185.72,148.43) --
	(184.22,148.44) --
	(182.72,148.47) --
	(181.22,148.49) --
	(179.72,148.53) --
	(178.22,148.55) --
	(176.72,148.55) --
	(175.22,148.51) --
	(173.72,148.45) --
	(172.21,148.42) --
	(170.71,148.44) --
	(169.21,148.44) --
	(167.71,148.42) --
	(166.21,148.36) --
	(164.71,148.35) --
	(163.21,148.27) --
	(161.71,148.26) --
	(160.21,148.27) --
	(158.71,148.29) --
	(157.21,148.33) --
	(155.70,148.37) --
	(154.20,148.45) --
	(152.70,148.43) --
	(151.20,148.38) --
	(149.70,148.35) --
	(148.20,148.39) --
	(146.70,148.39) --
	(145.20,148.33) --
	(143.70,148.41) --
	(142.20,148.38) --
	(140.70,148.43) --
	(139.19,148.43) --
	(137.69,148.42) --
	(136.19,148.38) --
	(134.69,148.41) --
	(133.19,148.41) --
	(131.69,148.35) --
	(130.19,148.33) --
	(128.69,148.33) --
	(127.19,148.36) --
	(125.69,148.37) --
	(124.19,148.37) --
	(122.68,148.40) --
	(121.18,148.35) --
	(119.68,148.38) --
	(118.18,148.28) --
	(116.68,148.22) --
	(115.18,148.23) --
	(113.68,148.28) --
	(112.18,148.29) --
	(110.68,148.31) --
	(109.18,148.34) --
	(107.68,148.33) --
	(106.17,148.26) --
	(104.67,148.28) --
	(103.17,148.30) --
	(101.67,148.25) --
	(100.17,148.16) --
	( 98.67,148.08) --
	( 97.17,147.99) --
	( 95.67,147.95) --
	( 94.17,147.90) --
	( 92.67,147.74) --
	( 91.17,147.61) --
	( 89.66,147.43) --
	( 88.16,147.21) --
	( 86.66,146.94) --
	( 85.16,146.58) --
	( 83.66,146.21) --
	( 82.16,145.66) --
	( 80.66,144.96) --
	( 79.16,144.18) --
	( 77.66,143.21) --
	( 76.16,142.04) --
	( 74.66,140.51) --
	( 73.15,138.66) --
	( 71.65,136.49) --
	( 70.15,133.75) --
	( 68.65,130.50) --
	( 67.15,126.76) --
	( 65.65,122.40) --
	( 64.15,117.51) --
	( 62.65,112.28) --
	( 61.15,106.72) --
	( 59.65,101.23) --
	( 58.15, 95.69) --
	( 56.64, 90.20) --
	( 54.54, 84.33) --
        ( 53.3,69.9) --
	cycle;
\definecolor{drawColor}{cmyk}{.80,.29,.05,0}
\path[draw=drawColor, loosely dotted, line width=1pt,line join=round,line cap=round] (  0.00, 65.81) -- (234.88, 65.81);
\end{scope}

\begin{scope}
\definecolor{drawColor}{RGB}{0,0,0}
\path[draw=drawColor,line width= 0.4pt,line join=round,line cap=round] 
        ( 53.5,70) --
        ( 55.14, 84.18) --
	( 56.64, 88.43) --
	( 58.15, 93.32) --
	( 59.65, 98.28) --
	( 61.15,103.26) --
	( 62.65,108.29) --
	( 64.15,113.15) --
	( 65.65,117.72) --
	( 67.15,121.83) --
	( 68.65,125.43) --
	( 70.15,128.57) --
	( 71.65,131.26) --
	( 73.15,133.47) --
	( 74.66,135.26) --
	( 76.16,136.78) --
	( 77.66,137.97) --
	( 79.16,138.96) --
	( 80.66,139.77) --
	( 82.16,140.43) --
	( 83.66,140.97) --
	( 85.16,141.37) --
	( 86.66,141.71) --
	( 88.16,141.99) --
	( 89.66,142.21) --
	( 91.17,142.36) --
	( 92.67,142.52) --
	( 94.17,142.63) --
	( 95.67,142.72) --
	( 97.17,142.78) --
	( 98.67,142.85) --
	(100.17,142.93) --
	(101.67,143.02) --
	(103.17,143.02) --
	(104.67,143.02) --
	(106.17,143.02) --
	(107.68,143.07) --
	(109.18,143.08) --
	(110.68,143.06) --
	(112.18,143.02) --
	(113.68,143.01) --
	(115.18,143.00) --
	(116.68,143.00) --
	(118.18,143.04) --
	(119.68,143.10) --
	(121.18,143.09) --
	(122.68,143.11) --
	(124.19,143.07) --
	(125.69,143.07) --
	(127.19,143.10) --
	(128.69,143.09) --
	(130.19,143.08) --
	(131.69,143.12) --
	(133.19,143.16) --
	(134.69,143.15) --
	(136.19,143.14) --
	(137.69,143.15) --
	(139.19,143.16) --
	(140.70,143.13) --
	(142.20,143.08) --
	(143.70,143.08) --
	(145.20,143.05) --
	(146.70,143.09) --
	(148.20,143.10) --
	(149.70,143.06) --
	(151.20,143.11) --
	(152.70,143.15) --
	(154.20,143.14) --
	(155.70,143.08) --
	(157.21,143.05) --
	(158.71,143.02) --
	(160.21,142.99) --
	(161.71,143.00) --
	(163.21,142.99) --
	(164.71,143.06) --
	(166.21,143.09) --
	(167.71,143.14) --
	(169.21,143.16) --
	(170.71,143.15) --
	(172.21,143.14) --
	(173.72,143.18) --
	(175.22,143.26) --
	(176.72,143.27) --
	(178.22,143.29) --
	(179.72,143.28) --
	(181.22,143.22) --
	(182.72,143.22) --
	(184.22,143.16) --
	(185.72,143.12) --
	(187.22,143.08) --
	(188.72,143.07) --
	(190.23,143.03) --
	(191.73,143.03) --
	(193.23,143.05) --
	(194.73,143.02) --
	(196.23,143.03) --
	(197.73,143.07) --
	(199.23,143.05) --
	(200.73,143.05) --
	(202.23,143.01) --
	(203.73,143.04);
\end{scope}

\begin{scope}
\definecolor{drawColor}{RGB}{0,0,0}

\path[draw=drawColor,line width= 0.4pt,line join=round,line cap=round] ( 53.64, 61.20) -- (203.73, 61.20);

\path[draw=drawColor,line width= 0.4pt,line join=round,line cap=round] ( 53.64, 61.20) -- ( 53.64, 55.20);

\path[draw=drawColor,line width= 0.4pt,line join=round,line cap=round] ( 83.66, 61.20) -- ( 83.66, 55.20);

\path[draw=drawColor,line width= 0.4pt,line join=round,line cap=round] (113.68, 61.20) -- (113.68, 55.20);

\path[draw=drawColor,line width= 0.4pt,line join=round,line cap=round] (143.70, 61.20) -- (143.70, 55.20);

\path[draw=drawColor,line width= 0.4pt,line join=round,line cap=round] (173.72, 61.20) -- (173.72, 55.20);

\path[draw=drawColor,line width= 0.4pt,line join=round,line cap=round] (203.73, 61.20) -- (203.73, 55.20);

\node[text=drawColor,anchor=base,inner sep=0pt, outer sep=0pt, scale=  1.00] at ( 53.64, 39.60) {0};

\node[text=drawColor,anchor=base,inner sep=0pt, outer sep=0pt, scale=  1.00] at ( 83.66, 39.60) {20};

\node[text=drawColor,anchor=base,inner sep=0pt, outer sep=0pt, scale=  1.00] at (113.68, 39.60) {40};

\node[text=drawColor,anchor=base,inner sep=0pt, outer sep=0pt, scale=  1.00] at (143.70, 39.60) {60};

\node[text=drawColor,anchor=base,inner sep=0pt, outer sep=0pt, scale=  1.00] at (173.72, 39.60) {80};

\node[text=drawColor,anchor=base,inner sep=0pt, outer sep=0pt, scale=  1.00] at (203.73, 39.60) {100};

\path[draw=drawColor,line width= 0.4pt,line join=round,line cap=round] ( 49.20, 65.81) -- ( 49.20,181.07);

\path[draw=drawColor,line width= 0.4pt,line join=round,line cap=round] ( 49.20, 65.81) -- ( 43.20, 65.81);

\path[draw=drawColor,line width= 0.4pt,line join=round,line cap=round] ( 49.20, 88.86) -- ( 43.20, 88.86);

\path[draw=drawColor,line width= 0.4pt,line join=round,line cap=round] ( 49.20,111.91) -- ( 43.20,111.91);

\path[draw=drawColor,line width= 0.4pt,line join=round,line cap=round] ( 49.20,134.96) -- ( 43.20,134.96);

\path[draw=drawColor,line width= 0.4pt,line join=round,line cap=round] ( 49.20,158.02) -- ( 43.20,158.02);

\path[draw=drawColor,line width= 0.4pt,line join=round,line cap=round] ( 49.20,181.07) -- ( 43.20,181.07);

\node[text=drawColor,rotate= 90.00,anchor=base,inner sep=0pt, outer sep=0pt, scale=  1.00] at ( 34.80, 65.81) {0.0};

\node[text=drawColor,rotate= 90.00,anchor=base,inner sep=0pt, outer sep=0pt, scale=  1.00] at ( 34.80, 94.625) {0.1};

\node[text=drawColor,rotate= 90.00,anchor=base,inner sep=0pt, outer sep=0pt, scale=  1.00] at ( 34.80,123.44) {0.2};

\node[text=drawColor,rotate= 90.00,anchor=base,inner sep=0pt, outer sep=0pt, scale=  1.00] at ( 34.80,152.255) {0.3};

\node[text=drawColor,rotate= 90.00,anchor=base,inner sep=0pt, outer sep=0pt, scale=  1.00] at ( 34.80,181.07) {0.4};

\path[draw=drawColor,line width= 0.4pt,line join=round,line cap=round] ( 49.20, 61.20) --
	(209.68, 61.20) --
	(209.68,185.68) --
	( 49.20,185.68) --
	cycle;
\end{scope}

\end{tikzpicture}}};
    \node at (2,-7.5) {\scalebox{0.7}{\begin{tikzpicture}[x=1pt,y=1pt]
\definecolor{fillColor}{RGB}{255,255,255}

\begin{scope}
\path[clip] ( 49.20, 61.20) rectangle (209.68,185.68);
\definecolor{fillColor}{RGB}{190,190,190}
\path[fill=fillColor,dotted, line width= 0.8pt,fill opacity=0.60] ( 55.14, 65.81) --
	( 56.64, 65.81) --
	( 58.15, 65.81) --
	( 59.65, 65.81) --
	( 61.15, 65.81) --
	( 62.65, 65.81) --
	( 64.15, 65.81) --
	( 65.65, 65.81) --
	( 67.15, 65.81) --
	( 68.65, 65.81) --
	( 70.15, 65.81) --
	( 71.65, 65.81) --
	( 73.15, 65.81) --
	( 74.66, 65.81) --
	( 76.16, 65.81) --
	( 77.66, 65.81) --
	( 79.16, 65.81) --
	( 80.66, 65.81) --
	( 82.16, 65.81) --
	( 83.66, 65.81) --
	( 85.16, 65.81) --
	( 86.66, 65.81) --
	( 88.16, 65.81) --
	( 89.66, 65.81) --
	( 91.17, 65.81) --
	( 92.67, 65.81) --
	( 94.17, 65.81) --
	( 95.67, 65.81) --
	( 97.17, 65.81) --
	( 98.67, 65.81) --
	(100.17, 65.81) --
	(101.67, 65.81) --
	(103.17, 65.81) --
	(104.67, 65.81) --
	(106.17, 65.81) --
	(107.68, 65.81) --
	(109.18, 65.81) --
	(110.68, 65.81) --
	(112.18, 65.81) --
	(113.68, 65.81) --
	(115.18, 65.81) --
	(116.68, 65.81) --
	(118.18, 65.81) --
	(119.68, 65.81) --
	(121.18, 65.81) --
	(122.68, 65.81) --
	(124.19, 65.81) --
	(125.69, 65.81) --
	(127.19, 65.81) --
	(128.69, 65.81) --
	(130.19, 65.81) --
	(131.69, 65.81) --
	(133.19, 65.81) --
	(134.69, 65.81) --
	(136.19, 65.81) --
	(137.69, 65.81) --
	(139.19, 65.81) --
	(140.70, 65.81) --
	(142.20, 65.81) --
	(143.70, 65.81) --
	(145.20, 65.81) --
	(146.70, 65.81) --
	(148.20, 65.81) --
	(149.70, 65.81) --
	(151.20, 65.81) --
	(152.70, 65.81) --
	(154.20, 65.81) --
	(155.70, 65.81) --
	(157.21, 65.81) --
	(158.71, 65.81) --
	(160.21, 65.81) --
	(161.71, 65.81) --
	(163.21, 65.81) --
	(164.71, 65.81) --
	(166.21, 65.81) --
	(167.71, 65.81) --
	(169.21, 65.81) --
	(170.71, 65.81) --
	(172.21, 65.81) --
	(173.72, 65.81) --
	(175.22, 65.81) --
	(176.72, 65.81) --
	(178.22, 65.81) --
	(179.72, 65.81) --
	(181.22, 65.81) --
	(182.72, 65.81) --
	(184.22, 65.81) --
	(185.72, 65.81) --
	(187.22, 65.81) --
	(188.72, 65.81) --
	(190.23, 65.81) --
	(191.73, 65.81) --
	(193.23, 65.81) --
	(194.73, 65.81) --
	(196.23, 65.81) --
	(197.73, 65.81) --
	(199.23, 65.81) --
	(200.73, 65.81) --
	(202.23, 65.81) --
	(203.73, 65.81) --
	(203.73, 65.81) --
	(202.23, 65.81) --
	(200.73, 65.81) --
	(199.23, 65.81) --
	(197.73, 65.81) --
	(196.23, 65.81) --
	(194.73, 65.81) --
	(193.23, 65.81) --
	(191.73, 65.81) --
	(190.23, 65.81) --
	(188.72, 65.81) --
	(187.22, 65.81) --
	(185.72, 65.81) --
	(184.22, 65.81) --
	(182.72, 65.81) --
	(181.22, 65.81) --
	(179.72, 65.81) --
	(178.22, 65.81) --
	(176.72, 65.81) --
	(175.22, 65.81) --
	(173.72, 65.81) --
	(172.21, 65.81) --
	(170.71, 65.81) --
	(169.21, 65.81) --
	(167.71, 65.81) --
	(166.21, 65.81) --
	(164.71, 65.81) --
	(163.21, 65.81) --
	(161.71, 65.81) --
	(160.21, 65.81) --
	(158.71, 65.81) --
	(157.21, 65.81) --
	(155.70, 65.81) --
	(154.20, 65.81) --
	(152.70, 65.81) --
	(151.20, 65.81) --
	(149.70, 65.81) --
	(148.20, 65.81) --
	(146.70, 65.81) --
	(145.20, 65.81) --
	(143.70, 65.81) --
	(142.20, 65.81) --
	(140.70, 65.81) --
	(139.19, 65.81) --
	(137.69, 65.81) --
	(136.19, 65.81) --
	(134.69, 65.81) --
	(133.19, 65.81) --
	(131.69, 65.81) --
	(130.19, 65.81) --
	(128.69, 65.81) --
	(127.19, 65.81) --
	(125.69, 65.81) --
	(124.19, 65.81) --
	(122.68, 65.81) --
	(121.18, 65.81) --
	(119.68, 65.81) --
	(118.18, 65.81) --
	(116.68, 65.81) --
	(115.18, 65.81) --
	(113.68, 65.81) --
	(112.18, 65.81) --
	(110.68, 65.81) --
	(109.18, 65.81) --
	(107.68, 65.81) --
	(106.17, 65.81) --
	(104.67, 65.81) --
	(103.17, 65.81) --
	(101.67, 65.81) --
	(100.17, 65.81) --
	( 98.67, 65.81) --
	( 97.17, 65.81) --
	( 95.67, 65.81) --
	( 94.17, 65.81) --
	( 92.67, 65.81) --
	( 91.17, 65.81) --
	( 89.66, 65.81) --
	( 88.16, 65.81) --
	( 86.66, 65.81) --
	( 85.16, 65.81) --
	( 83.66, 65.81) --
	( 82.16, 65.81) --
	( 80.66, 65.81) --
	( 79.16, 65.81) --
	( 77.66, 65.81) --
	( 76.16, 65.81) --
	( 74.66, 65.81) --
	( 73.15, 65.81) --
	( 71.65, 65.81) --
	( 70.15, 65.81) --
	( 68.65, 65.81) --
	( 67.15, 65.81) --
	( 65.65, 65.81) --
	( 64.15, 65.81) --
	( 62.65, 65.81) --
	( 61.15, 65.81) --
	( 59.65, 65.81) --
	( 58.15, 65.81) --
	( 56.64, 65.81) --
	( 55.14, 65.81) --
	cycle;
\definecolor{drawColor}{cmyk}{.80,.29,.05,0}
\path[draw=drawColor, loosely dotted, line width= 1.2pt,line join=round,line cap=round] (  0.00, 65.81) -- (234.88, 65.81);
\end{scope}

\begin{scope}
\definecolor{drawColor}{RGB}{0,0,0}
\path[draw=drawColor, line width= 0.4pt,line join=round,line cap=round] ( 55.14, 65.81) --
	( 56.64, 65.81) --
	( 58.15, 65.81) --
	( 59.65, 65.81) --
	( 61.15, 65.81) --
	( 62.65, 65.81) --
	( 64.15, 65.81) --
	( 65.65, 65.81) --
	( 67.15, 65.81) --
	( 68.65, 65.81) --
	( 70.15, 65.81) --
	( 71.65, 65.81) --
	( 73.15, 65.81) --
	( 74.66, 65.81) --
	( 76.16, 65.81) --
	( 77.66, 65.81) --
	( 79.16, 65.81) --
	( 80.66, 65.81) --
	( 82.16, 65.81) --
	( 83.66, 65.81) --
	( 85.16, 65.81) --
	( 86.66, 65.81) --
	( 88.16, 65.81) --
	( 89.66, 65.81) --
	( 91.17, 65.81) --
	( 92.67, 65.81) --
	( 94.17, 65.81) --
	( 95.67, 65.81) --
	( 97.17, 65.81) --
	( 98.67, 65.81) --
	(100.17, 65.81) --
	(101.67, 65.81) --
	(103.17, 65.81) --
	(104.67, 65.81) --
	(106.17, 65.81) --
	(107.68, 65.81) --
	(109.18, 65.81) --
	(110.68, 65.81) --
	(112.18, 65.81) --
	(113.68, 65.81) --
	(115.18, 65.81) --
	(116.68, 65.81) --
	(118.18, 65.81) --
	(119.68, 65.81) --
	(121.18, 65.81) --
	(122.68, 65.81) --
	(124.19, 65.81) --
	(125.69, 65.81) --
	(127.19, 65.81) --
	(128.69, 65.81) --
	(130.19, 65.81) --
	(131.69, 65.81) --
	(133.19, 65.81) --
	(134.69, 65.81) --
	(136.19, 65.81) --
	(137.69, 65.81) --
	(139.19, 65.81) --
	(140.70, 65.81) --
	(142.20, 65.81) --
	(143.70, 65.81) --
	(145.20, 65.81) --
	(146.70, 65.81) --
	(148.20, 65.81) --
	(149.70, 65.81) --
	(151.20, 65.81) --
	(152.70, 65.81) --
	(154.20, 65.81) --
	(155.70, 65.81) --
	(157.21, 65.81) --
	(158.71, 65.81) --
	(160.21, 65.81) --
	(161.71, 65.81) --
	(163.21, 65.81) --
	(164.71, 65.81) --
	(166.21, 65.81) --
	(167.71, 65.81) --
	(169.21, 65.81) --
	(170.71, 65.81) --
	(172.21, 65.81) --
	(173.72, 65.81) --
	(175.22, 65.81) --
	(176.72, 65.81) --
	(178.22, 65.81) --
	(179.72, 65.81) --
	(181.22, 65.81) --
	(182.72, 65.81) --
	(184.22, 65.81) --
	(185.72, 65.81) --
	(187.22, 65.81) --
	(188.72, 65.81) --
	(190.23, 65.81) --
	(191.73, 65.81) --
	(193.23, 65.81) --
	(194.73, 65.81) --
	(196.23, 65.81) --
	(197.73, 65.81) --
	(199.23, 65.81) --
	(200.73, 65.81) --
	(202.23, 65.81) --
	(203.73, 65.81);
\end{scope}

\begin{scope}
\definecolor{drawColor}{RGB}{0,0,0}

\path[draw=drawColor,line width= 0.4pt,line join=round,line cap=round] ( 53.64, 61.20) -- (203.73, 61.20);

\path[draw=drawColor,line width= 0.4pt,line join=round,line cap=round] ( 53.64, 61.20) -- ( 53.64, 55.20);

\path[draw=drawColor,line width= 0.4pt,line join=round,line cap=round] ( 83.66, 61.20) -- ( 83.66, 55.20);

\path[draw=drawColor,line width= 0.4pt,line join=round,line cap=round] (113.68, 61.20) -- (113.68, 55.20);

\path[draw=drawColor,line width= 0.4pt,line join=round,line cap=round] (143.70, 61.20) -- (143.70, 55.20);

\path[draw=drawColor,line width= 0.4pt,line join=round,line cap=round] (173.72, 61.20) -- (173.72, 55.20);

\path[draw=drawColor,line width= 0.4pt,line join=round,line cap=round] (203.73, 61.20) -- (203.73, 55.20);

\node[text=drawColor,anchor=base,inner sep=0pt, outer sep=0pt, scale=  1.00] at ( 53.64, 39.60) {0};

\node[text=drawColor,anchor=base,inner sep=0pt, outer sep=0pt, scale=  1.00] at ( 83.66, 39.60) {20};

\node[text=drawColor,anchor=base,inner sep=0pt, outer sep=0pt, scale=  1.00] at (113.68, 39.60) {40};

\node[text=drawColor,anchor=base,inner sep=0pt, outer sep=0pt, scale=  1.00] at (143.70, 39.60) {60};

\node[text=drawColor,anchor=base,inner sep=0pt, outer sep=0pt, scale=  1.00] at (173.72, 39.60) {80};

\node[text=drawColor,anchor=base,inner sep=0pt, outer sep=0pt, scale=  1.00] at (203.73, 39.60) {100};

\path[draw=drawColor,line width= 0.4pt,line join=round,line cap=round] ( 49.20, 65.81) -- ( 49.20,181.07);

\path[draw=drawColor,line width= 0.4pt,line join=round,line cap=round] ( 49.20, 65.81) -- ( 43.20, 65.81);

\path[draw=drawColor,line width= 0.4pt,line join=round,line cap=round] ( 49.20, 88.86) -- ( 43.20, 88.86);

\path[draw=drawColor,line width= 0.4pt,line join=round,line cap=round] ( 49.20,111.91) -- ( 43.20,111.91);

\path[draw=drawColor,line width= 0.4pt,line join=round,line cap=round] ( 49.20,134.96) -- ( 43.20,134.96);

\path[draw=drawColor,line width= 0.4pt,line join=round,line cap=round] ( 49.20,158.02) -- ( 43.20,158.02);

\path[draw=drawColor,line width= 0.4pt,line join=round,line cap=round] ( 49.20,181.07) -- ( 43.20,181.07);

\node[text=drawColor,rotate= 90.00,anchor=base,inner sep=0pt, outer sep=0pt, scale=  1.00] at ( 34.80, 65.81) {0.0};

\node[text=drawColor,rotate= 90.00,anchor=base,inner sep=0pt, outer sep=0pt, scale=  1.00] at ( 34.80, 94.625) {0.1};

\node[text=drawColor,rotate= 90.00,anchor=base,inner sep=0pt, outer sep=0pt, scale=  1.00] at ( 34.80,123.44) {0.2};

\node[text=drawColor,rotate= 90.00,anchor=base,inner sep=0pt, outer sep=0pt, scale=  1.00] at ( 34.80,152.255) {0.3};

\node[text=drawColor,rotate= 90.00,anchor=base,inner sep=0pt, outer sep=0pt, scale=  1.00] at ( 34.80,181.07) {0.4};

\path[draw=drawColor,line width= 0.4pt,line join=round,line cap=round] ( 49.20, 61.20) --
	(209.68, 61.20) --
	(209.68,185.68) --
	( 49.20,185.68) --
	cycle;
\end{scope}

\end{tikzpicture}}};
    \node at (7,-3.5) {\scalebox{0.7}{\input{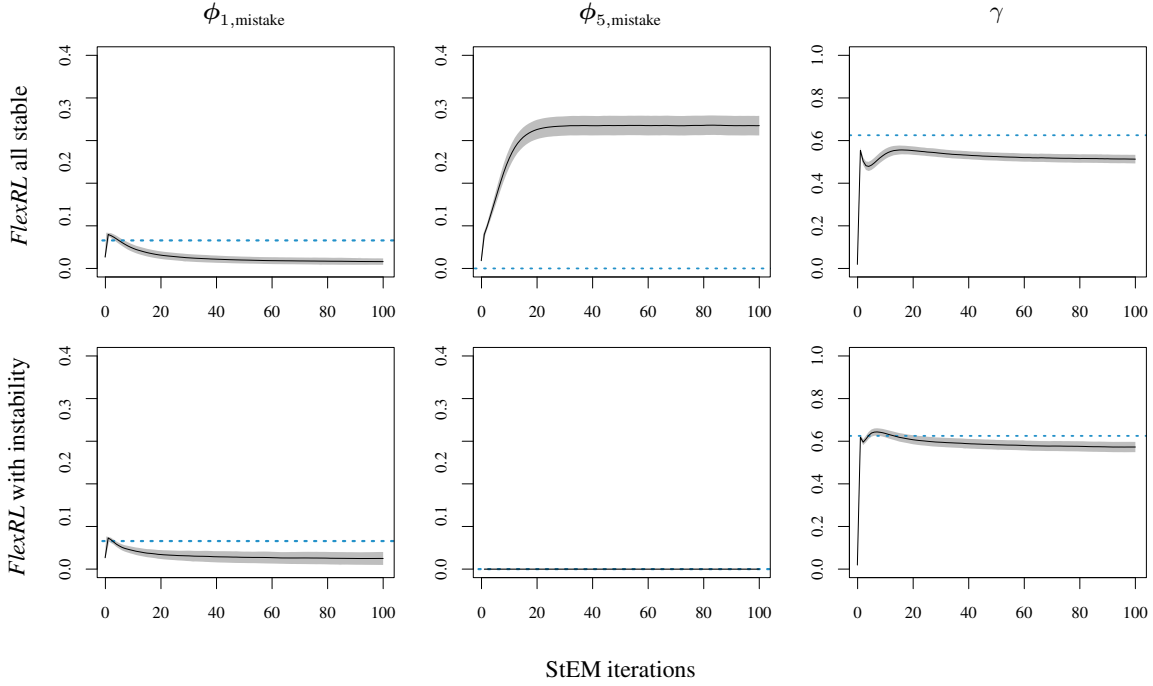}}};
    \node at (7,-7.5) {\scalebox{0.7}{\begin{tikzpicture}[x=1pt,y=1pt]
\definecolor{fillColor}{RGB}{255,255,255}

\begin{scope}
\path[clip] ( 49.20, 61.20) rectangle (209.68,185.68);
\definecolor{fillColor}{RGB}{190,190,190}
\path[fill=fillColor,fill opacity=0.60] ( 53.5,68) --
( 55.14,135.86) --
	( 56.64,132.89) --
	( 58.15,134.38) --
	( 59.65,136.23) --
	( 61.15,137.44) --
	( 62.65,137.92) --
	( 64.15,137.99) --
	( 65.65,137.81) --
	( 67.15,137.48) --
	( 68.65,137.07) --
	( 70.15,136.53) --
	( 71.65,136.09) --
	( 73.15,135.67) --
	( 74.66,135.24) --
	( 76.16,134.77) --
	( 77.66,134.48) --
	( 79.16,134.20) --
	( 80.66,133.92) --
	( 82.16,133.60) --
	( 83.66,133.31) --
	( 85.16,133.11) --
	( 86.66,132.91) --
	( 88.16,132.74) --
	( 89.66,132.62) --
	( 91.17,132.47) --
	( 92.67,132.30) --
	( 94.17,132.14) --
	( 95.67,132.06) --
	( 97.17,131.95) --
	( 98.67,131.87) --
	(100.17,131.79) --
	(101.67,131.67) --
	(103.17,131.61) --
	(104.67,131.53) --
	(106.17,131.46) --
	(107.68,131.39) --
	(109.18,131.32) --
	(110.68,131.21) --
	(112.18,131.10) --
	(113.68,131.02) --
	(115.18,130.93) --
	(116.68,130.89) --
	(118.18,130.81) --
	(119.68,130.75) --
	(121.18,130.67) --
	(122.68,130.64) --
	(124.19,130.58) --
	(125.69,130.57) --
	(127.19,130.48) --
	(128.69,130.43) --
	(130.19,130.38) --
	(131.69,130.33) --
	(133.19,130.30) --
	(134.69,130.25) --
	(136.19,130.24) --
	(137.69,130.20) --
	(139.19,130.17) --
	(140.70,130.11) --
	(142.20,130.02) --
	(143.70,130.00) --
	(145.20,129.93) --
	(146.70,129.85) --
	(148.20,129.78) --
	(149.70,129.77) --
	(151.20,129.77) --
	(152.70,129.73) --
	(154.20,129.70) --
	(155.70,129.65) --
	(157.21,129.65) --
	(158.71,129.65) --
	(160.21,129.64) --
	(161.71,129.64) --
	(163.21,129.59) --
	(164.71,129.58) --
	(166.21,129.53) --
	(167.71,129.50) --
	(169.21,129.52) --
	(170.71,129.49) --
	(172.21,129.48) --
	(173.72,129.43) --
	(175.22,129.38) --
	(176.72,129.35) --
	(178.22,129.30) --
	(179.72,129.28) --
	(181.22,129.24) --
	(182.72,129.24) --
	(184.22,129.20) --
	(185.72,129.18) --
	(187.22,129.17) --
	(188.72,129.12) --
	(190.23,129.09) --
	(191.73,129.04) --
	(193.23,129.04) --
	(194.73,129.03) --
	(196.23,129.04) --
	(197.73,129.01) --
	(199.23,129.02) --
	(200.73,129.03) --
	(202.23,129.05) --
	(203.73,129.05) --
	(203.73,134.61) --
	(202.23,134.61) --
	(200.73,134.64) --
	(199.23,134.63) --
	(197.73,134.62) --
	(196.23,134.62) --
	(194.73,134.61) --
	(193.23,134.62) --
	(191.73,134.60) --
	(190.23,134.63) --
	(188.72,134.66) --
	(187.22,134.71) --
	(185.72,134.73) --
	(184.22,134.78) --
	(182.72,134.75) --
	(181.22,134.75) --
	(179.72,134.78) --
	(178.22,134.82) --
	(176.72,134.88) --
	(175.22,134.95) --
	(173.72,134.95) --
	(172.21,134.98) --
	(170.71,134.97) --
	(169.21,134.99) --
	(167.71,135.03) --
	(166.21,135.03) --
	(164.71,135.01) --
	(163.21,135.08) --
	(161.71,135.06) --
	(160.21,135.09) --
	(158.71,135.06) --
	(157.21,135.01) --
	(155.70,135.05) --
	(154.20,135.05) --
	(152.70,135.07) --
	(151.20,135.10) --
	(149.70,135.15) --
	(148.20,135.18) --
	(146.70,135.24) --
	(145.20,135.24) --
	(143.70,135.32) --
	(142.20,135.40) --
	(140.70,135.44) --
	(139.19,135.48) --
	(137.69,135.49) --
	(136.19,135.50) --
	(134.69,135.50) --
	(133.19,135.56) --
	(131.69,135.62) --
	(130.19,135.65) --
	(128.69,135.67) --
	(127.19,135.77) --
	(125.69,135.80) --
	(124.19,135.90) --
	(122.68,135.90) --
	(121.18,135.98) --
	(119.68,136.04) --
	(118.18,136.08) --
	(116.68,136.13) --
	(115.18,136.16) --
	(113.68,136.22) --
	(112.18,136.28) --
	(110.68,136.41) --
	(109.18,136.49) --
	(107.68,136.50) --
	(106.17,136.57) --
	(104.67,136.65) --
	(103.17,136.70) --
	(101.67,136.76) --
	(100.17,136.86) --
	( 98.67,136.92) --
	( 97.17,136.99) --
	( 95.67,137.04) --
	( 94.17,137.22) --
	( 92.67,137.30) --
	( 91.17,137.42) --
	( 89.66,137.60) --
	( 88.16,137.73) --
	( 86.66,137.85) --
	( 85.16,138.01) --
	( 83.66,138.21) --
	( 82.16,138.38) --
	( 80.66,138.58) --
	( 79.16,138.85) --
	( 77.66,139.14) --
	( 76.16,139.38) --
	( 74.66,139.74) --
	( 73.15,140.14) --
	( 71.65,140.51) --
	( 70.15,140.89) --
	( 68.65,141.26) --
	( 67.15,141.62) --
	( 65.65,141.84) --
	( 64.15,141.97) --
	( 62.65,141.86) --
	( 61.15,141.33) --
	( 59.65,140.02) --
	( 58.15,137.93) --
	( 56.64,136.02) --
	( 55.14,138.45) --
        ( 53.5,68) --
	cycle;
\definecolor{drawColor}{cmyk}{.80,.29,.05,0}
\path[draw=drawColor, loosely dotted, line width=1pt, line join=round,line cap=round] ( 49.20,137.85) -- (209.68,137.85);
\end{scope}

\begin{scope}
\definecolor{drawColor}{RGB}{0,0,0}
\path[draw=drawColor,line width= 0.4pt,line join=round,line cap=round] ( 53.5,68) --
	  ( 55.14,137.15) --
	( 56.64,134.45) --
	( 58.15,136.16) --
	( 59.65,138.12) --
	( 61.15,139.39) --
	( 62.65,139.89) --
	( 64.15,139.98) --
	( 65.65,139.83) --
	( 67.15,139.55) --
	( 68.65,139.17) --
	( 70.15,138.71) --
	( 71.65,138.30) --
	( 73.15,137.90) --
	( 74.66,137.49) --
	( 76.16,137.08) --
	( 77.66,136.81) --
	( 79.16,136.52) --
	( 80.66,136.25) --
	( 82.16,135.99) --
	( 83.66,135.76) --
	( 85.16,135.56) --
	( 86.66,135.38) --
	( 88.16,135.24) --
	( 89.66,135.11) --
	( 91.17,134.95) --
	( 92.67,134.80) --
	( 94.17,134.68) --
	( 95.67,134.55) --
	( 97.17,134.47) --
	( 98.67,134.40) --
	(100.17,134.32) --
	(101.67,134.21) --
	(103.17,134.15) --
	(104.67,134.09) --
	(106.17,134.01) --
	(107.68,133.95) --
	(109.18,133.90) --
	(110.68,133.81) --
	(112.18,133.69) --
	(113.68,133.62) --
	(115.18,133.54) --
	(116.68,133.51) --
	(118.18,133.44) --
	(119.68,133.40) --
	(121.18,133.32) --
	(122.68,133.27) --
	(124.19,133.24) --
	(125.69,133.18) --
	(127.19,133.12) --
	(128.69,133.05) --
	(130.19,133.01) --
	(131.69,132.97) --
	(133.19,132.93) --
	(134.69,132.88) --
	(136.19,132.87) --
	(137.69,132.85) --
	(139.19,132.82) --
	(140.70,132.77) --
	(142.20,132.71) --
	(143.70,132.66) --
	(145.20,132.58) --
	(146.70,132.55) --
	(148.20,132.48) --
	(149.70,132.46) --
	(151.20,132.44) --
	(152.70,132.40) --
	(154.20,132.38) --
	(155.70,132.35) --
	(157.21,132.33) --
	(158.71,132.36) --
	(160.21,132.36) --
	(161.71,132.35) --
	(163.21,132.34) --
	(164.71,132.30) --
	(166.21,132.28) --
	(167.71,132.26) --
	(169.21,132.26) --
	(170.71,132.23) --
	(172.21,132.23) --
	(173.72,132.19) --
	(175.22,132.16) --
	(176.72,132.12) --
	(178.22,132.06) --
	(179.72,132.03) --
	(181.22,132.00) --
	(182.72,132.00) --
	(184.22,131.99) --
	(185.72,131.95) --
	(187.22,131.94) --
	(188.72,131.89) --
	(190.23,131.86) --
	(191.73,131.82) --
	(193.23,131.83) --
	(194.73,131.82) --
	(196.23,131.83) --
	(197.73,131.81) --
	(199.23,131.83) --
	(200.73,131.83) --
	(202.23,131.83) --
	(203.73,131.83);
\end{scope}

\begin{scope}
\definecolor{drawColor}{RGB}{0,0,0}

\path[draw=drawColor,line width= 0.4pt,line join=round,line cap=round] ( 53.64, 61.20) -- (203.73, 61.20);

\path[draw=drawColor,line width= 0.4pt,line join=round,line cap=round] ( 53.64, 61.20) -- ( 53.64, 55.20);

\path[draw=drawColor,line width= 0.4pt,line join=round,line cap=round] ( 83.66, 61.20) -- ( 83.66, 55.20);

\path[draw=drawColor,line width= 0.4pt,line join=round,line cap=round] (113.68, 61.20) -- (113.68, 55.20);

\path[draw=drawColor,line width= 0.4pt,line join=round,line cap=round] (143.70, 61.20) -- (143.70, 55.20);

\path[draw=drawColor,line width= 0.4pt,line join=round,line cap=round] (173.72, 61.20) -- (173.72, 55.20);

\path[draw=drawColor,line width= 0.4pt,line join=round,line cap=round] (203.73, 61.20) -- (203.73, 55.20);

\node[text=drawColor,anchor=base,inner sep=0pt, outer sep=0pt, scale=  1.00] at ( 53.64, 39.60) {0};

\node[text=drawColor,anchor=base,inner sep=0pt, outer sep=0pt, scale=  1.00] at ( 83.66, 39.60) {20};

\node[text=drawColor,anchor=base,inner sep=0pt, outer sep=0pt, scale=  1.00] at (113.68, 39.60) {40};

\node[text=drawColor,anchor=base,inner sep=0pt, outer sep=0pt, scale=  1.00] at (143.70, 39.60) {60};

\node[text=drawColor,anchor=base,inner sep=0pt, outer sep=0pt, scale=  1.00] at (173.72, 39.60) {80};

\node[text=drawColor,anchor=base,inner sep=0pt, outer sep=0pt, scale=  1.00] at (203.73, 39.60) {100};

\path[draw=drawColor,line width= 0.4pt,line join=round,line cap=round] ( 49.20, 65.81) -- ( 49.20,181.07);

\path[draw=drawColor,line width= 0.4pt,line join=round,line cap=round] ( 49.20, 65.81) -- ( 43.20, 65.81);

\path[draw=drawColor,line width= 0.4pt,line join=round,line cap=round] ( 49.20, 88.86) -- ( 43.20, 88.86);

\path[draw=drawColor,line width= 0.4pt,line join=round,line cap=round] ( 49.20,111.91) -- ( 43.20,111.91);

\path[draw=drawColor,line width= 0.4pt,line join=round,line cap=round] ( 49.20,134.96) -- ( 43.20,134.96);

\path[draw=drawColor,line width= 0.4pt,line join=round,line cap=round] ( 49.20,158.02) -- ( 43.20,158.02);

\path[draw=drawColor,line width= 0.4pt,line join=round,line cap=round] ( 49.20,181.07) -- ( 43.20,181.07);

\node[text=drawColor,rotate= 90.00,anchor=base,inner sep=0pt, outer sep=0pt, scale=  1.00] at ( 34.80, 65.81) {0.0};

\node[text=drawColor,rotate= 90.00,anchor=base,inner sep=0pt, outer sep=0pt, scale=  1.00] at ( 34.80, 88.86) {0.2};

\node[text=drawColor,rotate= 90.00,anchor=base,inner sep=0pt, outer sep=0pt, scale=  1.00] at ( 34.80,111.91) {0.4};

\node[text=drawColor,rotate= 90.00,anchor=base,inner sep=0pt, outer sep=0pt, scale=  1.00] at ( 34.80,134.96) {0.6};

\node[text=drawColor,rotate= 90.00,anchor=base,inner sep=0pt, outer sep=0pt, scale=  1.00] at ( 34.80,158.02) {0.8};

\node[text=drawColor,rotate= 90.00,anchor=base,inner sep=0pt, outer sep=0pt, scale=  1.00] at ( 34.80,181.07) {1.0};

\path[draw=drawColor,line width= 0.4pt,line join=round,line cap=round] ( 49.20, 61.20) --
	(209.68, 61.20) --
	(209.68,185.68) --
	( 49.20,185.68) --
	cycle;
\end{scope}

\end{tikzpicture}}};
    \node at (2.2,-10) {StEM iterations};
    \end{tikzpicture}
    \caption{Probability of mistake represented by the parameter $\boldsymbol{\phi}_{\text{mistake}}$ for a stable PIV (here the $1^{\text{st}}$ one) on the left and the unstable $5^{\text{th}}$ PIV on the middle. Probability for a record in $\mathcal{A}$ to form a link with a record in $\mathcal{B}$ represented by the parameter $\gamma$ on the right. On the top is \textit{FlexRL} considering all PIVs stable. On the bottom is \textit{FlexRL} taking into account the instability of the $5^{\text{th}}$ PIV (in which case $\boldsymbol{\phi}_{5,\text{mistake}}=0$ is known). The dotted thick line is the true probability. The solid thin line is the averaged estimated probability over the 500 simulations and the faded interval around corresponds to the probability standard deviation, showing the simulations noise. We use the 25 last values of the parameter to build the final estimate of the linkage (we discard 75 iterations as burn-in).}
    \label{fig_graph_cv_param_phi_gamma}
\end{figure}

\begin{figure}[!h]
    \centering
    \begin{tikzpicture}
    % \node[rotate=90] at (-3.8,-7.85) {\textit{FlexRL} with instability};
    % \node at (4.1,-5.55) {\textit{FlexRL} with instability};
    \node at (0,-7.85) {\scalebox{0.7}{\input{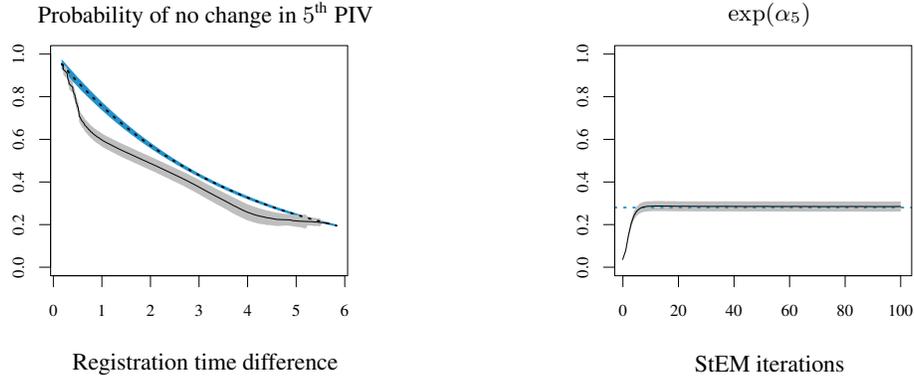}}};
    \node at (0.30,-10.3) {Registration time difference};
    \node at (7.5,-7.85) {\scalebox{0.7}{\begin{tikzpicture}[x=1pt,y=1pt]
\definecolor{fillColor}{RGB}{255,255,255}

\begin{scope}
\path[clip] ( 49.20, 61.20) rectangle (209.68,185.68);
\definecolor{fillColor}{RGB}{190,190,190}
\path[fill=fillColor,fill opacity=0.60] ( 53.5,70) --
( 55.14, 74.27) --
	( 56.64, 81.77) --
	( 58.15, 87.70) --
	( 59.65, 91.50) --
	( 61.15, 93.68) --
	( 62.65, 94.88) --
	( 64.15, 95.54) --
	( 65.65, 95.86) --
	( 67.15, 96.02) --
	( 68.65, 96.08) --
	( 70.15, 96.18) --
	( 71.65, 96.18) --
	( 73.15, 96.18) --
	( 74.66, 96.17) --
	( 76.16, 96.16) --
	( 77.66, 96.16) --
	( 79.16, 96.14) --
	( 80.66, 96.09) --
	( 82.16, 96.09) --
	( 83.66, 96.04) --
	( 85.16, 96.06) --
	( 86.66, 96.07) --
	( 88.16, 96.08) --
	( 89.66, 96.04) --
	( 91.17, 96.05) --
	( 92.67, 96.03) --
	( 94.17, 95.96) --
	( 95.67, 96.02) --
	( 97.17, 96.02) --
	( 98.67, 95.99) --
	(100.17, 96.01) --
	(101.67, 95.99) --
	(103.17, 95.96) --
	(104.67, 96.00) --
	(106.17, 95.99) --
	(107.68, 95.97) --
	(109.18, 95.97) --
	(110.68, 96.02) --
	(112.18, 95.98) --
	(113.68, 95.99) --
	(115.18, 95.98) --
	(116.68, 95.94) --
	(118.18, 95.97) --
	(119.68, 95.95) --
	(121.18, 95.93) --
	(122.68, 95.99) --
	(124.19, 95.97) --
	(125.69, 95.96) --
	(127.19, 95.96) --
	(128.69, 95.97) --
	(130.19, 95.96) --
	(131.69, 95.99) --
	(133.19, 95.94) --
	(134.69, 95.94) --
	(136.19, 95.93) --
	(137.69, 95.93) --
	(139.19, 95.94) --
	(140.70, 95.96) --
	(142.20, 95.95) --
	(143.70, 95.94) --
	(145.20, 95.95) --
	(146.70, 95.96) --
	(148.20, 95.94) --
	(149.70, 95.91) --
	(151.20, 95.91) --
	(152.70, 95.90) --
	(154.20, 95.88) --
	(155.70, 95.90) --
	(157.21, 95.91) --
	(158.71, 95.90) --
	(160.21, 95.86) --
	(161.71, 95.88) --
	(163.21, 95.88) --
	(164.71, 95.89) --
	(166.21, 95.89) --
	(167.71, 95.89) --
	(169.21, 95.88) --
	(170.71, 95.87) --
	(172.21, 95.88) --
	(173.72, 95.90) --
	(175.22, 95.91) --
	(176.72, 95.93) --
	(178.22, 95.92) --
	(179.72, 95.92) --
	(181.22, 95.92) --
	(182.72, 95.94) --
	(184.22, 95.89) --
	(185.72, 95.89) --
	(187.22, 95.90) --
	(188.72, 95.86) --
	(190.23, 95.87) --
	(191.73, 95.86) --
	(193.23, 95.88) --
	(194.73, 95.90) --
	(196.23, 95.87) --
	(197.73, 95.90) --
	(199.23, 95.89) --
	(200.73, 95.90) --
	(202.23, 95.92) --
	(203.73, 95.89) --
	(203.73,101.35) --
	(202.23,101.36) --
	(200.73,101.36) --
	(199.23,101.36) --
	(197.73,101.32) --
	(196.23,101.33) --
	(194.73,101.35) --
	(193.23,101.36) --
	(191.73,101.35) --
	(190.23,101.33) --
	(188.72,101.36) --
	(187.22,101.35) --
	(185.72,101.38) --
	(184.22,101.37) --
	(182.72,101.38) --
	(181.22,101.36) --
	(179.72,101.37) --
	(178.22,101.40) --
	(176.72,101.38) --
	(175.22,101.33) --
	(173.72,101.37) --
	(172.21,101.40) --
	(170.71,101.39) --
	(169.21,101.36) --
	(167.71,101.38) --
	(166.21,101.39) --
	(164.71,101.33) --
	(163.21,101.31) --
	(161.71,101.31) --
	(160.21,101.33) --
	(158.71,101.39) --
	(157.21,101.39) --
	(155.70,101.39) --
	(154.20,101.36) --
	(152.70,101.38) --
	(151.20,101.40) --
	(149.70,101.36) --
	(148.20,101.37) --
	(146.70,101.42) --
	(145.20,101.39) --
	(143.70,101.36) --
	(142.20,101.34) --
	(140.70,101.36) --
	(139.19,101.40) --
	(137.69,101.35) --
	(136.19,101.38) --
	(134.69,101.38) --
	(133.19,101.36) --
	(131.69,101.43) --
	(130.19,101.43) --
	(128.69,101.41) --
	(127.19,101.41) --
	(125.69,101.41) --
	(124.19,101.40) --
	(122.68,101.37) --
	(121.18,101.38) --
	(119.68,101.39) --
	(118.18,101.39) --
	(116.68,101.42) --
	(115.18,101.43) --
	(113.68,101.38) --
	(112.18,101.39) --
	(110.68,101.42) --
	(109.18,101.48) --
	(107.68,101.44) --
	(106.17,101.44) --
	(104.67,101.43) --
	(103.17,101.41) --
	(101.67,101.38) --
	(100.17,101.37) --
	( 98.67,101.42) --
	( 97.17,101.41) --
	( 95.67,101.42) --
	( 94.17,101.39) --
	( 92.67,101.46) --
	( 91.17,101.41) --
	( 89.66,101.45) --
	( 88.16,101.44) --
	( 86.66,101.47) --
	( 85.16,101.47) --
	( 83.66,101.44) --
	( 82.16,101.47) --
	( 80.66,101.50) --
	( 79.16,101.55) --
	( 77.66,101.54) --
	( 76.16,101.53) --
	( 74.66,101.63) --
	( 73.15,101.62) --
	( 71.65,101.61) --
	( 70.15,101.56) --
	( 68.65,101.48) --
	( 67.15,101.47) --
	( 65.65,101.25) --
	( 64.15,100.91) --
	( 62.65,100.12) --
	( 61.15, 98.70) --
	( 59.65, 96.11) --
	( 58.15, 91.53) --
	( 56.64, 84.33) --
	( 55.14, 75.27) --
        ( 53.5,70) --
	cycle;
\definecolor{drawColor}{cmyk}{.80,.29,.05,0}
\path[draw=drawColor, loosely dotted, line width=1pt, line join=round,line cap=round] ( 49.20, 98.08) -- (209.68, 98.08);
\end{scope}

\begin{scope}
\definecolor{drawColor}{RGB}{0,0,0}
\path[draw=drawColor,line width= 0.4pt,line join=round,line cap=round] ( 53.5,70) --
        ( 55.14, 74.77) --
	( 56.64, 83.05) --
	( 58.15, 89.62) --
	( 59.65, 93.80) --
	( 61.15, 96.19) --
	( 62.65, 97.50) --
	( 64.15, 98.23) --
	( 65.65, 98.56) --
	( 67.15, 98.74) --
	( 68.65, 98.78) --
	( 70.15, 98.87) --
	( 71.65, 98.90) --
	( 73.15, 98.90) --
	( 74.66, 98.90) --
	( 76.16, 98.84) --
	( 77.66, 98.85) --
	( 79.16, 98.84) --
	( 80.66, 98.80) --
	( 82.16, 98.78) --
	( 83.66, 98.74) --
	( 85.16, 98.77) --
	( 86.66, 98.77) --
	( 88.16, 98.76) --
	( 89.66, 98.75) --
	( 91.17, 98.73) --
	( 92.67, 98.74) --
	( 94.17, 98.68) --
	( 95.67, 98.72) --
	( 97.17, 98.72) --
	( 98.67, 98.70) --
	(100.17, 98.69) --
	(101.67, 98.69) --
	(103.17, 98.69) --
	(104.67, 98.71) --
	(106.17, 98.72) --
	(107.68, 98.71) --
	(109.18, 98.73) --
	(110.68, 98.72) --
	(112.18, 98.69) --
	(113.68, 98.69) --
	(115.18, 98.71) --
	(116.68, 98.68) --
	(118.18, 98.68) --
	(119.68, 98.67) --
	(121.18, 98.66) --
	(122.68, 98.68) --
	(124.19, 98.68) --
	(125.69, 98.69) --
	(127.19, 98.68) --
	(128.69, 98.69) --
	(130.19, 98.69) --
	(131.69, 98.71) --
	(133.19, 98.65) --
	(134.69, 98.66) --
	(136.19, 98.66) --
	(137.69, 98.64) --
	(139.19, 98.67) --
	(140.70, 98.66) --
	(142.20, 98.65) --
	(143.70, 98.65) --
	(145.20, 98.67) --
	(146.70, 98.69) --
	(148.20, 98.66) --
	(149.70, 98.64) --
	(151.20, 98.66) --
	(152.70, 98.64) --
	(154.20, 98.62) --
	(155.70, 98.64) --
	(157.21, 98.65) --
	(158.71, 98.64) --
	(160.21, 98.60) --
	(161.71, 98.59) --
	(163.21, 98.59) --
	(164.71, 98.61) --
	(166.21, 98.64) --
	(167.71, 98.63) --
	(169.21, 98.62) --
	(170.71, 98.63) --
	(172.21, 98.64) --
	(173.72, 98.63) --
	(175.22, 98.62) --
	(176.72, 98.66) --
	(178.22, 98.66) --
	(179.72, 98.64) --
	(181.22, 98.64) --
	(182.72, 98.66) --
	(184.22, 98.63) --
	(185.72, 98.64) --
	(187.22, 98.62) --
	(188.72, 98.61) --
	(190.23, 98.60) --
	(191.73, 98.60) --
	(193.23, 98.62) --
	(194.73, 98.63) --
	(196.23, 98.60) --
	(197.73, 98.61) --
	(199.23, 98.62) --
	(200.73, 98.63) --
	(202.23, 98.64) --
	(203.73, 98.62);
\end{scope}

\begin{scope}
\definecolor{drawColor}{RGB}{0,0,0}

\path[draw=drawColor,line width= 0.4pt,line join=round,line cap=round] ( 53.64, 61.20) -- (203.73, 61.20);

\path[draw=drawColor,line width= 0.4pt,line join=round,line cap=round] ( 53.64, 61.20) -- ( 53.64, 55.20);

\path[draw=drawColor,line width= 0.4pt,line join=round,line cap=round] ( 83.66, 61.20) -- ( 83.66, 55.20);

\path[draw=drawColor,line width= 0.4pt,line join=round,line cap=round] (113.68, 61.20) -- (113.68, 55.20);

\path[draw=drawColor,line width= 0.4pt,line join=round,line cap=round] (143.70, 61.20) -- (143.70, 55.20);

\path[draw=drawColor,line width= 0.4pt,line join=round,line cap=round] (173.72, 61.20) -- (173.72, 55.20);

\path[draw=drawColor,line width= 0.4pt,line join=round,line cap=round] (203.73, 61.20) -- (203.73, 55.20);

\node[text=drawColor,anchor=base,inner sep=0pt, outer sep=0pt, scale=  1.00] at ( 53.64, 39.60) {0};

\node[text=drawColor,anchor=base,inner sep=0pt, outer sep=0pt, scale=  1.00] at ( 83.66, 39.60) {20};

\node[text=drawColor,anchor=base,inner sep=0pt, outer sep=0pt, scale=  1.00] at (113.68, 39.60) {40};

\node[text=drawColor,anchor=base,inner sep=0pt, outer sep=0pt, scale=  1.00] at (143.70, 39.60) {60};

\node[text=drawColor,anchor=base,inner sep=0pt, outer sep=0pt, scale=  1.00] at (173.72, 39.60) {80};

\node[text=drawColor,anchor=base,inner sep=0pt, outer sep=0pt, scale=  1.00] at (203.73, 39.60) {100};

\path[draw=drawColor,line width= 0.4pt,line join=round,line cap=round] ( 49.20, 65.81) -- ( 49.20,181.07);

\path[draw=drawColor,line width= 0.4pt,line join=round,line cap=round] ( 49.20, 65.81) -- ( 43.20, 65.81);

\path[draw=drawColor,line width= 0.4pt,line join=round,line cap=round] ( 49.20, 88.86) -- ( 43.20, 88.86);

\path[draw=drawColor,line width= 0.4pt,line join=round,line cap=round] ( 49.20,111.91) -- ( 43.20,111.91);

\path[draw=drawColor,line width= 0.4pt,line join=round,line cap=round] ( 49.20,134.96) -- ( 43.20,134.96);

\path[draw=drawColor,line width= 0.4pt,line join=round,line cap=round] ( 49.20,158.02) -- ( 43.20,158.02);

\path[draw=drawColor,line width= 0.4pt,line join=round,line cap=round] ( 49.20,181.07) -- ( 43.20,181.07);

\node[text=drawColor,rotate= 90.00,anchor=base,inner sep=0pt, outer sep=0pt, scale=  1.00] at ( 34.80, 65.81) {0.0};

\node[text=drawColor,rotate= 90.00,anchor=base,inner sep=0pt, outer sep=0pt, scale=  1.00] at ( 34.80, 88.86) {0.2};

\node[text=drawColor,rotate= 90.00,anchor=base,inner sep=0pt, outer sep=0pt, scale=  1.00] at ( 34.80,111.91) {0.4};

\node[text=drawColor,rotate= 90.00,anchor=base,inner sep=0pt, outer sep=0pt, scale=  1.00] at ( 34.80,134.96) {0.6};

\node[text=drawColor,rotate= 90.00,anchor=base,inner sep=0pt, outer sep=0pt, scale=  1.00] at ( 34.80,158.02) {0.8};

\node[text=drawColor,rotate= 90.00,anchor=base,inner sep=0pt, outer sep=0pt, scale=  1.00] at ( 34.80,181.07) {1.0};

\path[draw=drawColor,line width= 0.4pt,line join=round,line cap=round] ( 49.20, 61.20) --
	(209.68, 61.20) --
	(209.68,185.68) --
	( 49.20,185.68) --
	cycle;
\end{scope}

\end{tikzpicture}}};
    \node at (7.8,-10.3) {StEM iterations};
    \node at (0.30,-5.65) {Probability of no change in $5^{\text{th}}$ PIV};
    %\node[rotate=90] at (11.6,-7.85) {};
    \node at (7.8,-5.65) {$\exp(\alpha_5)$};
    \end{tikzpicture}
    \caption{Baseline hazard $\exp(\alpha_5)$ involved in the survival function parametrizing the instability of the $5^{\text{th}}$ PIV on the right and survival function $S_{\alpha_5}$ modelling the instability on the left, in \textit{FlexRL} taking account of instability. The dotted thick line represent the truth. The solid thin line is the averaged estimated probability over the 500 simulations. The faded interval around corresponds to the probability standard deviation, showing the simulations noise. We use the 25 last values of the parameter to build the final estimate of the linkage (we discard 75 iterations as burn-in).}
    \label{fig_graph_cv_param_alpha}
\end{figure}

The simulated data summary in the upper \cref{tab_Simu_Story_And_Results} assesses the weakness of the PIVs. They all are categorical (numeric format), with a few unique values and registration errors (missing values and mistakes). As we can observe from \cref{fig_graph_cv_param_phi_gamma} and \cref{fig_graph_cv_param_alpha}, the parameters converge throughout our algorithm. We mentioned the parametrizations used to run the different methodologies at the beginning of the \cref{sec6: simulations and applications}. Though, we modified the prior for the distortion of the $5^{\text{th}}$ unstable PIV in \textit{Exchanger} and shifted it towards 0 since we assume a known null probability of mistake.

\subsubsection{Results} 
\label{subsubsec611: results}

The difficulty of the record linkage task simulated here is illustrated by the performance of the simplistic approach in the results---see the lower \cref{tab_Simu_Story_And_Results}. This simulation scenario reflects a real-life situation of low data quality, where the PIVs are non-uniformly distributed, categorical with few unique values, potentially unstable, and have missing values and mistakes. Note that, as the task becomes easier (more uniform distribution of the PIVs, more unique values, fewer registration errors), the performance gap between the methods narrows. 

\textit{BRL} and \textit{Exchanger} are able to link some pairs for which the $5^{\text{th}}$ PIV changed, though \textit{FlexRL} is able to detect more $TP$ with changes. \textit{BRL} for the rest is conservative and link the pairs that agree together while \textit{Exchanger} is more flexible and ventures in less certain areas, although not to benefit the $TP$. \textit{FlexRL} is more liberal than \textit{BRL} but makes more sensible links than \textit{Exchanger}.

When the instability of a PIV indexed by $k$ is not detected by the researcher or analyst i.e.\ not taken into account in the process, the algorithm explains the changes as mistakes. Indeed, we do blocking on the true latent values generated, which are going to correct for the changes as if they were mistakes so that the PIV dynamics are going to be incorporated in the parameters $\phi_{k,\text{mistake}}^{\mathcal{A}}, \phi_{k,\text{mistake}}^{\mathcal{B}}$ instead of $S_{\alpha_k}$. In practice we see on the lower \cref{tab_Simu_Story_And_Results} that this reduces the performance of the linkage estimate with moderation; the interval of performance variability are overlapping as we can notice with the standard deviation of the metrics. 

Moreover, when considering all PIVs stable, \textit{FlexRL} results in more $TP$ without too many more $FP$, leading to a good balance between FDR and sensitivity, as indicated by a high F1-score. This shows that modelling the data generation process (hence avoiding information reduction and inconsistencies due to comparison vectors, \citep{tancredi_bay_RL_2011}) overrides \textit{BRL} in situations where the record linkage task is not straightforward. Both \textit{Exchanger} and \textit{BRL} in such context do not perform better than the naive record linkage approach linking the pairs for which all PIVs agree, though they minimise the number of $FP$ and hence the FDR.

\subsection{Application: The National Long Term Care Survey (NLTCS)} 
\label{subsec62: application}

The NLTCS data are available under request to the National Archive of Computerized Data on Aging (NACDA); the survey was sponsored by the National Institute of Aging and was conducted by the Duke University Center for Demographic Studies under Grant No. U01-AG007198, \citep{NLTCS_NACDA}. It consist of six waves conducted between 1982 and 2004 with a sample size of about 20 000 per wave. A unique identifier is provided, allowing to report the performance of record linkage procedures. We use data from 1982 and 1994 to illustrate the record linkage task. These data sets gather six PIVs which can be used to link the data: sex, birth date (day, month and year), state code and regional code. They contain approximately 20 500 and 9 500 records respectively (after filtering the data to their common support), of which 7 500 are common to both files. Data from the NLTCS are often used in the record linkage literature, \citep{rl_nltcs_medical, smered2016, dblink_2021, bayRLFabl}. Record linkage was previously performed using the birth day, which makes the task rather trivial and is not realistic in a context where open-source data sets are pseudonymised following privacy regulations. Therefore, we only use birth month and birth year in our application.

\begin{table}[!ht]
    \centering
    \begin{tabular}{ccccccc}
        \hline
        \noalign{\vskip\doublerulesep \vskip-\arrayrulewidth}
        \multicolumn{2}{c}{\textbf{Registrations}} & \textbf{Sex} & \textbf{Birth month} & \textbf{Birth year} & \textbf{State code} & \textbf{Regional code} \\
        \hline
        \noalign{\vskip\doublerulesep \vskip-\arrayrulewidth}
        \multirow{2}{*}{\textbf{Data}} & \textbf{Unique} & $2$ & $12$ & $57$ & $58$ & $12$ \\
        & \textbf{Type} & categorical & categorical & categorical & categorical & categorical \\
        \cline{2-2}
        \noalign{\vskip\doublerulesep \vskip-\arrayrulewidth}
        \textbf{True Links} & \textbf{Agree} & $1$ & $1$ & $1$ & $.91$ & $.92$ \\
        \hline
        \noalign{\vskip\doublerulesep \vskip-\arrayrulewidth}
    \end{tabular}
    \caption{Summary of the full NLTCS data of 1982 and 1994. Characteristics of the PIVs and level of agreement among the 7519 links referring to the same individuals. There are a few missing values in the PIVs sex and state code but their proportion appear to be null, $2\%$ of regional code values are missing.}
    \label{tab_NLTCS_story}
\end{table}

We describe the data in \cref{tab_NLTCS_story} in which the number of unique values quantifies the discriminating strength of the PIVs. The proportion of agreements among true links refers to cases where the recorded value in $\mathcal{A}$ matches its counterpart in $\mathcal{B}$. If the values differ, it may indicate a mistake in the recorded information or a change in the value between the registration times of file $\mathcal{A}$ and file $\mathcal{B}$. We should claim that a PIV is unstable when the proportion of disagreements is too high to only be attributed to registration errors. In practice, this decision is based on common sense and requires to have access to explanatory variables to build a model alike $S_{\alpha_k}$ in our methodology. With the real data application, we cannot be certain about the classification of a disagreement (between change or mistake) and, given the lack of registration times in the data, we consider all PIVs stable. In view of the level of disagreements in the PIVs on \cref{tab_NLTCS_story} (less than $10\%$), this is a reasonable assumption.

When running on the complete data of 1982 and 1994, \textit{BRL} and \textit{Exchanger} encountered memory errors. Thus we compare the different methods on subsets of the data.

\subsubsection{Comparison with the literature on regional subsets}

In order to compare \textit{FlexRL} with the methods developed in the literature we divide the data sets into regional subsets. We show the variability in performance over 12 subsets of the data defined by their regional office: Boston, New York, Philadelphia, Detroit, Chicago, Kansas city, Seattle, Charlotte, Atlanta, Dallas, Denver, Los Angeles.

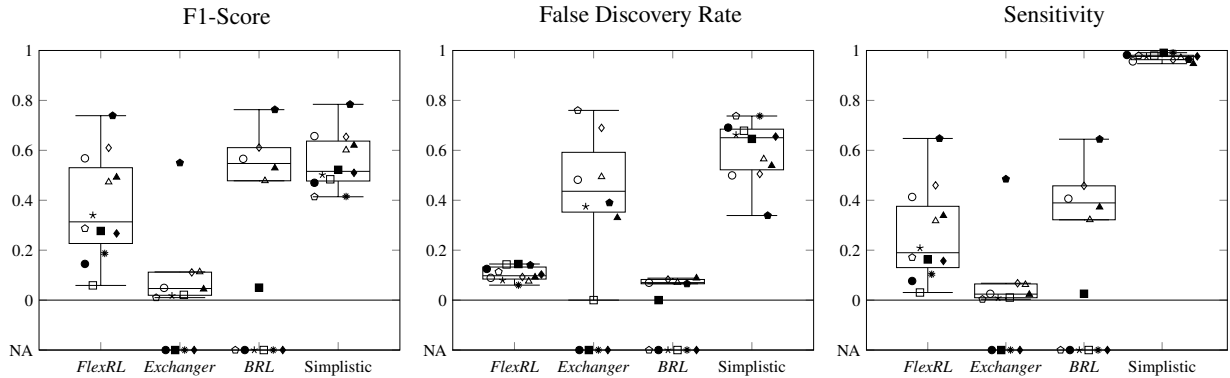
\begin{figure}[!h]
    \centering
    \begin{tikzpicture}
    \node at (0.3,-5.27) {F1-Score};
    \node at (5.8,-5.27) {False Discovery Rate};
    \node at (11.3,-5.27) {Sensitivity};
    \node at (0,-7.85) {\scalebox{0.7}{\begin{tikzpicture}[x=1pt,y=1pt]
      \begin{axis}
        [
        boxplot/draw direction=y,
        ymin=-0.2,ymax=1,
        ytick={-0.2,0,0.2,0.4,0.6,0.8,1},
        yticklabels={NA,0,0.2,0.4,0.6,0.8,1},
        xtick={1,2,3,4},
        xticklabels={\vphantom{\scalebox{1.3}{$^\star$}}\textit{FlexRL}, \vphantom{\scalebox{1.3}{$^\star$}}\textit{Exchanger}, \vphantom{\scalebox{1.3}{$^\star$}}\textit{BRL}, \vphantom{\scalebox{1.3}{$^\star$}}Simplistic},
        cycle list={{black},{black},{black},{black}},
        scatter/classes={
        a={mark=o, black},
        b={mark=otimes*, drawcolor},
        c={mark=star, drawcolor},
        d={mark=square, drawcolor},
        e={mark=square*, drawcolor},
        f={mark=triangle, black},
        g={mark=triangle*, black},
        h={mark=diamond, black},
        i={mark=diamond*, drawcolor},
        j={mark=pentagon, drawcolor},
        k={mark=pentagon*, black},
        l={mark=10-pointed star, drawcolor}
        }
        % scatter/classes={
        % a={mark=o},
        % b={mark=otimes*, black},
        % c={mark=star},
        % d={mark=square},
        % e={mark=square*, black},
        % f={mark=triangle},
        % g={mark=triangle*, black},
        % h={mark=diamond},
        % i={mark=diamond*, black},
        % j={mark=pentagon},
        % k={mark=pentagon*, black},
        % l={mark=10-pointed star}
        % }
        ]
        % FLEXRL
        \addplot+[,
        boxplot prepared={
          lower whisker  =0.0588,
          lower quartile =0.2268,
          median         =0.3135,
          upper quartile =0.5304,
          upper whisker  =0.7388
        },
        ] coordinates {};
        \addplot+[scatter, only marks, scatter src=explicit symbolic,
        ] coordinates {(0.8,0.568)[a](0.8,0.145)[b](0.9,0.340)[c](0.9,0.059)[d](1,0.277)[e](1.1,0.473)[f](1.2,0.492)[g](1.1,0.610)[h](1.2,0.267)[i](0.8,0.287)[j](1.15,0.739)[k](1.05,0.187)[l]};
        % EXCHANGER
        \addplot+[
        boxplot prepared={
          lower whisker  =0.01,
          lower quartile =0.01935,
          median         =0.04645,
          upper quartile =0.112,
          upper whisker  =0.113
        },
        ] coordinates {};
        \addplot+[scatter, only marks, scatter src=explicit symbolic,
        ] coordinates {
        (1.8,0.049)[a](1.82,-0.2)[b](1.9,0.0177)[c](2.05,0.021)[d](1.94,-0.2)[e](2.25,0.113)[f](2.3,0.0439)[g](2.15,0.111)[h](2.18,-0.2)[i](1.7,0.01)[j](2,0.55)[k](2.06,-0.2)[l]};
        % BRL
        \addplot+[
        boxplot prepared={
          lower whisker  =0.4780279,
          lower quartile =0.4780279,
          median         =0.5472934,
          upper quartile =0.6107635,
          upper whisker  =0.7629234
        },
        ] coordinates {};
        \addplot+[scatter, only marks, scatter src=explicit symbolic,
        ] coordinates {
        (2.8,0.566)[a](2.82,-0.2)[b](2.94,-0.2)[c](3.06,-0.2)[d](3,0.050)[e](3.075,0.478)[f](3.2,0.529)[g](3,0.611)[h](3.30,-0.2)[i](2.7,-0.2)[j](3.2,0.763)[k](3.18,-0.2)[l]};
        % SIMPLISTIC
        \addplot+[
        boxplot prepared={
          lower whisker  =0.4140,
          lower quartile =0.4771,
          median         =0.5158,
          upper quartile =0.6370,
          upper whisker  =0.7843
        },
        ] coordinates {};
        \addplot+[scatter, only marks, scatter src=explicit symbolic,
        ] coordinates {
        (3.7,0.657)[a](3.7,0.470)[b](3.8,0.502)[c](3.9,0.484)[d](4,0.522)[e](4.1,0.601)[f](4.2,0.62)[g](4.1,0.654)[h](4.2,0.51)[i](3.7,0.414)[j](4.15,0.784)[k](4.1,0.415)[l]};
        \path[draw] (-50, 0) rectangle (450,20);
      \end{axis}
      \end{tikzpicture}}};
    \node at (5.5,-7.85) {\scalebox{0.7}{\begin{tikzpicture}[x=1pt,y=1pt]
      \begin{axis}
        [
        boxplot/draw direction=y,
        ymin=-0.2,ymax=1,
        ytick={-0.2,0,0.2,0.4,0.6,0.8,1},
        yticklabels={NA,0,0.2,0.4,0.6,0.8,1},
        xtick={1,2,3,4},
        xticklabels={\vphantom{\scalebox{1.3}{$^\star$}}\textit{FlexRL}, \vphantom{\scalebox{1.3}{$^\star$}}\textit{Exchanger}, \vphantom{\scalebox{1.3}{$^\star$}}\textit{BRL}, \vphantom{\scalebox{1.3}{$^\star$}}Simplistic},
        cycle list={{black},{black},{black},{black}},
        scatter/classes={
        a={mark=o, black},
        b={mark=otimes*, drawcolor},
        c={mark=star, drawcolor},
        d={mark=square, drawcolor},
        e={mark=square*, drawcolor},
        f={mark=triangle, black},
        g={mark=triangle*, black},
        h={mark=diamond, black},
        i={mark=diamond*, drawcolor},
        j={mark=pentagon, drawcolor},
        k={mark=pentagon*, black},
        l={mark=10-pointed star, drawcolor}
        }
        % scatter/classes={
        % a={mark=o},
        % b={mark=otimes*, black},
        % c={mark=star},
        % d={mark=square},
        % e={mark=square*, black},
        % f={mark=triangle},
        % g={mark=triangle*, black},
        % h={mark=diamond},
        % i={mark=diamond*, black},
        % j={mark=pentagon},
        % k={mark=pentagon*, black},
        % l={mark=10-pointed star}
        % }
        % bajcdelhfkgi
        ]
        % FLEXRL
        \addplot+[
        boxplot prepared={
          lower whisker  =0.0600,
          lower quartile =0.0843,
          median         =0.0976,
          upper quartile =0.1325,
          upper whisker  =0.1446
        },
        ] coordinates {};
        \addplot+[scatter, only marks, scatter src=explicit symbolic,
        ] coordinates {(0.7,0.08921933)[a](0.65,0.12500000)[b](0.85,0.07936508)[c](0.9,0.14285714)[d](1.05,0.14457831)[e](1.18,0.07563025)[f](1.26,0.09183673)[g](1.1,0.09259259)[h](1.34,0.10256410)[i](0.8,0.11309524)[j](1.2,0.14000000)[k](1.05,0.06000000)[l]};
        % EXCHANGER
        \addplot+[
        boxplot prepared={
          lower whisker  =0.00,
          lower quartile =0.3525,
          median         =0.4360,
          upper quartile =0.5920,
          upper whisker  =0.7600
        },
        ] coordinates {};
        \addplot+[scatter, only marks, scatter src=explicit symbolic,
        ] coordinates {(1.8,0.482)[a](1.82,-0.2)[b](1.9,0.375)[c](2,0.000)[d](1.94,-0.2)[e](2.1,0.494)[f](2.3,0.330)[g](2.1,0.690)[h](2.18,-0.2)[i](1.8,0.760)[j](2.2,0.390)[k](2.06,-0.2)[l]};
        % BRL
        \addplot+[
        boxplot prepared={
          lower whisker  =0.06550218,
          lower quartile =0.06550218,
          median         =0.07016570,
          upper quartile =0.08270677,
          upper whisker  =0.08796296
        },
        ] coordinates {};
        \addplot+[scatter, only marks, scatter src=explicit symbolic,
        ] coordinates {(2.7,0.06949807)[a](2.82,-0.2)[b](2.94,-0.2)[c](3.06,-0.2)[d](2.82,0.0)[e](3.06,0.07083333)[f](3.3,0.08796296)[g](2.94,0.08270677)[h](3.30,-0.2)[i](2.7,-0.2)[j](3.18,0.06550218)[k](3.18,-0.2)[l]};
        % SIMPLISTIC
        \addplot+[
        boxplot prepared={
          lower whisker  =0.3388,
          lower quartile =0.5220,
          median         =0.6505,
          upper quartile =0.6848,
          upper whisker  =0.7375
        },
        ] coordinates {};
        \addplot+[scatter, only marks, scatter src=explicit symbolic,
        ] coordinates {(3.75,0.4995587)[a](3.7,0.6909993)[b](3.8,0.6619018)[c](3.9,0.6785119)[d](4,0.6458853)[e](4.15,0.5652455)[f](4.25,0.5386740)[g](4.1,0.5053038)[h](4.3,0.6550633)[i](3.8,0.7375385)[j](4.2,0.3388430)[k](4.1,0.7373030)[l]};
        \path[draw] (-50, 0) rectangle (450,20);
      \end{axis}
      \end{tikzpicture}}};
    \node at (11,-7.85) {\scalebox{0.7}{\begin{tikzpicture}[x=1pt,y=1pt]
      \begin{axis}
        [
        boxplot/draw direction=y,
        ymin=-0.2,ymax=1,
        ytick={-0.2,0,0.2,0.4,0.6,0.8,1},
        yticklabels={NA,0,0.2,0.4,0.6,0.8,1},
        xtick={1,2,3,4},
        xticklabels={\vphantom{\scalebox{1.3}{$^\star$}}\textit{FlexRL}, \vphantom{\scalebox{1.3}{$^\star$}}\textit{Exchanger}, \vphantom{\scalebox{1.3}{$^\star$}}\textit{BRL}, \vphantom{\scalebox{1.3}{$^\star$}}Simplistic},
        cycle list={{black},{black},{black},{black}},
        scatter/classes={
        a={mark=o, black},
        b={mark=otimes*, drawcolor},
        c={mark=star, drawcolor},
        d={mark=square, drawcolor},
        e={mark=square*, drawcolor},
        f={mark=triangle, black},
        g={mark=triangle*, black},
        h={mark=diamond, black},
        i={mark=diamond*, drawcolor},
        j={mark=pentagon, drawcolor},
        k={mark=pentagon*, black},
        l={mark=10-pointed star, drawcolor}
        }
        % scatter/classes={
        % a={mark=o},
        % b={mark=otimes*, black},
        % c={mark=star},
        % d={mark=square},
        % e={mark=square*, black},
        % f={mark=triangle},
        % g={mark=triangle*, black},
        % h={mark=diamond},
        % i={mark=diamond*, black},
        % j={mark=pentagon},
        % k={mark=pentagon*, black},
        % l={mark=10-pointed star}
        % % bajcdelhfkgi
        % }
        ]
        % FLEXRL
        \addplot+[
        boxplot prepared={
          lower whisker  =0.03045685,
          lower quartile =0.13023745,
          median         =0.18985042,
          upper quartile =0.37576387,
          upper whisker  =0.64759036
        },
        ] coordinates {};
        \addplot+[scatter, only marks, scatter src=explicit symbolic,
        ] coordinates {(0.8,0.413153456998314)[a](0.8,0.0765765765765766)[b](0.9,0.20863309352518)[c](0.9,0.0304568527918782)[d](1,0.162790697674419)[e](1.1,0.317460317460317)[f](1.2,0.338374291115312)[g](1.1,0.459662288930582)[h](1.2,0.15695067264574)[i](0.8,0.171067738231917)[j](1.15,0.647590361445783)[k](1.05,0.10352422907489)[l]};
        % EXCHANGER
        \addplot+[
        boxplot prepared={
          lower whisker  =0.003444317,
          lower quartile =0.009556429,
          median         =0.023989710,
          upper quartile =0.064795638,
          upper whisker  =0.067542214
        },
        ] coordinates {};
        \addplot+[scatter, only marks, scatter src=explicit symbolic,
        ] coordinates {(1.8,0.0252951096121417)[a](1.82,-0.2)[b](1.9,0.00896057347670251)[c](2.05,0.0101522842639594)[d](1.94,-0.2)[e](2.25,0.062049062049062)[f](2.3,0.0226843100189036)[g](2.15,0.0675422138836773)[h](2.18,-0.2)[i](1.7,0.0034443168771527)[j](2,0.484939759036145)[k](2.06,-0.2)[l]};
        % BRL
        \addplot+[
        boxplot prepared={
          lower whisker  =0.3217893,
          lower quartile =0.3217893,
          median         =0.3894044,
          upper quartile =0.4577861,
          upper whisker  =0.6445783
        },
        ] coordinates {};
        \addplot+[scatter, only marks, scatter src=explicit symbolic,
        ] coordinates {(2.8,0.406408094435076)[a](2.82,-0.2)[b](2.94,-0.2)[c](3.06,-0.2)[d](3,0.0255813953488372)[e](3.075,0.321789321789322)[f](3.2,0.372400756143667)[g](3,0.457786116322702)[h](3.30,-0.2)[i](2.7,-0.2)[j](3.2,0.644578313253012)[k](3.18,-0.2)[l]};
        % SENSITIVITY
        \addplot+[
        boxplot prepared={
          lower whisker  =0.9470699,
          lower quartile =0.9631660,
          median         =0.9756971,
          upper quartile =0.9808387,
          upper whisker  =0.9911894
        },
        ] coordinates {};
        \addplot+[scatter, only marks, scatter src=explicit symbolic,
        ] coordinates {(3.625,0.956155143338954)[a](3.55,0.981981981981982)[b](3.8,0.974910394265233)[c](3.9,0.979695431472081)[d](4.02,0.990697674418605)[e](4.24,0.971139971139971)[f](4.4,0.947069943289225)[g](4.14,0.962476547842402)[h](4.45,0.976483762597984)[i](3.7,0.979334098737084)[j](4.34,0.963855421686747)[k](4.14,0.991189427312775)[l]};
        \path[draw] (-50, 0) rectangle (450,20);
      \end{axis}
      \end{tikzpicture}}};
    \end{tikzpicture}
    \caption{Boxplots showing the variability of the F1-score, the FDR and the sensitivity of the compared methods on regional subsets of the NLTCS data sets: Boston, New York, Philadelphia, Detroit, Chicago, Kansas city, Seattle, Charlotte, Atlanta, Dallas, Denver, Los Angeles. \\ \textit{Exchanger} failed to link records in 4 regional offices and \textit{BRL} failed in 6; these points are defined as `NA' (Not Available) on the boxplots.
}
    \label{fig: Regional Boxplots}
\end{figure}

The simplistic approach performance enables us to evaluate the difficulty of the task. As expected it has a high FDR, due to numerous pairs for which all the PIVs match although the records do not concern the same people and in contrast, a high sensitivity, since most of the links have their PIVs matching. 

\textit{FlexRL} is as good as \textit{BRL} for those regions for which BRL finds links, we observe that the corresponding symbols are on the same height. \textit{FlexRL} still renders results for the other more difficult regions.

There is a high proportion of agreements in the pairs linked by \textit{BRL}, the method links sensible pairs (few $FP$, low FDR). \textit{Exchanger} is less conservative and links more uncertain pairs, to the detriment of its performance. \textit{FlexRL} is positioned between those two methods and capture more links than \textit{BRL} while making more sensible choices than \textit{Exchanger}.

The same conclusions can be drawn from the SHIW data sets available in the supplementary material.

\subsubsection{Performance on the complete data sets}

On the complete data sets, we use the PIVs described in \cref{tab_NLTCS_story} including the regional code. However, there is a strong association between state and regional codes due to their hierarchical relationship. We discuss the impact of dependencies among PIVs on the linkage in Appendix \cref{app3: assumptions deviations} and recommend to merge pairwise highly associated PIVs (see \cref{tab_extra_dependencies} and \cref{tab_NLTCSFull_results_dependencies}). Accordingly, we present results for a scenario where both codes are combined as a single PIV. The task is not straightforward as we can judge based on the number of $FP$ detected by the simplistic approach in \cref{tab_NLTCSFull_results}. On our machine, \textit{BRL} and \textit{Exchanger} encountered memory errors when running on those large data sets.

\begin{table}[!ht]
    \centering
    \begin{tabular}{ccccccc}
      \hline
      \noalign{\vskip\doublerulesep \vskip-\arrayrulewidth}
      \multicolumn{1}{c}{\textbf{Methods}} & \multicolumn{2}{c}{\textbf{Linked Records}} & \multirow{2}{*}{\textbf{FN}} & \multirow{2}{*}{\textbf{F1-Score}} & \multirow{2}{*}{\textbf{FDR}} & \multirow{2}{*}{\textbf{Sensitivity}} \\
      \cline{2-3}
      \noalign{\vskip\doublerulesep \vskip-\arrayrulewidth}
      \multicolumn{1}{c}{} & \textbf{TP} & \textbf{FP} & & & & \\
      \hline
      \noalign{\vskip\doublerulesep \vskip-\arrayrulewidth}
      \textbf{\textit{FlexRL} (0.5)} & $1611$ & $198$ & $5831$ & $.35$ & $.11$ & $.22$ \\
      \textbf{Simplistic approach} & $6734$ & $12069$ & $878$ & $.51$ & $.64$ & $.88$ \\
      \hline
      \noalign{\vskip\doublerulesep \vskip-\arrayrulewidth}
      \textbf{\textit{FlexRL} (0.6)} & $1209$ & $123$ & $6233$ & $.28$ & $.09$ & $.16$ \\
      \textbf{\textit{FlexRL} (0.7)} & $842$ & $76$ & $6600$ & $.20$ & $.08$ & $.11$ \\
      \textbf{\textit{FlexRL} (0.8)} & $511$ & $33$ & $6931$ & $.13$ & $.06$ & $.07$ \\
      \textbf{\textit{FlexRL} (0.9)} & $157$ & $12$ & $7285$ & $.04$ & $.07$ & $.02$ \\
      \hline
      \noalign{\vskip\doublerulesep \vskip-\arrayrulewidth}
    \end{tabular}
    \caption{Performance of FlexRL on the complete NLTCS data for several linkage probability thresholds (larger than 0.5 to ensure a one-to-one assignment) and of the simplistic approach.}
    \label{tab_NLTCSFull_results}
\end{table}

The simplistic approach is run on all PIVs as usual as it is not affected by dependencies among PIVs. On the other hand, \textit{FlexRL} is affected and has a high FDR ($0.49$) when considering both state and regional codes without regard for their hierarchical relationship, although it remains lower than the simplistic approach FDR---see Appendix \cref{tab_NLTCSFull_results_dependencies}. Excluding the regional code from the PIVs leads to a better set of linked records in term of FDR and merging the associated PIVs as presented in \cref{tab_NLTCSFull_results} appears to be a good solution as it increases the $TP$ detected in comparison, without increasing the FDR level.

As mentioned earlier we built a final set of linked records based on a probability threshold $\xi$ at $0.5$, though a natural way to select a final set of linked records could be to choose a threshold such that the estimated FDR (which may differ from the actual one) would be inferior to a certain level. We show on \cref{tab_NLTCSFull_results} the influence of the linkage probabilities threshold on the confusion matrix and the performance metrics. The task is harder in the SHIW application where the FDR levels obtained are higher.

This study demonstrates the scalability of \textit{FlexRL} on large data sets and its efficiency in a real setting where the record linkage task is not easy. \textit{FlexRL} achieves a good FDR level, around $10\%$, and the estimated FDR is unbiased. In the case where the state and the regional codes are used as PIVs without regard for their association, \textit{FlexRL} overestimates the proportion of links, leading to an underestimation of the FDR, while the actual FDR is higher than what we present here.

One may judge the level of difficulty of the record linkage task using the description of the data in \cref{tab_NLTCS_story}, and the amount of $FP$ detected by the simplistic approach. Moreover, we may characterise the level of distortion of the data using the sum of medians of disagreements and missing values among the PIVs of true links. The natural level of distortion of the NLTCS data is low ($0.2\%$). In the supplementary material we artificially distorted the data by changing and removing some values in the PIVs to create registration errors. We observed the evolution of the performance metrics of \textit{FlexRL} compared to the simplistic approach under increasing distortion levels. Both approaches have their performance decreasing with the increasing distortion, however we noticed that the decrease in performance is more controlled for \textit{FlexRL}. This result is consistent with the robustness of the method to the data quality, as stated in the simulation study---see \cref{subsubsec611: results}.

\section{Conclusion and Discussion}
\label{sec7: conclusion and discussion}

%innovations
Our paper introduced a novel approach to record linkage, using a Stochastic Expectation Maximisation on a latent variable model to combine records spread over two files without a unique identifier. By modelling dynamics of the PIVs we provided an accurate way to handle unstable PIVs, which offers an alternative to the usual blocking strategies employed to connect records together. This is particularly critical for survival analysis, where handling unstable PIVs is paramount for the long-term monitoring. We demonstrated the scalability of the method on real large data sets, facilitated by the low memory footprint of the developed algorithm. Furthermore, we asserted the robustness of the method to the quality of the linkage variables and the improved performance of the estimator in complex undertaking as one can encounter in healthcare data. The innovations discussed herein characterise the flexibility of our methodology, which adapts to diverse data complexities.

%analysis
Our analysis revealed that accommodating the instability of PIVs maximises the amount of correctly linked records, which holds importance for later inference on the linked records. PIVs dynamics may be taken into account as registration errors, though our novelty allows to detect more true positives, and in particular more links which are not detected otherwise due to changes of the information over time. We advise modelling the instability of the PIVs when there exist sensible explanatory variables for that. In that case, an external model could easily be incorporated in our methodology as an alternative to the survival function we used in our approach. It is important to note that the parameters monitoring PIVs dynamics and errors may not be jointly identifiable and one may have to decide between modelling the instability or the error processes of an unstable PIV. 

%results
Our approach performs particularly well in weak scenarios characterised by categorical data with few unique values, missing data and mistakes in the registrations. It stands between traditional methods and graphical entity resolution models, being less conservative than the former but less audacious than the latter, thereby building more sensible links. The real data applications showcased the scalability of our method, demonstrating that it can be applied to large data sets without the need for substantial computational resources. This scalability is achieved without compromising on unrealistic assumptions nor reduction of information. It is worth noting that some assumptions may be violated in real settings, and more conservative approaches may be preferred sometimes. The dependencies among PIVs, the registration errors, the instability of certain PIVs, the unknown size of the overlapping set of records, as well as the potential false positives due to weak PIVs inducing too similar characteristics, may have a strong impact on the process.

%further
Further research could be focused on adapting the method to more nuanced error modelling by considering non-equally probable mistakes and by including character type data to distinguish between substantive mistakes and typographical errors. Incorporating pairwise associations between PIVs, such as marital status and age, could also refine the model. A Dirichlet prior in a Bayesian setting could be used to model such associations, \citep{tancredi_bay_RL_2011}. Otherwise, merging the values of the associated PIVs into a single variable to be used instead appears as a straightforward solution. In addition, the linkage model could easily be extended to incorporate more knowledge for later inference or to handle time-to-event data. Finally, our Stochastic Expectation Maximisation approach could be adapted to the Bayesian framework using empirical Bayes, \citep{emp_bayes_casella}, which may allow for more robust linkage estimation, especially with small sample size, while maintaining the flexibility of our methodology.

%%%%%%%%%%%%%%

\begin{appendices}

\section{Rows sums independence}\label{app1: rows sums independence}

\textbf{This appendix refers to the linkage matrix model---see \cref{sec4: linkage}}

\subsubsection*{Concerns}

When applying record linkage, the size of the data sets $\mathcal{A}$ and $\mathcal{B}$ and the size of the overlapping set of links are realisation of some random variables. Indeed, there is a random sampling process specific to each research topic requiring a record linkage procedure, in which individuals have a certain probability to enter study $\mathcal{A}$ and study $\mathcal{B}$. Thus, while the linkage statuses of pairs of records are independent of each other, they may not be when conditioning on the size of the sets. We need to investigate that, for those realisations we are given in practice, the independence assumption still holds. 

\subsubsection*{Case study}

Let us consider a toy example where we `capture' units from a population $\mathcal{A}$ and wonder if they are registered in $\mathcal{B}$. There may be dependencies between captures conditionally on the random size of file $\mathcal{B}$. To study these dependencies, we introduce some notations: \smash{$C_k \coloneqq \ind{\big\{\text{the $k^{th}$ capture from $\mathcal{A}$ is in $\mathcal{B}$}\big\}}$}, \smash{$\widetilde{C}_k \coloneqq \sum_{\ell=1}^k C_{\ell}$} is the count of captured units that are in $\mathcal{B}$ after $k$ captures, \smash{$\gamma \coloneqq \mathbb{P}(C_k = 1)$} is the prior probability to capture a unit from $\mathcal{A}$ that is registered in $\mathcal{B}$, and $N^{\mathcal{B}}$ is the random size of $\mathcal{B}$.

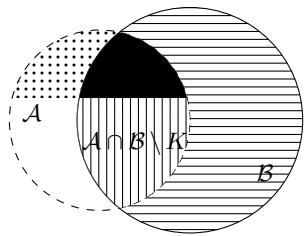
\begin{figure}[!h]
    \centering
    \begin{tikzpicture}
        \draw [dashed] (-0.5,0) circle (1.2cm); % BLOB
        \node at (-1.4,0.1) {$\mathcal{A}$};
        \draw [] (0.7,0) circle (1.5cm); % BLOB
        \node at (1.7,-0.7) {$\mathcal{B}$};
        \fill[pattern=horizontal lines] (0.7,0) circle (1.5cm); 
        \begin{scope}
            \clip (-0.5,0) circle (1.2cm);
            \fill[pattern=dots] (-2,0.3) rectangle (1,1.5);
        \end{scope}
        \begin{scope}
            \clip (-0.5,0) circle (1.2cm);
            \clip (0.7,0) circle (1.5cm);
            \fill[white] (-0.5,0) circle (1.2cm);
            \node at (-0.03,-0.3) {$\mathcal{A} \cap \mathcal{B} \, \backslash \, K$};
        \end{scope}
        \begin{scope}
            \clip (-0.5,0) circle (1.2cm);
            \clip (0.7,0) circle (1.5cm);
            \fill[] (-.8,0.3) rectangle (.8,1.5);
            \fill[pattern=vertical lines] (-.8,0.3) rectangle (.8,-1.5);
            \draw (0.7,0) circle (1.5cm);
        \end{scope}
    \end{tikzpicture}
    \caption{Illustration of the context.}
    \label{fig_rarediseasevscold}
\end{figure}
% north west lines

We distinguish various populations on \cref{fig_rarediseasevscold} representing sets $\mathcal{A}$ and $\mathcal{B}$ overlapping. At some point of the process we captured $k$ units gathered in the area \smash{$K \subset \mathcal{A}$} (upper minor segment of the set $\mathcal{A}$) composed of \smash{$\widetilde{C}_k$} units that were registered in $\mathcal{B}$ (fully coloured area \smash{$K \cap \mathcal{B}$}) and \smash{$k - \widetilde{C}_k$} other units (dotted area \smash{$K \, \backslash \, \mathcal{B}$}). There are \smash{$N^{\mathcal{A} \cap \mathcal{B} \, \backslash \, K}$} interesting units not yet captured (vertically striped area \smash{$\mathcal{A} \cap \mathcal{B} \, \backslash \, K$}) and \smash{$N^{\mathcal{B} \, \backslash \, \mathcal{A}}$} other units (horizontally striped area \smash{$\mathcal{B} \, \backslash \, \mathcal{A}$}).

To explore how likely those captures are to be independent when conditioning on the random size of $\mathcal{B}$ we study how comparable are \smash{$\mathbb{P}(C_k = 1 \mid \widetilde{C}_{k-1} = c, N^{\mathcal{B}} = \nB)$} and \smash{$\mathbb{P}(C_k = 1)$}. \begin{align} \label{eq: indepPoiss}
    \mathbb{P}(C_k = 1 \mid \widetilde{C}_{k-1} = c, N^{\mathcal{B}} = \nB) = \: & \frac{\mathbb{P}(C_k = 1, \widetilde{C}_{k-1} = c, N^{\mathcal{B}} = \nB)}{\mathbb{P}(\widetilde{C}_{k-1} = c, N^{\mathcal{B}} = \nB)}\\
    = \: & \frac{\mathbb{P}(N^{\mathcal{B}} = \nB \mid C_k = 1, \widetilde{C}_{k-1} = c) \cdot \mathbb{P}(C_k = 1, \widetilde{C}_{k-1} = c)}{\mathbb{P}(N^{\mathcal{B}} = \nB \mid \widetilde{C}_{k-1} = c) \cdot \mathbb{P}(\widetilde{C}_{k-1} = c)} \nonumber\\
    = \: & \frac{\mathbb{P}(N^{\mathcal{B}} = \nB \mid C_k = 1, \widetilde{C}_{k-1} = c)}{\mathbb{P}(N^{\mathcal{B}} = \nB \mid \widetilde{C}_{k-1} = c)} \cdot \mathbb{P}(C_k = 1) \text{ since } C_k \indep \widetilde{C}_{k-1} \nonumber\\
    = \: & \frac{\mathbb{P}(N^{\mathcal{B}} = \nB \mid \widetilde{C}_{k} = c+1)}{\mathbb{P}(N^{\mathcal{B}} = \nB \mid \widetilde{C}_{k-1} = c)} \cdot \gamma \nonumber\\
    = \: & \frac{ \sum_{\ell = 0}^{\nB - (c+1)} \mathbb{P}(N^{\mathcal{A} \cap \mathcal{B} \, \backslash \, K) = \ell} \cdot \mathbb{P}(N^{\mathcal{B} \, \backslash \, \mathcal{A}} = \nB - (c+1) - \ell) }{ \sum_{\ell = 0}^{\nB - c} \mathbb{P}(N^{\mathcal{A} \cap \mathcal{B} \, \backslash \, K-1} = \ell) \cdot \mathbb{P}(N^{\mathcal{B} \, \backslash \, \mathcal{A}} = \nB - c - \ell) } \cdot \gamma.
\end{align}

To obtain the last line, we note that \smash{$\mathbb{P}(N^{\mathcal{B}} = \nB \mid \widetilde{C}_{k} = c+1) = \mathbb{P}( N^{\mathcal{B} \, \backslash \, \mathcal{A}} + N^{\mathcal{A} \cap \mathcal{B} \, \backslash \, K} = \nB - (c+1) )$} since \smash{$N^{\mathcal{B}} = N^{\mathcal{B} \, \backslash \, \mathcal{A}} + \widetilde{C}_k + N^{\mathcal{A} \cap \mathcal{B} \, \backslash \, K}$}, which in turn is equal to \smash{$\sum_{\ell = 0}^{\nB - (c+1)} \mathbb{P}(N^{\mathcal{A} \cap \mathcal{B} \, \backslash \, K} = \ell) \cdot \mathbb{P}(N^{\mathcal{B} \, \backslash \, \mathcal{A}} = \nB - (c+1) - \ell)$} by discrete convolution. 

% Therefore we have: \begin{align} \label{eq: indep}
%     \mathbb{P}(C_k = 1 \mid \widetilde{C}_{k-1} = c, N^{\mathcal{B}} = \nB) = \: & \frac{\mathbb{P}(N^{\mathcal{B}} = \nB \mid \widetilde{C}_{k} = c+1)}{\mathbb{P}(N^{\mathcal{B}} = \nB \mid \widetilde{C}_{k-1} = c)} \cdot \gamma\\
%     = \: & \frac{ \sum_{\ell = 0}^{\nB - (c+1)} \mathbb{P}(N^{\mathcal{A} \cap \mathcal{B} \, \backslash \, K) = \ell} \cdot \mathbb{P}(N^{\mathcal{B} \, \backslash \, \mathcal{A}} = \nB - (c+1) - \ell) }{ \sum_{\ell = 0}^{\nB - c} \mathbb{P}(N^{\mathcal{A} \cap \mathcal{B} \, \backslash \, K-1} = \ell) \cdot \mathbb{P}(N^{\mathcal{B} \, \backslash \, \mathcal{A}} = \nB - c - \ell) } \cdot \gamma. \: \: \: \nonumber
% \end{align}

\subsubsection*{Sensitivity analysis}

We examine the sensitivity of the ratio of probabilities in \cref{eq: indepPoiss} to the random size of the sets by modelling the captures mechanism. For one realisation $\nA = 200$ and different realisations $\nB$ such that $\nB \geq \nA$ we simulate $N^{\mathcal{A} \cap \mathcal{B} \, \backslash \, K} \sim \text{Bin}(\nA - k, 0.5)$, $N^{\mathcal{A} \cap \mathcal{B} \, \backslash \, K-1} \sim \text{Bin}(\nA - (k-1), 0.5)$ and $N^{\mathcal{B} \, \backslash \, \mathcal{A}} \sim \text{Poisson}(\nB)$.

\begin{figure}[!h]
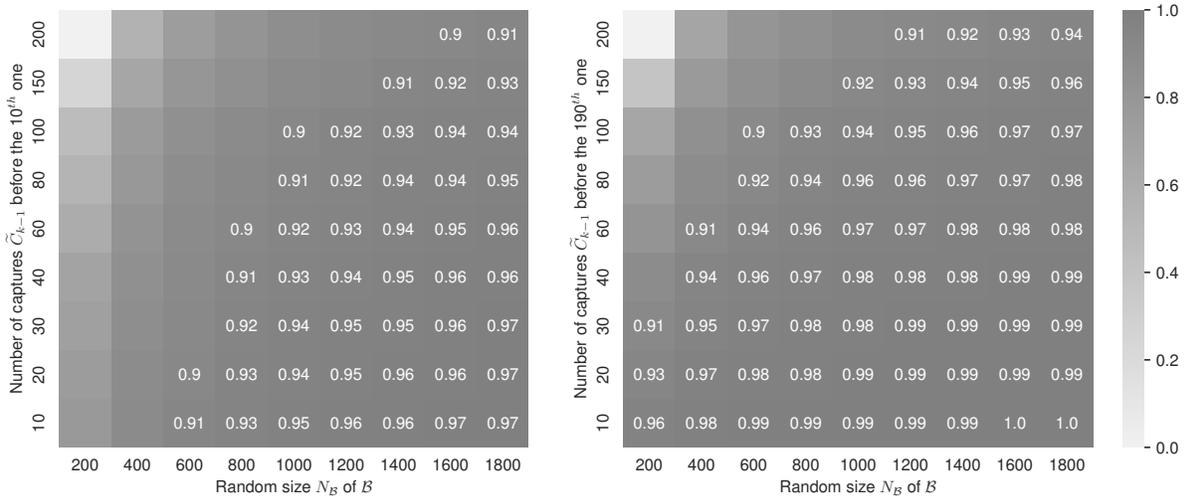

    \centering
    \begin{tikzpicture}
    \node at (0,0) {\scalebox{0.62}{\input{Figures/FigAppendix_Sensitivity_Indep_CaptureRecapture/10th_capture_recapture}}};
    \node at (7.5,0) {\scalebox{0.62}{\input{Figures/FigAppendix_Sensitivity_Indep_CaptureRecapture/190th_capture_recapture}}};
    % \scalebox{0.65}{\input{Figures/FigAppendix Sensitivity Indep CaptureRecapture/10th capture recapture}}\scalebox{0.65}{\input{Figures/FigAppendix Sensitivity Indep CaptureRecapture/190th capture recapture}}
    \end{tikzpicture}
    \caption{Values of the ratio of probabilities \smash{${\mathbb{P}(N^{\mathcal{B}} = \nB \mid \widetilde{C}_{k} = c+1)}/{\mathbb{P}(N^{\mathcal{B}} = \nB \mid \widetilde{C}_{k-1} = c)}$} from \cref{eq: indepPoiss} for different realisations $c$ (y-axis) and $\nB$ (x-axis) at two time points of the capture process $k = 5\% \cdot \nA = 10$ (left) and $k = 95\% \cdot \nA = 190$ (right) with $\nA = 200$. We only annotated cases in which the ratio is superior to $0.90$ to highlight the situations in which \smash{$\mathbb{P}(C_k = 1 \mid \widetilde{C}_{k-1} = c, N^{\mathcal{B}} = \nB) \approx \mathbb{P}(C_k = 1)$}.}
    \label{fig:sens_analysis}
\end{figure}

All in all the independence assumption of the rows sums configuration of $\boldsymbol{\Delta}$ is reasonable for scenarios where data set $\mathcal{B}$ is large relative to the set of units shared with data set $\mathcal{A}$. This corresponds to situations where the set of candidates in $\mathcal{B}$ to form a link with remains large enough along the sequential record linkage process. Specifically, in the above simulation, as long as the potential overlapping set between $\mathcal{A}$ and $\mathcal{B}$ is smaller than 20\% of file $\mathcal{B}$ (i.e.\ is 80\% smaller than $\mathcal{B}$), the ratio is larger than 0.90 so that the conditional and unconditional probabilities of a link are similar.

To assess the sensitivity of our method to such assumption, we applied our record linkage method on extreme scenarios where both data sets $\mathcal{A}$ and $\mathcal{B}$ have comparable size with almost all records shared. The results show no impact of the theoretical lack of independence among rows sums configuration on the performance of the model.

\section{Update of the latent variables}\label{app2.0: update of the latent variables}

\textbf{This appendix refers to the E-step of the StEM algorithm---see \cref{subsec51: stE-step}}

\subsubsection*{True values for non-linked records} \begin{align} \label{model:h|g}
    & \mathbb{P} \Big( \HAik(v,z) = \hAik \bigm| \GAik = \gAik, \sum_{j=1}^{\nB} \Delta_{i,j}(v,z-1)=0 ; \boldsymbol{\theta}(v-1) \Big)\\
    = \: & \frac{\mathbb{P} \Big( \HAik(v,z) = \hAik, \GAik = \gAik \bigm| \sum_{j=1}^{\nB} \Delta_{i,j}(v,z-1)=0 ; \boldsymbol{\theta}(t-1) \Big)}{\mathbb{P} \Big( \GAik = \gAik \bigm| \sum_{j=1}^{\nB} \Delta_{i,j}(v,z-1)=0 ; \boldsymbol{\theta}(v-1) \Big)} \nonumber \\
    = \: & \frac{\mathbb{P} \Big( \GAik = \gAik \bigm| \HAik(v,z) = \hAik ; \boldsymbol{\theta}(v-1) \Big) \cdot \mathbb{P} \Big( \HAik(v,z) = \hAik ; \boldsymbol{\theta}(v-1) \Big)}{\sum_{\hAik} \mathbb{P} \Big( \GAik = \gAik \bigm| \HAik(v,z) = \hAik ; \boldsymbol{\theta}(v-1) \Big) \cdot \mathbb{P} \Big( \HAik(v,z) = \hAik ; \boldsymbol{\theta}(v-1) \Big)} \text{ since $\textbf{G} \indep \boldsymbol{\Delta} \mid \textbf{H} $} \nonumber,
\end{align} and similarly for $\mathbb{P} \big( \HBjk(v,z) = \hBjk \bigm| \GBjk = \gAik, \sum_{j=1}^{\nB} \Delta_{i,j}(v,z-1)=0 ; \boldsymbol{\theta}(v-1) \big)$.  

\subsubsection*{True values for linked records} \begin{align*}
    & \mathbb{P} \Big( \HAik(v,z) = \hAik, \HBjk(v,z) = \hBjk \bigm| \GAik = \gAik, \GBjk = \gBjk, {t}_{i,j}, \Delta_{i,j}(v,z-1) = 1 ; \boldsymbol{\theta}(v-1) \Big)\\
    = \: & \frac{\mathbb{P} \Big( \HAik(v,z) = \hAik, \HBjk(v,z) = \hBjk, \GAik = \gAik, \GBjk = \gBjk \bigm| {t}_{i,j}, \Delta_{i,j}(v,z-1) = 1 ; \boldsymbol{\theta}(v-1) \Big)}{\mathbb{P} \Big( \GAik = \gAik, \GBjk = \gBjk \bigm| {t}_{i,j}, \Delta_{i,j}(v,z-1) = 1 ; \boldsymbol{\theta}(v-1) \Big)} \\
    = \: & \frac{\mathbb{P} \Big( \HAik(v,z) = \hAik, \HBjk(v,z) = \hBjk, \GAik = \gAik, \GBjk = \gBjk \bigm| {t}_{i,j}, \Delta_{i,j}(v,z-1) = 1 ; \boldsymbol{\theta}(v-1) \Big)}{\sum_{\hAik,\hBjk} \mathbb{P} \Big( \HAik(v,z) = \hAik, \HBjk(v,z) = \hBjk, \GAik = \gAik, \GBjk = \gBjk \bigm| {t}_{i,j}, \Delta_{i,j}(v,z-1) = 1 ; \boldsymbol{\theta}(v-1) \Big)}
\end{align*} We decompose the joint probability of the numerator into: \begin{align} \label{model:pivs|delta}
    & \mathbb{P} \Big( \HAik(v,z) = \hAik, \HBjk(v,z) = \hBjk, \GAik = \gAik, \GBjk = \gBjk \bigm| {t}_{i,j}, \Delta_{i,j}(v,z-1) = 1 ; \boldsymbol{\theta}(v-1) \Big)\\
    = \: & \mathbb{P} \Big( \GAik = \gAik, \GBjk = \gBjk \bigm| \HAik(v,z) = \hAik, \HBjk(v,z) = \hBjk, \Delta_{i,j}(v,z-1) = 1 ; \boldsymbol{\theta}(v-1) \Big) \nonumber \\
    & \qquad \qquad \qquad \qquad \cdot \mathbb{P} \Big( \HAik(v,z) = \hAik, \HBjk(v,z) = \hBjk \bigm| {t}_{i,j}, \Delta_{i,j}(v,z-1) = 1 ; \boldsymbol{\theta}(v-1) \Big) \nonumber \\
    = \: & \mathbb{P} \Big( \GAik = \gAik \bigm| \HAik(v,z) = \hAik ; \boldsymbol{\theta}(v-1) \Big) \nonumber \cdot \mathbb{P} \Big( \GBjk = \gBjk \bigm| \HBjk(v,z) = \hBjk ; \boldsymbol{\theta}(v-1) \Big) \nonumber\\
    & \qquad \qquad \qquad \qquad \cdot \mathbb{P} \Big( \HAik(v,z) = \hAik, \HBjk(v,z) = \hBjk \bigm| {t}_{i,j}, \Delta_{i,j}(v,z-1) = 1 ; \boldsymbol{\theta}(v-1) \Big) \nonumber.
\end{align} 

\subsubsection*{Linkage indicators} The linkage matrix is updated sequentially so that we give an explicit formula to update value $\Delta_{i,j}(v,z)$ given the elements of the matrix updated so far $$\big\{ \Delta_{1,1}(v,z), \dots, \Delta_{1,\nB}(v,z), \dots, \Delta_{i-1,1}(v,z), \dots, \Delta_{i-1,\nB}(v,z), \Delta_{i,1}(v,z), \dots, \Delta_{i,j-1}(v,z) \big\}$$ (all precedent rows and, on the row we are focusing on, all precedent columns) and given the elements of the matrix which are not yet updated $$\big\{ \Delta_{i,j+1}(v,z-1), \dots, \Delta_{i,\nB}(v,z-1), \Delta_{i+1,1}(v,z-1), \dots, \Delta_{i+1,\nB}(v,z-1), \dots, \Delta_{\nA,1}(v,z-1), \dots, \Delta_{\nA,\nB}(v,z-1) \big\}$$ (on the row we are focusing on, all subsequent columns and, all subsequent rows). We gather those sets into $\boldsymbol{\Delta}_{-( i,j)}(v,z-1,z)$ where we highlight the iterative updating process of the linkage matrix in which precedent elements are new while subsequent ones are old, hence the dependence on $z-1$ and on $z$: \begin{align*}
    & \mathbb{P} \Big( \Delta_{i,j}(v,z) = 1 \bigm| \boldsymbol{\Delta}_{-(i,j)}(v,z-1,z), \textbf{H}_{i}^{\mathcal{A}}(v,z), \textbf{H}_{j}^{\mathcal{B}}(v,z), \textbf{G}_{i}^{\mathcal{A}}, \textbf{G}_{j}^{\mathcal{B}}, {t}_{i,j}; \boldsymbol{\theta}(v-1) \Big)\\
    = \: & \frac{\mathbb{P} \Big( \Delta_{i,j}(v,z) = 1, \boldsymbol{\Delta}_{-(i,j)}(v,z-1,z), \textbf{H}_{i}^{\mathcal{A}}(v,z), \textbf{H}_{j}^{\mathcal{B}}(v,z), \textbf{G}_{i}^{\mathcal{A}}, \textbf{G}_{j}^{\mathcal{B}}, {t}_{i,j}; \boldsymbol{\theta}(v-1) \Big)}{\mathbb{P} \Big(\boldsymbol{\Delta}_{-(i,j)}(v,z-1,z), \textbf{H}_{i}^{\mathcal{A}}(v,z), \textbf{H}_{j}^{\mathcal{B}}(v,z), \textbf{G}_{i}^{\mathcal{A}}, \textbf{G}_{j}^{\mathcal{B}}, {t}_{i,j}; \boldsymbol{\theta}(v-1) \Big)}.
\end{align*} The joint probability of the denominator can be decomposed as a sum of probabilities over the partition \smash{$\big\{\Delta_{i,j}(v,z)=0, \Delta_{i,j}(v,z)=1\big\}$}, where the joint with $\Delta_{i,j}(v,z) = 0$ derives from: \begin{align*}
    & \mathbb{P} \Big( \Delta_{i,j}(v,z) = 0, \boldsymbol{\Delta}_{-(i,j)}(v,z-1,z), \textbf{H}_{i}^{\mathcal{A}}(v,z), \textbf{H}_{j}^{\mathcal{B}}(v,z), \textbf{G}_{i}^{\mathcal{A}}, \textbf{G}_{j}^{\mathcal{B}}; \boldsymbol{\theta}(v-1) \Big)\\
    = \: & \mathbb{P} \Big( \textbf{H}_{i}^{\mathcal{A}}(v,z), \textbf{H}_{j}^{\mathcal{B}}(v,z), \textbf{G}_{i}^{\mathcal{A}}, \textbf{G}_{j}^{\mathcal{B}} \bigm| \Delta_{i,j}(v,z) = 0 ; \boldsymbol{\theta}(v-1) \Big) \cdot \mathbb{P} \Big( \Delta_{i,j}(v,z) = 0, \boldsymbol{\Delta}_{-(i,j)}(v,z-1,z) ; \boldsymbol{\theta}(v-1) \Big),
\end{align*} and in a similar fashion the joint with $\Delta_{i,j}(v,z) = 1$ derives from: \begin{align*}
    & \mathbb{P} \Big( \Delta_{i,j}(v,z) = 1, \boldsymbol{\Delta}_{-(i,j)}(v,z-1,z), \textbf{H}_{i}^{\mathcal{A}}(v,z), \textbf{H}_{j}^{\mathcal{B}}(v,z), \textbf{G}_{i}^{\mathcal{A}}, \textbf{G}_{j}^{\mathcal{B}}, {t}_{i,j}; \boldsymbol{\theta}(v-1) \Big)\\
    = \: & \mathbb{P} \Big( \textbf{H}_{i}^{\mathcal{A}}(v,z), \textbf{H}_{j}^{\mathcal{B}}(v,z), \textbf{G}_{i}^{\mathcal{A}}, \textbf{G}_{j}^{\mathcal{B}} \bigm| \Delta_{i,j}(v,z) = 1, {t}_{i,j}; \boldsymbol{\theta}(v-1) \Big) \cdot \mathbb{P} \Big( \Delta_{i,j}(v,z) = 1, \boldsymbol{\Delta}_{-(i,j)}(v,z-1,z) ; \boldsymbol{\theta}(v-1) \Big).
\end{align*} The joint distribution of registered and true values for non-linked records is discussed earlier in \cref{model:h|g}, the joint distribution of registered and true values for linked records in \cref{model:pivs|delta} and, the joint distribution of linkage indicators corresponds to the model in \cref{model: delta}. 

\section{Parameters update}\label{app2: parameters update}

\textbf{This appendix refers to the M-step of the StEM algorithm---see \cref{subsec52: M-step}}

\subsubsection*{Registration errors $\boldsymbol{\phi}$}

% PHI
In practice, we obtain $\phi_{k,\text{mistake}}^{\mathcal{A}}(v)$---and similarly $\phi_{k,\text{mistake}}^{\mathcal{B}}(v)$---by averaging the proportion of disagreements between registered and true values obtained in each iteration of the Gibbs sampler (excluding the missing values):
\begin{align*}
    \phi_{k,\text{mistake}}^{\mathcal{A}}(v) = \: & \underset{\phi_{k,\text{mistake}}^{\mathcal{A}}}{\text{argmax}} \, \sum_{z=Z_0+1}^{Z_0+Z_1} \, \log \mathcal{L}_{\boldsymbol{\phi}} \big( \textbf{G}^{\mathcal{A}} \bigm| \textbf{H}^{\mathcal{A}}(v,z) \big)\\
    = \: & \frac{1}{Z_1} \sum_{z=Z_0+1}^{Z_0+Z_1} \frac{ \sum_{i} \ind{\{\gAik \neq 0\}} \cdot \ind{\{\gAik \neq \hAik(v,z)\}} }{\sum_{i} \ind{\{\gAik \neq 0\}} }.
\end{align*}

\subsubsection*{PIVs dynamics $\boldsymbol{\alpha}$}

% ALPHA
Computing the update $\alpha_{k}(v)$ is more complex due to the survival transformation; for an unstable PIV indexed by $k$, the optimisation problem is the following: \begin{align*}
    \alpha_{k}(v) = \: & \underset{\alpha_k}{\text{argmax}} \, \sum_{z=Z_0+1}^{Z_0+Z_1} \, \log \mathcal{L}_{\boldsymbol{\alpha}} \big( \textbf{H}^{\mathcal{B}}(v,z) \bigm| \textbf{H}^{\mathcal{A}}(v,z), \textbf{t}^{\mathcal{A}}, \textbf{t}^{\mathcal{B}}, \boldsymbol{\Delta}(v,z) \big)\\
    = \: & \underset{\alpha_{k}}{\text{argmax}} \, \sum_{z=Z_0+1}^{Z_0+Z_1} \, \sum_{i,j} \Delta_{i,j}(v,z) \cdot n_k \cdot \bigg[ \ind{\{ \hAik(v,z) \neq \hBjk(v,z) \}} \cdot \log\big[\exp\big\{ \exp(\alpha_k)\, t_{i,j} \big\} - 1 \big] - \exp(\alpha_{k})\, t_{i,j} \bigg].
\end{align*} We make use of a computational optimisation method to find a solution. In our modelling, the dynamics of an unstable PIV indexed by $k$ for a linked pair of records depend on the time elapsed between the two data collection. We could extend this model with covariates as mentioned in the \cref{remark: survival with covariates} (age, family, work opportunities, are different arguments to move for example) or use a different modelling based on external information from a national statistics institute for instance.

\subsubsection*{PIVs distribution $\boldsymbol{\eta}$}

% ETA
\begin{align*}
    \boldsymbol{\eta}_{k}(v) = \: & \underset{{\boldsymbol{\eta}}_{k}}{\text{argmax}} \, \sum_{z=Z_0+1}^{Z_0+Z_1} \, \log \mathcal{L}_{\boldsymbol{\eta}} \big( \textbf{H}^{\mathcal{A}}(v,z) \big) \text{ subject to } \sum_\ell \eta_{k,\ell}(v) = 1\\
    \eta_{k,\ell}(v) = \: & \frac{ \splitfrac{ \sum_z \big( \sum_{i, j} \Delta_{i,j}(v,z) \cdot \ind{\{\ell=\hAik(v,z)=\hBjk(v,z)\}} }{ \splitfrac{ \qquad \qquad + \sum_{i} \big( 1 - \sum_{j} \Delta_{i,j}(v,z) \big) \cdot \ind{\{\hAik(v,z) = \ell\}} }{ \qquad \qquad \qquad \qquad + \sum_{j} \big( 1 - \sum_{i} \Delta_{i,j}(v,z) \big) \cdot \ind{\{\hBjk(v,z) = \ell\}} \big) }}}{{ \splitfrac{ \sum_\ell \sum_z \big( \sum_{i, j} \Delta_{i,j}(v,z) \cdot \ind{\{\ell=\hAik(v,z)=\hBjk(v,z)\}} }{ \qquad \qquad \splitfrac{ + \sum_{i} \big( 1 - \sum_{j} \Delta_{i,j}(v,z) \big) \cdot \ind{\{\hAik(v,z)=\ell\}} }{ \qquad \qquad \qquad + \sum_{j} \big( 1 - \sum_{i} \Delta_{i,j}(v,z) \big) \cdot \ind{\{\hBjk(v,z)=\ell\}} \big) }}}}.
\end{align*}
The update of each coordinate $\eta_{k,\ell}(v)$ of $\boldsymbol{\eta}_{k}(v)$ is obtained by constrained maximum likelihood thanks to a basic Lagrangian optimisation.

\subsubsection*{Links proportion $\gamma$}

% GAMMA
We finally update $\gamma(v)$ with the average portion of links made through the iterations of the Gibbs sampler as a fraction of the smallest file:
\begin{align*}
    \gamma(v) = \underset{\gamma}{\text{argmax}} \, \sum_{z=Z_0+1}^{Z_0+Z_1} \, \log \mathcal{L}_{\gamma} \big( \boldsymbol{\Delta}(v,z) \big) = \frac{1}{Z_1} \sum_{z=Z_0+1}^{Z_0+Z_1} \frac{\sum_{i,j} \Delta_{i,j}(v,z)}{\nA}.
\end{align*}

\section{Robustness to assumptions deviations}\label{app3: assumptions deviations}

\textbf{This appendix presents sensitivity analyses for our model assumptions: independence of the PIVs, uniform weights for temporal changes in latent values and, estimability issue between instability and mistakes---see \cref{sec3: pivs}}

\subsubsection*{Dependencies among PIVs}

As a measure of association between two PIVs, we use the Cramér's V. This metric ranges from 0 to 1, from no association to perfect association. In the simulations \cref{subsec61: simulations}, we generated five independent PIVs; the Cramér's V for each pair of PIVs is approximately 0.1.

We analyse the impact of violations of the independence assumption between PIVs by introducing different levels of dependencies in one pair of PIVs. The remaining PIVs are simulated according to the initial simulation study scenario of \cref{subsec61: simulations}. The Cramér's V for the associated pair is shown on the left in \cref{tab_extra_dependencies}, while the Cramér's V for the other pairs of PIVs remained approximately 0.1. We observe a gradual decline in performance as the association level between PIVs increases, particularly when instability is not accounted for. We obtain reasonable performance, comparable to the literature methods, for moderated association levels (Cramér's V smaller than 0.5). Unless PIVs dynamics are accounted for, we recommend combining highly associated PIVs (Cramér's V larger than 0.8) to preserve the FDR.

\cref{tab_extra_dependencies} below presents the results. The first lines recall those from the independent case (simulation study \cref{subsec61: simulations}, \cref{tab_Simu_Story_And_Results}) while the last lines show the results from combining the pair of highly associated PIVs into a single PIV.

\cref{tab_NLTCSFull_results_dependencies} presents the results obtained for the NLTCS data sets when both the state and regional codes are used as PIVs, when they are combined into a single PIV to be used instead (as done in \cref{subsec62: application}, \cref{tab_NLTCSFull_results}) and, when only the state code is used.

\subsubsection*{Non-uniform weights for changes across time}

In our model, when values of PIVs change across time, we assume that they uniformly substitute for one of the other possible values in the support. To investigate the robustness of our model to deviations from uniform substitutions, we perform the following sensitivity analysis.

For a pair of records $(i,j)$, instead of uniform substitutions we define \smash{$\mathbb{P}(\HBjk=\tilde{h} | \HAik=h, h \neq \tilde{h}, \Delta_{i,j} = 1)$}, the new distribution of the unstable $k^{\text{th}}$ PIV as \smash{$\exp\{0.25 \tilde{h}\} / \sum_{\ell=1,\ell \neq h}^{n_k} \exp\{0.25 \ell\}$}. Using uniform weights for more complex temporal changes reduces the number of linked pairs without increasing the errors made. With such a rough approximation, we loose some of the advantages of modelling dynamics, sensitivity is lower, without the FDR being affected.

\cref{tab_extra_nonunif} below show the results. The first lines recall those obtained with uniform changes in the unstable PIV (simulation study \cref{subsec61: simulations}, \cref{tab_Simu_Story_And_Results}) for comparison.

\subsubsection*{Estimability of $\boldsymbol{\phi}$ and $\boldsymbol{\alpha}$}

Both parameters for PIVs dynamics and registration errors in our model are not guaranteed to be estimable. We suggest to fix the parameter for mistake when modelling instability, which we do in the simulation study of \cref{subsec61: simulations}.

We analyse the impact of fixing the parameter that controls mistakes for the unstable PIV in order to avoid estimability issues. In the simulations of \cref{subsec61: simulations} we fix the mistakes probability to its true value, in that case: 0. However unstable PIVs may also contain errors; we therefore generate 2\% of mistakes in the $5^{\text{th}}$ PIV here and, we fix \smash{$\phi_{5,\text{mistake}}$} to different plausible levels. The parameter controlling changes over time adapts; the limiting value of the probability of changes decreases as the probability of mistakes increases, maintaining the performance stable.

The results, shown in \cref{tab_extra1}, cannot be compared with the one from the simulation study in \cref{subsec61: simulations}, \cref{tab_Simu_Story_And_Results} since the $5^{\text{th}}$ PIV is here generated with mistakes. On the left of \cref{tab_extra1} we show the fixed value of the mistakes parameter and the averaged limiting value of the parameter controlling PIVs dynamics.

\begin{table}
    \centering
    \begin{tabular}{cccccccc}
      \hline
      \noalign{\vskip\doublerulesep \vskip-\arrayrulewidth}
      \multirow{2}{*}{\textbf{Cramér's V}} & \multirow{2}{*}{\textbf{Methods}} & \multicolumn{2}{c}{\textbf{Linked Records}} & \multirow{2}{*}{\textbf{FN}} & \multirow{2}{*}{\textbf{F1-Score}} & \multirow{2}{*}{\textbf{FDR}} & \multirow{2}{*}{\textbf{Sensitivity}} \\
      \cline{3-4}
      \noalign{\vskip\doublerulesep \vskip-\arrayrulewidth}
      \multicolumn{2}{c}{} & \textbf{TP} & \textbf{FP} & & & & \\
      \hline
      \noalign{\vskip\doublerulesep \vskip-\arrayrulewidth}
      \multirow{5}{*}{$0.1$} & {\textbf{\textit{FlexRL} with instability}} & $290 (11)$ & $74 (10)$ & $209 (11)$ & $.67 (.02)$ & $.20 (.02)$ & $.58 (.02)$\\
      & {\textbf{\textit{FlexRL} all stable}} & $272 (12)$ & $72 (10)$ & $227 (12)$ & $.64 (.02)$ & $.21 (.02)$ & $.55 (.02)$\\
      & {\textbf{\textit{Exchanger}}} & $152 (9)$ & $61 (15)$ & $348 (9)$ & $.42(.03)$ & $.29 (.04)$ & $.30 (.02)$\\
      & {\textbf{\textit{BRL}}} & $203 (36)$ & $43 (14)$ & $297 (36)$ & $.54 (.07)$ & $.17 (.03)$ & $.41 (.07)$\\
      & {\textbf{Simplistic approach}} & $204 (9)$ & $110 (14)$ & $296 (9)$ & $.50 (.02)$ & $.35 (.03)$ & $.41 (.02)$\\
      \hline
      \noalign{\vskip\doublerulesep \vskip-\arrayrulewidth}
      \multirow{2}{*}{$0.2$} & \textbf{\textit{FlexRL} with instability} & $282 (14)$ & $78 (11)$ & $218 (14)$ & $.66 (.02)$ & $.22 (.02)$ & $.56 (.03)$ \\
      & \textbf{\textit{FlexRL} all stable} & $266 (15)$ & $82 (13)$ & $234 (15)$ & $.63 (.02)$ & $.23 (.03)$ & $.53 (.03)$ \\
      \multirow{2}{*}{$0.4$} & \textbf{\textit{FlexRL} with instability} & $271 (15)$ & $90 (14)$ & $229 (15)$ & $.63 (.03)$ & $.25 (.03)$ & $.54 (.03)$ \\
      & \textbf{\textit{FlexRL} all stable} & $252 (17)$ & $116 (26)$ & $248 (17)$ & $.58 (.04)$ & $.31 (.06)$ & $.50 (.03)$ \\
      \multirow{2}{*}{$0.6$} & \textbf{\textit{FlexRL} with instability} & $250 (21)$ & $107 (21)$ & $250 (21)$ & $.58 (.04)$ & $.30 (.05)$ & $.50 (.04)$ \\
      & \textbf{\textit{FlexRL} all stable} & $208 (34)$ & $173 (55)$ & $292 (34)$ & $.47 (.09)$ & $.45 (.12)$ & $.42 (.07)$ \\
      \multirow{2}{*}{$0.8$} & \textbf{\textit{FlexRL} with instability} & $220 (33)$ & $130 (32)$ & $280 (33)$ & $.52 (.07)$ & $.37 (.09)$ & $.44 (.07)$ \\
      & \textbf{\textit{FlexRL} all stable} & $133 (69)$ & $256 (96)$ & $367 (69)$ & $.30 (.17)$ & $.64 (.22)$ & $.27 (.14)$ \\
      \hline
      \noalign{\vskip\doublerulesep \vskip-\arrayrulewidth}
      \multirow{1}{*}{$>0.8$} & \textbf{\textit{FlexRL} with instability} & $150 (12)$ & $48 (9)$ & $347 (12)$ & $.43 (.03)$ & $.24 (.03)$ & $.30 (.02)$ \\
      \multirow{1}{*}{PIVs merged} & \textbf{\textit{FlexRL} all stable} & $139 (12)$ & $47 (10)$ & $358 (12)$ & $.41 (.103)$ & $.25 (.04)$ & $.28 (.02)$ \\
      \hline
      \noalign{\vskip\doublerulesep \vskip-\arrayrulewidth}
    \end{tabular}
    \caption{Robustness of \textit{FlexRL} (over 500 simulations) to dependencies among pairs of PIVs as assessed by the maximal Cramér's V. The first rows recall the results of \cref{tab_Simu_Story_And_Results} with independent PIVs, the subsequent rows show the results for higher levels of association, the last rows show the results when the highly associated PIVs are combined.}
    \label{tab_extra_dependencies}
\end{table}

\begin{table}
    \centering
    \begin{tabular}{cccccccc}
      \hline
      \noalign{\vskip\doublerulesep \vskip-\arrayrulewidth}
      \multirow{2}{*}{\textbf{Cramér's V}} & \multirow{2}{*}{\textbf{Methods}} & \multicolumn{2}{c}{\textbf{Linked Records}} & \multirow{2}{*}{\textbf{FN}} & \multirow{2}{*}{\textbf{F1-Score}} & \multirow{2}{*}{\textbf{FDR}} & \multirow{2}{*}{\textbf{Sensitivity}} \\
      \cline{3-4}
      \noalign{\vskip\doublerulesep \vskip-\arrayrulewidth}
      \multicolumn{2}{c}{} & \textbf{TP} & \textbf{FP} & & & & \\
      \hline
      \noalign{\vskip\doublerulesep \vskip-\arrayrulewidth}
      \multirow{1}{*}{$0.11$} & \textbf{\textit{FlexRL}} & $1611$ & $198$ & $5831$ & $.35$ & $.11$ & $.22$ \\
      \multirow{1}{*}{state + region merged} & \textbf{Simplistic approach} & $6740$ & $12091$ & $710$ & $.51$ & $.64$ & $.90$ \\
      \hline
      \noalign{\vskip\doublerulesep \vskip-\arrayrulewidth}
      \multirow{1}{*}{$0.11$} & \textbf{\textit{FlexRL}} & $1405$ & $159$ & $6207$ & $.31$ & $.10$ & $.18$ \\
      \multirow{1}{*}{state only} & \textbf{Simplistic approach} & $6896$ & $15120$ & $716$ & $.47$ & $.69$ & $.91$ \\
      \hline
      \noalign{\vskip\doublerulesep \vskip-\arrayrulewidth}
      \multirow{1}{*}{$0.92$} & \textbf{\textit{FlexRL}} & $3250$ & $3085$ & $4362$ & $.47$ & $.49$ & $.43$ \\
      \multirow{1}{*}{state and region} & \textbf{Simplistic approach} & $6812$ & $12291$ & $800$ & $.51$ & $.64$ & $.89$ \\
      \hline
      \noalign{\vskip\doublerulesep \vskip-\arrayrulewidth}
    \end{tabular}
    \caption{Performance of FlexRL and of the simplistic approach on the complete NLTCS data for different set of PIVs. First rows use state and regional codes combined as a single PIV (as presented in \cref{tab_NLTCSFull_results}), second rows use the state code only, third rows use both state and regional codes. The level of dependencies among pairs of PIVs is assessed by the maximal Cramér's V on the left. Changes in the results of the simplistic approach are due to filtering the PIVs to their common support.}
    \label{tab_NLTCSFull_results_dependencies}
\end{table}

\begin{table}
    \centering
    \begin{tabular}{cccccccc}
      \hline
      \noalign{\vskip\doublerulesep \vskip-\arrayrulewidth}
      \multirow{2}{*}{$\mathbb{P}\big(\HBjk=\tilde{h} \mid \dots \big)$} & \multirow{2}{*}{\textbf{Methods}} & \multicolumn{2}{c}{\textbf{Linked Records}} & \multirow{2}{*}{\textbf{FN}} & \multirow{2}{*}{\textbf{F1-Score}} & \multirow{2}{*}{\textbf{FDR}} & \multirow{2}{*}{\textbf{Sensitivity}} \\
      \cline{3-4}
      \noalign{\vskip\doublerulesep \vskip-\arrayrulewidth}
      \multicolumn{2}{c}{} & \textbf{TP} & \textbf{FP} & & & & \\
      \hline
      \noalign{\vskip\doublerulesep \vskip-\arrayrulewidth}
      \multirow{5}{*}{\textbf{Uniform}} & {\textbf{\textit{FlexRL} with instability}} & $290 (11)$ & $74 (10)$ & $209 (11)$ & $.67 (.02)$ & $.20 (.02)$ & $.58 (.02)$\\
      & {\textbf{\textit{FlexRL} all stable}} & $272 (12)$ & $72 (10)$ & $227 (12)$ & $.64 (.02)$ & $.21 (.02)$ & $.55 (.02)$\\
      & {\textbf{\textit{Exchanger}}} & $152 (9)$ & $61 (15)$ & $348 (9)$ & $.42(.03)$ & $.29 (.04)$ & $.30 (.02)$\\
      & {\textbf{\textit{BRL}}} & $203 (36)$ & $43 (14)$ & $297 (36)$ & $.54 (.07)$ & $.17 (.03)$ & $.41 (.07)$\\
      & {\textbf{Simplistic approach}} & $204 (9)$ & $110 (14)$ & $296 (9)$ & $.50 (.02)$ & $.35 (.03)$ & $.41 (.02)$\\
      \hline
      \noalign{\vskip\doublerulesep \vskip-\arrayrulewidth}
      \multirow{2}{*}{\textbf{Non-uniform}} & \textbf{\textit{FlexRL} with instability} & $234 (16)$ & $58 (9)$ & $266 (16)$ & $.59 (.03)$ & $.20 (.03)$ & $.47 (.03)$ \\
      & \textbf{\textit{FlexRL} all stable} & $231 (17)$ & $62 (10)$ & $269 (17)$ & $.58 (.03)$ & $.21 (.03)$ & $.46 (.03)$ \\
      \hline
      \noalign{\vskip\doublerulesep \vskip-\arrayrulewidth}
    \end{tabular}
    \caption{Robustness of \textit{FlexRL} (over 500 simulations) to non-uniformly distributed changes across time. The first rows recall the results of \cref{tab_Simu_Story_And_Results} with uniform changes in the unstable PIV, the subsequent rows show the results when actual changes are non-uniformly distributed. In the table, \smash{$\mathbb{P}\big(\HBjk=\tilde{h} \mid \dots \big)$} denotes the new distribution \smash{$\mathbb{P}(\HBjk=\tilde{h} | \HAik=h, h \neq \tilde{h}, \Delta_{i,j} = 1) = \exp\{0.25 \tilde{h}\} / \sum_{\ell=1,\ell \neq h}^{n_k} \exp\{0.25 \ell\}$}.}
    \label{tab_extra_nonunif}
\end{table}

\begin{table}
    \centering
    \begin{tabular}{ccccccc}
      \hline
      \noalign{\vskip\doublerulesep \vskip-\arrayrulewidth}
      \multirow{2}{*}{\textbf{\textit{FlexRL}}} & \multicolumn{2}{c}{\textbf{Linked Records}} & \multirow{2}{*}{\textbf{FN}} & \multirow{2}{*}{\textbf{F1-Score}} & \multirow{2}{*}{\textbf{FDR}} & \multirow{2}{*}{\textbf{Sensitivity}} \\
      \cline{2-3}
      \noalign{\vskip\doublerulesep \vskip-\arrayrulewidth}
      \multicolumn{1}{c}{} & \textbf{TP} & \textbf{FP} & & & & \\
      \hline
      \noalign{\vskip\doublerulesep \vskip-\arrayrulewidth}
      $\phi_{5,\text{mistake}} = 0.00, \exp(\hat\alpha_5) = 0.30$ & $279 (14)$ & $72 (10)$ & $220 (14)$ & $.66 (.02)$ & $.20 (.02)$ & $.56 (.03)$ \\
      $\phi_{5,\text{mistake}} = 0.02, \exp(\hat\alpha_5) = 0.24$ & $275 (13)$ & $67 (10)$ & $225 (13)$ & $.65 (.02)$ & $.20 (.02)$ & $.55 (.03)$ \\
      $\phi_{5,\text{mistake}} = 0.04, \exp(\hat\alpha_5) = 0.20$ & $269 (14)$ & $64 (8)$ & $231 (14)$ & $.65 (.02)$ & $.19 (.02)$ & $.54 (.03)$ \\
      $\phi_{5,\text{mistake}} = 0.06, \exp(\hat\alpha_5) = 0.16$ & $269 (14)$ & $64 (9)$ & $231 (14)$ & $.65 (.02)$ & $.19 (.02)$ & $.54 (.03)$ \\
      $\phi_{5,\text{mistake}} = 0.08, \exp(\hat\alpha_5) = 0.13$ & $265 (14)$ & $62 (8)$ & $235 (14)$ & $.64 (.02)$ & $.19 (.02)$ & $.53 (.03)$ \\
      \hline
      \noalign{\vskip\doublerulesep \vskip-\arrayrulewidth}
    \end{tabular}
    \caption{Robustness of \textit{FlexRL} (over 500 simulations) to a fixed parameter for mistakes \smash{$\phi_{5,\text{mistake}}=\phi_{5,\text{mistake}}^{\mathcal{A}}=\phi_{5,\text{mistake}}^{\mathcal{B}}$} while 2\% of mistakes are actually generated in the $5^{\text{th}}$ PIV. The estimation for changes through time \smash{$\exp(\hat\alpha_5)$} adapts to the fixed probability of mistakes.}
    \label{tab_extra1}
\end{table}

\end{appendices}

\section{Author contributions statement}
KR, MH developed the statistical model and algorithm, conceptualised by MH. All authors drew up the simulation framework. KR conducted the simulation study and the analysis. She prepared the first manuscript draft, which was reviewed and edited by SP, MH and MW. SP and MW provided feedback that enhanced the content included in the Appendix. All authors read and approved this manuscript.

\section{Acknowledgments}
The authors thank the anonymous reviewers for their valuable suggestions.

\section{Data availability}
The implementation of \textit{FlexRL} in \texttt{R} is available on \href{https://cran.r-project.org/web/packages/FlexRL/index.html}{CRAN}. The experiments and data sets are available on \href{https://github.com/robachowyk/FlexRL-experiments}{GitHub}. The online supporting information contains a supplementary material document with additional details and applications.

\section{Funding}
This research received no specific grant from any funding agency in the public, commercial, or not-for-profit sectors.

\section{Conflict of interest}
None declared.

%\bibliographystyle{plain}
%\bibliography{reference}

%USE THE BELOW OPTIONS IN CASE YOU NEED AUTHOR YEAR FORMAT.
\bibliographystyle{abbrvnat}
\bibliography{reference}

\end{document}